\documentclass[fleqn,usenatbib, useAMS]{mnras}
\usepackage{graphicx}
\usepackage{multicol}
\usepackage{enumerate}
\usepackage{amsmath}
\usepackage{pdflscape}
\usepackage{amssymb}
\usepackage{appendix}
\usepackage{longtable}
\usepackage[flushleft]{threeparttable}
\usepackage{journal_names}
\def\farcs{\hbox{$.\!\!^{\prime\prime}$}}
\def\flx{erg cm$^{-2}$ s$^{-1}$}
\title[Optical spectroscopy of GBS X-ray sources] {Spectroscopic classification of X-ray sources in the Galactic Bulge Survey}
\author[Wevers et al.]{T. Wevers$^{1}$\thanks{Email: t.wevers@astro.ru.nl}, M. A. P. Torres$^{1,2,3,4}$, P. G. Jonker$^{1,2}$, G. Nelemans$^{1,5}$, C. Heinke$^{6}$, \newauthor 
D. Mata S{\'a}nchez$^{3,4}$, C. B. Johnson$^{7,8}$, R. Gazer$^{6}$, D. T. H. Steeghs$^{9}$, T. J. Maccarone$^{10}$, \newauthor R. I. Hynes$^{7}$, J. Casares$^{3,4,11}$, A. Udalski$^{12}$, J. Wetuski$^7$, C. T. Britt$^{7,13}$, \newauthor Z. Kostrzewa--Rutkowska$^{1,2}$ and \L. Wyrzykowski$^{12}$ \\\\
$^{1}$Department of Astrophysics/IMAPP, Radboud University, P O  Box 9010, NL-6500 GL Nijmegen, The Netherlands\\
$^{2}$SRON, Netherlands Institute for Space Research, Sorbonnelaan 2, NL-3584 CA Utrecht, The Netherlands\\
$^{3}$Instituto de Astrof\'isica de Canarias, E-38205 La Laguna, Santa Cruz de Tenerife, Spain \\
$^{4}$Departamento de Astrof\'isica, Universidad de La Laguna, E-38206 La Laguna, Santa Cruz de Tenerife, Spain\\
$^{5}$Institute for Astronomy, KU Leuven, Celestijnenlaan 200D, B-3001 Leuven, Belgium\\
$^{6}$Dept. of Physics, University of Alberta, CCLS 4-183, Edmonton, AB T6G 2E1, Canada\\
$^{7}$Department of Physics and Astronomy, Louisiana State University, Baton Rouge, LA 70803-4001, USA\\
$^{8}$Space Telescope Science Institute, 3700 San Martin Drive, Baltimore, MD, 21218, USA\\
$^{9}$Department of Physics, University of Warwick, Coventry CV4 7AL, UK\\
$^{10}$Department of Physics and Astronomy, Texas Tech University, box 41051, Lubbock, TX 79409-1051, USA\\
$^{11}$Department of Physics, Astrophysics, University of Oxford, Denys Wilkinson Building, Keble Road, Oxford OX1 3RH, UK\\
$^{12}$Warsaw University Astronomical Observatory, Al. Ujazdowskie 4, PL-00-478 Warszawa, Poland\\
$^{13}$Department of Physics and Astronomy, Michigan State University, 5678 Wilson Road, Lansing, MI 48824, USA\\
}

\begin{document}
\date{}
\pagerange{\pageref{firstpage}--\pageref{lastpage}} \pubyear{2017}
\maketitle
\label{firstpage}

\begin{abstract}
We present the classification of 26 optical counterparts to X-ray sources discovered in the Galactic Bulge Survey. We use (time-resolved) photometric and spectroscopic observations to classify the X-ray sources based on their multi-wavelength properties. We find a variety of source classes, spanning different phases of stellar/binary evolution. We classify CX21 as a quiescent cataclysmic variable (CV) below the period gap, and CX118 as a high accretion rate (nova-like) CV. CXB12 displays excess UV emission, and could contain a compact object with a giant star companion, making it a candidate symbiotic binary or quiescent low mass X-ray binary (although other scenarios cannot be ruled out). CXB34 is a magnetic CV (polar) that shows photometric evidence for a change in accretion state. The magnetic classification is based on the detection of X-ray pulsations with a period of 81\,$\pm$\,2 min. CXB42 is identified as a young stellar object, namely a weak-lined T Tauri star exhibiting (to date unexplained) UX Ori-like photometric variability. The optical spectrum of CXB43 contains two (resolved) unidentified double-peaked emission lines. No known scenario, such as an AGN or symbiotic binary, can easily explain its characteristics. We additionally classify 20 objects as likely active stars based on optical spectroscopy, their X-ray to optical flux ratios and photometric variability. In 4 cases we identify the sources as binary stars. 
\end{abstract}

%%%%%%%%%%%%%%%%%%%%%%%%%%%%%%%%%%%%%%%%%%%%%%%%%%%%%%%%%%%%%%%%%%%%%%%%%%%%%%%%%%%%%%%%%%%%%%%%%%%%%%%%%%%%%%%%%%%%%%%%%%%%%%%%%%%%%%%%%%%%%%%%%%%%%%%%%%%%%%%%%%%%%%%%%%%%%%%%%%%%%%%%%%%%%%%%%%%%%%%%%%%%%%%%%%%%%%%%%%
%%%%%%%%%%%%%%%%%%%%%%%%%%%%%%%%%%%%%%%%%%%%%%%%%%%%%%%%%%%%%%%%%%%%%%%%%%%%%%%%%%%%%%%%%%%%%%%%%%%%%%%%%%%%%%%%%%%%%%%%%%%%%%%%%%%%%%%%%%%%%%%%%%%%%%%%%%%%%%%%%%%%%%%%%%%%%%%%%%%%%%%%%%%%%%%%%%%%%%%%%%%%%%%%%%%%%%%%%%

\section{Introduction}
\label{sec:introduction}
The study of emission lines, and in particular hydrogen lines, provides a key source of information about optical counterparts to X-ray sources.
Matter falling into the deep potential well of a compact object such as a white dwarf (WD), neutron star (NS) or black hole (BH) becomes hot due to the release of potential energy. This leads to the emission of photons at UV and/or X-ray wavelengths. Such a radiation field can ionize the hydrogen atoms in the environment, leading to an H\,$\alpha$ spectral feature at optical wavelengths. The hydrogen rich material is usually confined to an accretion flow around the compact object. H emission lines can originate from the optically thin outer part of the disk \citep{Williams1980}, or from the surface of the disk if there is temperature inversion due to irradiation by the hot central object \citep{Warner1976}. Absorption lines can potentially arise from the compact object if it is a H-rich WD, or from an optically thick accretion disk (e.g. \citealt{Kiplinger1978}). Binary star systems with a WD, NS or BH component are transient H\,$\alpha$ emitters and can produce either emission or absorption lines, depending on the specific conditions in the system. Alternatively to this mechanism, late type stars produce transient H\,$\alpha$ emission through collisional ionization of H atoms in their chromospheres. Irradiation of the donor star by the compact object may also excite H atoms and lead to H\,$\alpha$ emission \citep{Bassa2009, Rodriguez2015}. On the other hand, an H\,$\alpha$ absorption line can be observed if the (typically late type) companion star or WD in a cataclysmic variable (CV) system dominates the optical spectrum.

The properties of the H Balmer lines provide information about the system being observed. For example, the presence of single and/or double-peaked lines allow us to constrain geometrical properties (e.g. \citeauthor{Horne1986} \citeyear{Horne1986}), as well as the presence/absence of an accretion disc \citep{Schwope2000,Ratti2012} and the nature of the accreting object and/or companion star (\citeauthor{Torres2014} \citeyear{Torres2014}, \citeauthor{Casares2015} \citeyear{Casares2015}, \citeyear{Casares2016}).

The identification and study of X-ray sources strongly relies on identifying the correct UV, optical or IR counterpart. Once this counterpart is known, we can resort to multiwavelength (time-resolved) photometric and spectroscopic observations to further characterize and study the nature and evolution of the X-ray emitter.

The Galactic Bulge Survey (GBS; \citeauthor{Jonker2011} \citeyear{Jonker2011}, \citeyear{Jonker2014}) is a combined X-ray, optical and NIR survey of part of the Galactic Bulge. It avoids the very crowded regions near the Galactic Centre, which are unfavourable for identifying the optical counterparts because of the very high dust extinction (up to 30 mag in A$_V$). \newline
The main science goals of the GBS are i) to measure (in a model independent way) the masses of NSs and BHs to constrain their equation of state and formation channels, and ii) to constrain the formation and evolution of low mass X-ray binaries (LMXBs), in particular the nature and efficiency of the common envelope phase. As such the GBS survey depth was chosen to optimize the number of discovered LMXBs over CVs. Moreover, the Bulge was selected in particular because it still contains a large number of sources (compared to the Galactic Centre) but avoids the high dust extinction, and it is an excellent place to look for LMXBs as they are thought to be part of an old stellar population. Evidently the 1640 X-ray sources discovered in the GBS belong to a wide range of source classes (see e.g. \citeauthor{Jonker2011} \citeyear{Jonker2011} for quantitative estimates), and studying them can provide insight into different (binary) stellar evolutionary phases. 

The classification of subsets of GBS X-ray sources has been done based on photometric information obtained from other wavelength regimes, such as the radio \citep{Maccarone2012} or the NIR \citep{Greiss2014}. \citet{Udalski2012} and \citet{Hynes2012} have focused their efforts on the brightest optical counterparts, while \citet{Britt2014} used time-resolved optical observations to constrain the nature of X-ray sources.

Alternatively, X-ray source classification can be performed using spectroscopic observations. For example, \citet{Torres2014} revealed a sample of accreting binaries in H\,$\alpha$ emission line selected sources. \citet{Britt2013} and \citet{Wu2015} also used spectroscopic observations to reveal and classify accreting binaries. Several interesting systems have already been identified in the GBS. For example, \citet{Ratti2013} found a quiescent long orbital period CV, while \citet{Hynes2014} reported a candidate symbiotic X-ray binary associated with a carbon star. \citet{Britt2016} identified a high-amplitude IR transient which they interpret as a young stellar object (YSO) with an accretion disk instability event, and \citet{Wevers2016b} discovered a high state AM CVn (compact double WD) system. These examples illustrate the large diversity of (sometimes rare and peculiar) objects that are being discovered in the GBS.

In this work, we continue the effort of classifying GBS X-ray sources by combining time-resolved photometry with optical spectroscopic observations and archival multiwavelength data. In Sections \ref{sec:data}, \ref{sec:phot} and \ref{sec:xraydata} we present respectively the optical spectroscopy, photometry and X-ray data, and their processing for scientific analysis. In Section \ref{sec:discussion} we discuss the classification of the X-ray sources, and we summarize in Section \ref{sec:summary}.

%%%%%%%%%%%%%%%%%%%%%%%%%%%%%%%%%%%%%%%%%%%%%%%%%%%%%%%%%%%%%%%%%%%%%%%%%%%%%%%%%%%%%%%%%%%%%%%%%%%%%%%%%%%%%%%%%%%%%%%%%%%%%%%%%%%%%%%%%%%%%%%%%%%%%%%%%%%%%%%%%%%%%%%%%%%%%%%%%%%%%%%%%%%%%%%%%%%%%%%%%%%%%%%%%%%%%%%%%%
%%%%%%%%%%%%%%%%%%%%%%%%%%%%%%%%%%%%%%%%%%%%%%%%%%%%%%%%%%%%%%%%%%%%%%%%%%%%%%%%%%%%%%%%%%%%%%%%%%%%%%%%%%%%%%%%%%%%%%%%%%%%%%%%%%%%%%%%%%%%%%%%%%%%%%%%%%%%%%%%%%%%%%%%%%%%%%%%%%%%%%%%%%%%%%%%%%%%%%%%%%%%%%%%%%%%%%%%%%
\section{Spectroscopic observations}
\label{sec:data}
We have obtained spectroscopic observations of a sample of optical counterparts to unclassified X-ray sources. These sources, which are listed in Table \ref{tab:observations},  were chosen using several selection criteria:
\begin{itemize}
\item photometrically selected H\,$\alpha$ emitters or blue outliers
\item optically bright ($r^{\prime}$\,$\leq$\,18) or variable counterpart
\item bright (H\,$\leq$\,14) variable NIR counterpart
\end{itemize}
The H\,$\alpha$ emitters and blue outliers with respect to field stars were selected by \citet{Wevers2016c} using ($r^{\prime}$\,--\,$i^{\prime}$, $r^{\prime}$\,--\,$H\alpha$) colour-colour diagrams. The sources of which we obtained spectroscopic observations are CX21, CX118, CX266, CX695 and CXB34.

In optically variable sources the visible light is potentially dominated by the accretion flow. The presence of NIR variability could signal a system containing a compact object where the optical light is dominated by the companion, but variable irradiation of the donor causes the NIR variations (see e.g. \citeauthor{Froning1999} \citeyear{Froning1999}). 

The bright counterparts, either in the optical or NIR, are easily accessible for 8m class telescopes even in poor weather conditions, and we will show that these contain intriguing X-ray sources. Below we describe the instrumental setups used to perform the observations (an overview is given in Table \ref{tab:observations}). 

\subsection{VLT/VIMOS}
We have obtained spectroscopic observations using the VIsible MultiObject Spectrograph (VIMOS, \citeauthor{Lefevre2003} \citeyear{Lefevre2003}) mounted on VLT-UT3 (Melipal). The spectra were obtained in MOS mode, using the medium resolution MR grism combined with the GG475 order sorting filter, providing data in the wavelength range between 4800 and 10000\,\AA. Slits with a width of 1 arcsec were cut on the masks. The dispersion of the spectra is 2.5\,\AA\ pix$^{-1}$ and the spectral resolution is R\,$\sim$\,600, corresponding to an instrumental FWHM resolution of 500 km s$^{-1}$. We obtained two 875\,s exposures of CX266. In addition three flatfield exposures and a helium-argon lamp flat for wavelength calibration were acquired. The data were reduced, combined and extracted following the steps in \citet{Torres2014}.

\subsection{VLT/FORS2}
The visual and near UV FOcal Reducer and low dispersion Spectrograph (FORS2, \citeauthor{Appenzeller1998} \citeyear{Appenzeller1998}) is mounted on VLT-UT1 (Antu). We obtained a single long-slit spectrum for CX118 on 2011 July 4 using the GRIS-600RI grism with a 1 arcsec~slit. The integration time was 2700\,s. The wavelength covered for this setup ranges from 5500\,--\,8600\,\AA, with a spectral resolution of R\,$\sim$\,1250 and a dispersion of 0.83\,\AA\ pix$^{-1}$. Using the sky spectrum, we measure the instrumental FWHM resolution to be 240 km s$^{-1}$.

We also obtained spectroscopic observations of bright and/or optically variable counterparts to X-ray sources. These targets were observed using the GRIS-600RI grism with a 1\farcs5 slit as a queue filler programme, but generally the observations are seeing-limited (Table \ref{tab:observations}). The exposure times range from 5\,s to 360\,s. We observed the optical counterparts of the X-ray sources CX1165, CXB8, CXB11, CXB12, CXB14, CXB41, CXB42, CXB49, CXB123, CXB321, CXB332, CXB336, CXB352 and CXB405 using this setup. The nominal instrumental FWHM resolution is 450 km s$^{-1}$ (R\,$\sim$\,660), but will be better in case of seeing-limited observing conditions.

Biases, flat fields and arc lamp frames were obtained as part of the standard ESO calibration programme. The spectra are bias-subtracted, flatfield corrected and wavelength calibrated using \textsc{iraf}. Cosmic rays are removed using the \textit{lacos} package \citep{vanDokkum2012} in \textsc{iraf}.

\subsection{VLT/X-shooter}
We also obtained spectroscopic observations using X-shooter \citep{Vernet2011}, mounted on VLT-UT2 (Kueyen). The spectra were obtained as a queue filler programme. We used a 1\farcs5 slit, giving a standard resolution of R\,=\,5400 in the visible arm (instrumental FWHM\,=\,56 km s$^{-1}$). However, the spectra have higher resolution as they were all taken in seeing-limited conditions. We obtained X-shooter spectra of the optical counterparts to CX152, CX204, CX237, CX265, CX391, CX403 and CX695. In this work, we present the ESO Phase 3 pipeline\footnote{http://www.eso.org/observing/dfo/quality/XSHOOTER/pipeline} reduced spectra obtained with the visible arm of the instrument.

\subsection{SOAR/Goodman High Throughput spectrograph}
CX21 was observed on 2012 May 24 using the Goodman high-resolution spectrograph \citep{Clemens2004} mounted on the Southern Astrophysical Research (SOAR) 4.1m telescope. We used the 400 lines/mm grating in combination with a 1 arcsec~slit, covering the wavelength range between 3000 and 7000\,\AA. The exposure time of the observation was 1200\,s. Using this configuration, the spectrum has a dispersion of 1\,\AA\ pix$^{-1}$ and a spectral resolution of R\,$\sim$\,1150. The spectrum was debiased and flatfield corrected, and a wavelength calibration was applied using HgAr arc lamp exposures; these steps were performed using standard \textsc{iraf} routines. We measure the instrumental FWHM resolution from the sky lines, and find that it is 260 km s$^{-1}$.

\subsection{Blanco/Hydra}
CX118 was observed on 2008 July 1 with the Hydra multi-fiber instrument mounted on the Victor M. Blanco telescope. We used a large fiber (2 arcsec diameter) in combination with the KPGL3 400 lines mm$^{-1}$ grating to feed the instrument, which delivers a resolution of R\,$\sim$\,1600 and a dispersion of 0.7\,\AA\ pix$^{-1}$. The wavelength covered is 3600\,--\,7400\,\AA, and we integrated on source for 900\,s. The spectrum was debiased and a flatfield correction has been applied using \textsc{iraf} routines tailored to the reduction of Hydra data ({\it dohydra}). The sky spectrum was measured using dedicated fibers, placed in empty fields on the sky. The wavelength calibration was performed using two HeNeAr arc lamp exposures. The setup used yields an instrumental FHWM resolution of 190 km s$^{-1}$.

\subsection{GTC/OSIRIS}
We observed CXB34 on 2016 August 4 using the Optical System for Imaging and low-Intermediate-Resolution Integrated Spectroscopy (OSIRIS) mounted on the Gran Telescopio Canarias (GTC). We obtained two 900\,s long-slit spectra using the R1000R grism with a 0\farcs8 slit in 1\farcs05 seeing conditions. This grating delivers a dispersion of 2.62\,\AA\ pix$^{-1}$ and a spectral resolution R\,$\sim$\,900. The data were reduced using standard \textsc{iraf} routines, and cosmic rays were removed using the {\it lacos} package. The instrumental FWHM resolution is 340 km s$^{-1}$, as measured from skylines.\\[4mm]
After the standard spectroscopic data reduction steps, we normalize the spectra to the continuum by fitting cubic splines in \textsc{molly}. Emission and prominent absorption lines and telluric features are masked to fit the continuum. We determine the equivalent width (EW) and Gaussian full-width at half-maximum (FHWM) of the emission/absorption lines using the \textit{splot} task in \textsc{iraf}. The uncertainties on these measurements are estimated by determining the standard deviation of repeat measurements assuming different continuum values. We correct the FWHM measurements for the effect of instrumental broadening by subtracting the measured instrumental FWHM in quadrature. Finally, finding charts of the sources that have more than 1 candidate counterpart within the X-ray error region are provided in Appendix \ref{sec:appendix}. 

\begin{table*}
 \centering
  \caption{Overview of optical spectroscopy of the targets and instrumental setups used to classify the sources. The position of the optical counterpart is given in decimal degrees. We provide the slit width and the mean seeing of each observation, to allow the calculation of the effective resolution of the spectrum. $\lambda$ gives the approximate wavelength coverage of the spectrum. T$_{\text{exp}}$ signifies the exposure time in seconds. We also give the date of the observation, and the ESO observing block ID (when available). For CX266 we present spectra of two sources in the original GBS X-ray error circle, which we label sources A and B.} 
  \begin{tabular}{ccccccccccc}
  \hline
CXID &RA ($^{\circ}$) & Dec ($^{\circ}$) &  Instrument & Grating & Slit/seeing &$\lambda$ (\AA) & T$_{\text{exp}}$ (s) & Date & Obs. ID\\\hline
CX21&265.39075&--28.67625&GHTS & 400 l/mm &1\farcs0/1\farcs0 & 3000--7000  & 1200& 2012--05--24 \\
CX118&264.70926&--28.80243 &FORS2 &600RI &1\farcs0/1\farcs1&5500--8600 & 2700& 2011--07--04 & 200217523 \\
&& &Hydra & 400 l/mm& -- /0\farcs7 &3600--7400 & 900&2008--07--01 & \\
CX152 & 267.23627&--29.64331&X-shooter& VIS & 1\farcs5/0\farcs85& 5500--10000 & 70 &2016--06--05 & 1346626 \\
CX204 & 268.30188&--28.79625&X-shooter& VIS & 1\farcs5/0\farcs7&5500--10000 &140 & 2016--06--05 & 1346629 \\
CX237 & 264.56662&--29.66107&X-shooter&VIS & 1\farcs5/0\farcs7&5500--10000 & 60 & 2016--06--05 & 1346635\\
CX265 & 266.79041&--30.47304&X-shooter&VIS &1\farcs5/0\farcs65& 5500--10000& 40 & 2016--08--23 & 1346638 \\
CX266 A&266.58663&--31.82957&VIMOS & MR & 1\farcs0/1\farcs3& 4800--10000 & 2\,$\times$\,875 & 2013--08--11 & 978669 \\
CX266 B&266.58633&--31.83099&VIMOS & MR & 1\farcs0/1\farcs3& 4800--10000 & 2\,$\times$\,875 & 2013--08--11 & 978669\\
CX391 & 264.02676&--30.05309&X-shooter&VIS &  1\farcs5/0\farcs8&5500--10000 & 40 & 2016--06--05 & 1346641 \\
CX403 & 268.27725&--29.09167&X-shooter&VIS & 1\farcs5/0\farcs75& 5500--10000& 80  &  2016--06--05 & 1346644\\
CX695&268.16754&--29.34934&X-shooter&VIS & 1\farcs5/0\farcs75& 5500--10000 & 70 &2016--08--23 &1346650  \\
CX1165 & 264.14032&--28.50108&FORS2& 600RI& 1\farcs5/0.65& 5500--8600 & 300 & 2016--04--03&1346173\\
CXB8&268.63375&--29.47339&FORS2&600RI & 1\farcs5/0\farcs55&5500--8600 & 3 & 2016--04--29 & 1346227 \\
CXB11&267.58625&--30.44778&FORS2&600RI & 1\farcs5/0\farcs70&5500--8600 & 5 & 2016--04--03 & 1346176 \\
CXB12&268.13886&--29.66276 &FORS2&600RI & 1\farcs5/0\farcs65&5500--8600 & 60 & 2016--04--11 & 1346179 \\
CXB14&268.48660&--29.01947&FORS2&600RI & 1\farcs5/0\farcs7&5500--8600 & 90 & 2016--04--11 & 1346182 \\
CXB34&266.87045&--32.24495&OSIRIS &R1000R & 0\farcs8/1\farcs05& 5000--10000 & 2\,$\times$\,900 & 2016--08--24 &  \\
CXB41&267.31738&--31.20661&FORS2 &600RI & 1\farcs5/2\farcs0&5500--8600 & 180 & 2016--04--17 & 1346209 \\
CXB42&268.38574&--29.78009&FORS2 &600RI & 1\farcs5/1\farcs8&5500--8600 & 20 & 2016--04--17 &1346215 \\
CXB43&267.66446&--30.34894&FORS2 &600RI & 1\farcs5/0\farcs6&5500--8600 & 180 & 2016--04--29 &1346233  \\
CXB49&267.37042&--31.30678&FORS2 &600RI & 1\farcs5/0\farcs75&5500--8600 & 60 & 2016--04--29 & 1346224\\
CXB123&268.34216&--29.40014&FORS2 &600RI & 1\farcs5/0\farcs65&5500--8600 & 200 & 2016--04--03 &1346230 \\
CXB321&269.22617&--28.59960&FORS2 &600RI & 1\farcs5/0\farcs75&5500--8600 & 180 & 2016--04--03 & 1346185 \\
CXB332&268.90949&--28.52188& FORS2 &600RI & 1\farcs5/0\farcs9&5500--8600 & 180 & 2016--04--03 & 1346188 \\
CXB336&268.83500& --28.78117&FORS2 &600RI & 1\farcs5/0\farcs75&5500--8600 & 180 & 2016--04--03 &1346191\\
CXB352&268.42706&--29.83106&FORS2 &600RI & 1\farcs5/1\farcs4&5500--8600 & 360 & 2016--04--22& 1346194\\
CXB405&266.55905&--32.25306&FORS2 &600RI & 1\farcs5/0\farcs55&5500--8600 & 90 & 2016--04--29 & 1346201 \\
\end{tabular}
 \label{tab:observations}
\end{table*}

\begin{table*}
 \centering
  \caption{Properties of the optical counterparts. Optical magnitudes are in the Vega system, taken from \citet{Wevers2016}; when marked with a $^{*}$ they are saturated and unreliable; when marked with $^{**}$ they are derived from the DECam observations. IR magnitudes in the H-band are taken from UKIDDS, VVV or 2MASS (when available). Variable IR sources are marked with a $\dagger$. EW is the equivalent width of H\,$\alpha$ in \AA, where a negative value indicates emission. The FWHM of the H\,$\alpha$ line is given in km s$^{-1}$ if measured reliably, and is corrected for the instrumental broadening. The numbers in brackets correspond to the uncertainty in the last digit. In the comments we give a classification if possible. WTTS stands for weak-lined T Tauri star. The capital letters are remarks from \citet{Udalski2012}: E stands for eclipsing, SP for spotted star, SRV for semi-regular variable. We list the periodicity from OGLE data if available.} 
 \begin{threeparttable}  
  \begin{tabular}{ccccccccccccc}
  \hline
CXID & $r^{\prime}$ & $i^{\prime}$ & H\,$\alpha$ & $r^{\prime}$--$i^{\prime}$ & H & EW (\AA) & FWHM (km\,s$^{-1}$) & $\frac{F_X}{F_{\text{opt}}}$ & Comments\\\hline
CX21&19.14(2)&18.46(1)&17.37(2)&0.68&16.9(1) & --136(5) & 893\,$\pm$\,5 &11.5& Quiescent CV\\
CX118&17.90(2)&16.94(1)&17.04(2)&0.96& &--17.0(1)& 940\,$\pm$\,3 & 0.81&Nova-like CV\\
CX152 & 16.22$^{*}$&15.09$^{*}$&15.83(1)  &  &13.3$^{\dagger}$ &1.0(1) & & 0.012& E, P\,=\,0.38 d \\
CX204 &16.42(1)&15.41$^{*}$&15.83(1)& & 13.8$^{\dagger}$&  --1.1(1) & &  0.012& H\,$\alpha$ emission, SP, P\,=\,2.01 d \\
CX237 &14.10$^{*}$&13.81$^{*}$&14.34$^{*}$& &12.8$^{\dagger}$ &  0.60(2)& &  0.003& E, P\,=\,0.2735 d\\
CX265 & 15.69$^{*}$&14.52$^{*}$&15.20(2)& & 12.3$^{\dagger}$ &--0.06(1) &  & 0.005& H\,$\alpha$ emission? \\
CX266 A&16.18(2)&14.75(1)$^{*}$&15.41(2)&&  &--2.2(1) & & 0.005&H\,$\alpha$ emission\\
CX266 B&19.39(2)&16.66(1)&18.10(2)&2.73& 12.27(1) &--19.9(7)&  & 0.005 & M star, H\,$\alpha$ emission\\
CX391 & 14.53$^{*}$&13.32$^{*}$&14.34(1)&  &12.2$^{\dagger}$  & 0.32(2)& & 0.001& E, SP, P\,=\,11.47 d \\
CX403 & 15.59(5)&14.97(5)&15.39(5)& 0.61  & 13.1$^{\dagger}$ &2.4(2) & & 0.005& \\
CX695&17.96(5)&16.21(5)&17.21(5)&1.75&13.0$^{\dagger}$  &--0.64(2)&  & 0.01& M star, H\,$\alpha$ emission, P\,=\,0.3099 d\\
CX1165 &16.91(5)&13.61$^{*}$&15.98(5) & & 9.22(3) & & & 0.001 & M star, SRV, P\,=\,74.1 d\\
CXB8 & 11.0$^{**}$ &  &  & & 8.78(3) &1.0(3) &  & 0.001&  \\
CXB11 & 11.6$^{**}$ &  &  & & 9.64(3)  &  & &0.002 & H\,$\alpha$ filled in\\
CXB12 &12.35$^{**}$ &  &  & &  &  1.04(3)& &0.002 & Symbiotic binary / qLMXB?\\
CXB14& 14.5$^{*}$&12.62$^{*}$& & 13.37$^{*}$  & &1.0(2) & & 0.01& \\
CXB34&21.39(4)&20.97(7)&21.34(1)&0.42& & --48.4(4) & 873\,$\pm$\,11& 3.3& Polar, P\,=\,81\,$\pm$\,2 min\\
CXB41&15.93(2)&14.91$^{*}$&15.36(2)& &12.42(1) & --1.57(1) &  & 0.005& H\,$\alpha$ emission\\
CXB42&13.28$^{*}$&12.66$^{*}$&13.21$^{*}$ &  &9.86(3) & --2.42(1) &  &0.01 & WTTS \\
CXB43&18.32(5)	&16.57(5)&17.64(5) & 1.75  & 12.42(4)&& &0.05 & Unidentified emission lines\\
CXB49& 14.6$^{**}$ &  &  &  &  11.38(4)&--17.5(2) &  & 0.004& H\,$\alpha$ emission\\
CXB123&13.54$^{*}$&13.08$^{*}$&13.88$^{*}$  &  & & 1.1(1)& & 0.002& \\
CXB321&15.11$^{*}$&13.68$^{*}$&15.90(5)&  & 12.04(1) &1.12(5) & & 0.001& \\
CXB332&15.71$^{*}$&14.79$^{*}$&15.71(5)&   & & 2.45(3) & & 0.002& \\
CXB336& 16.3$^{**}$ &  &  &  & & 0.42(6)& &  0.006& \\
CXB352&15.88(5)&15.02$^{*}$&15.45(5)& & & 1.3(1)& & 0.004& \\
CXB405&14.60$^{*}$ &15.12$^{*}$&15.24$^{*}$ &  &13.57(1) & & &0.002 &H\,$\alpha$ filled in \\
  \hline
  \end{tabular}
  \end{threeparttable}
  \label{tab:outlierctparts}
\end{table*}

\section{Optical photometric observations}
\label{sec:phot}
The southern part of the GBS footprint (e.g. fig. 1 in \citeauthor{Wevers2016} \citeyear{Wevers2016}) was observed using the Dark Energy Camera (DECam; \citealt{Abbott2012}) mounted on the Victor M. Blanco telescope. DECam uses a 62 CCD camera with 2048\,$\times$\,4096 pixels per chip, a pixel scale of 0\farcs27 pix$^{-1}$ and a 2.2 square degree field of view. We obtained photometric observations in the SDSS $r^{\prime}$-band on 2013 June 10-11, with an average cadence of $\sim$\,20\,--\,30 min, while the seeing conditions varied between 0\farcs9 and 1\farcs5 during the 2 nights. The observations comprised of 2\,$\times$\,90 second images followed by 2\,$\times$\,1 second exposures to account for the faint and bright sources in the field, along with bias and flat field images for calibration. The images were reduced with the NOAO DECam pipeline\footnote{http://ast.noao.edu/sites/default/files/NOAO$\_$DHB$\_$v2.2.pdf}. The American Association of Variable Star Observers (AAVSO) Photometric All-Sky Survey DR7 (APASS) was used to photometrically calibrate the images; the average photometric uncertainty is 0.05 mag. We used between 3\,--\,5 comparison stars in the field of view with uncertainties on the order of 0.001 magnitudes and propagated the error for each data point. The full dataset will be published elsewhere (Johnson et al. in prep.).

\section{X-ray observations}
\label{sec:xraydata}
\subsection{CXB12}
CXB12 was in the field of view of XMM--{\it Newton} on 2005, September 18. The source position was about 6.5\arcmin\ away from the optical axis of the X-ray mirrors. The observation start time was 09:52 (UTC). We used the Scientific Analysis Software (SAS) version 15.0.1 for the data analysis. The on-source time is 31.8 ks but the net observing time is 13.5 ks for the EPIC PN detector and 22 ks for the MOS1 and MOS2 detectors after we filtered out epochs of high background. The PN was operated in PrimeFullWindowExtended while the MOS1 and MOS2 were operated in observing mode PrimeFullWindow. All X-ray instruments were employed with the medium filter. In each of the three X-ray detectors counts were extracted from a circular region
with radius of 20\arcsec\ centred on the position of the optical source. Background counts were extracted from a circular region of
radius 1\arcmin\ located on the same CCD as the source. We merged the extracted MOS1, MOS2 and PN spectra into one using the SAS task {\sc epicspeccombine}.

\subsection{CXB34}
CXB34 was about 1.5\arcmin\ off-axis in an XMM--{\it Newton} observation that started at 06:15 (UTC) on 2014 August 31. We used the same SAS version indicated above for CXB12 for the data analysis. The on-source time is 31.9 ks but the net observing time is 26.8 ks for the EPIC PN detector and 31.3 ks for the MOS1 and MOS2 detectors after we filtered out epochs of high background. The PN, MOS1, and MOS2 were operated in PrimeFullWindow mode. All X-ray instruments were employed with the medium filter. We extracted the source light curve from a circular region with radius 20\arcsec\ centred on the position of the optical counterpart. We corrected the arrival times of the X-ray photons to the solar system barycenter using the SAS task {\sc barycen} and the coordinates of the optical source. 

\subsection{Improved X-ray source positions}
Because the X-ray part of the GBS consists of shallow 2 ks Chandra ACIS--I observations, the positional accuracy of the X-ray detections can be improved if deeper Chandra data are available for sources that were detected with only a few counts or far from the optical axis of the telescope. This is useful in particular when multiple optical sources are present within the X-ray error circle and it is unclear which object is the true optical counterpart. We therefore search the Chandra archive for ACIS--I observations overlapping with our sample of sources. When available, an analysis similar to the one described in \citet{Jonker2011} is performed to process the X-ray observations, which includes data processing with the \textsc{ciao} software tools. We then use \textsc{wavdetect} to perform the source detection, and we follow \citet{Evans2010} to calculate the positional uncertainties. The results of this analysis for the whole GBS area will be presented in a separate article (Wetuski et al. in prep); for now we note that for our sample, it typically yields similar results as the original analysis \citep{Jonker2011, Jonker2014}. The exceptions are CX266, CXB12 and CXB49, where we find improvements in the source localisation, which in turn helps to identify the correct optical counterpart and interpretation. We will discuss these results in Sections \ref{sec:cx266}, \ref{sec:cxb12} and \ref{sec:cxb49}, respectively.

\section{Discussion}
\label{sec:discussion}
We now turn our attention to the classification of the X-ray sources based on their optical lightcurves, colours and spectrosocopic observations as well as their X-ray properties. We will quote the number of detected X-ray photons detected in the 0.3\,–-\,8 keV band during the discovery observations with Chandra \citep{Jonker2011, Jonker2014}. Following \citet{Jonker2011} we adopt a conversion factor from X-ray count rate in the 0.3\,--\,8 keV band to (rough but useful) unabsorbed fluxes in the 0.5\,--\,10 keV band. Assuming a source spectrum with a power law with photon index $\Gamma$\,=\,2 absorbed by a Galactic hydrogen column density N$_{\rm H}$\,=\,10$^{22}$ cm$^{-2}$, this conversion factor is 7.76 $\times$ 10$^{-15}$ erg cm$^{-2}$ s$^{-1}$ photon$^{-1}$. However, if we can constrain the $E(B-V)$ using optical spectroscopy or the spectral energy distribution (SED), we will compute 0.5\,--\,10 keV unabsorbed X-ray fluxes adopting a power law model with $\Gamma$\,=\,2 modified by N$_{\rm H}$, which will be derived from the relationship in \citet{Bohlin1978}: 
\begin{equation}
{\rm N}_{\rm H} = 5.8 \times 10^{21} \times E(B-V)\ \text{cm$^{-2}$}.
\end{equation}
In some cases, the presence of diffuse interstellar bands (DIBs) can allow us to constrain $E(B-V)$, because their EW correlates with the amount of interstellar extinction. In particular, the DIBs at $\lambda5780$ \citep{Herbig1993} and $\lambda6284$ \citep{Cordiner2011} are well-known tracers of reddening, and are within the wavelength range of all our spectroscopic observations. Alternatively, the interstellar Na\,\textsc{i} D doublet at $\lambda\lambda5890,5895$ can also be used to constrain $E(B-V)$ \citep{Munari1997}.
We can also derive a distance from the $E(B-V)$ values using 3D reddening maps. We use the optical extinction map by \citet{Green2015} when available; if not, we use the IR extinction map by \citet{Schultheis2014}. If both are available, we can compare the obtained distances to estimate the error budget.

As an estimator for the optical flux we use the $i^{\prime}$-band magnitudes, as they are less sensitive to interstellar reddening than $r^{\prime}$-band magnitudes. We calculate the ratio of the X-ray to optical flux as 
\begin{equation}
\text{log} \left( \frac{F_{\text{X}}}{F_{\text{opt}}}\right) = \text{log}(F_X) + \frac{i^{\prime}}{2.5} + 5.01
\end{equation}
where --5.01 is the $i^{\prime}$-band zeropoint. The $i^{\prime}$-band measurements (Table \ref{tab:outlierctparts}) are taken from \citet{Wevers2016}. For sources with only an $r^{\prime}$-band magnitude available, we use --5.37 as the zeropoint \citep{Britt2014}.

\subsection{Individual sources}
\subsubsection{CX21 = CXOGBS J174133.7--284033, a likely short orbital period CV}
\label{subsec:cx21}
\begin{figure} 
  \includegraphics[height=5cm, keepaspectratio]{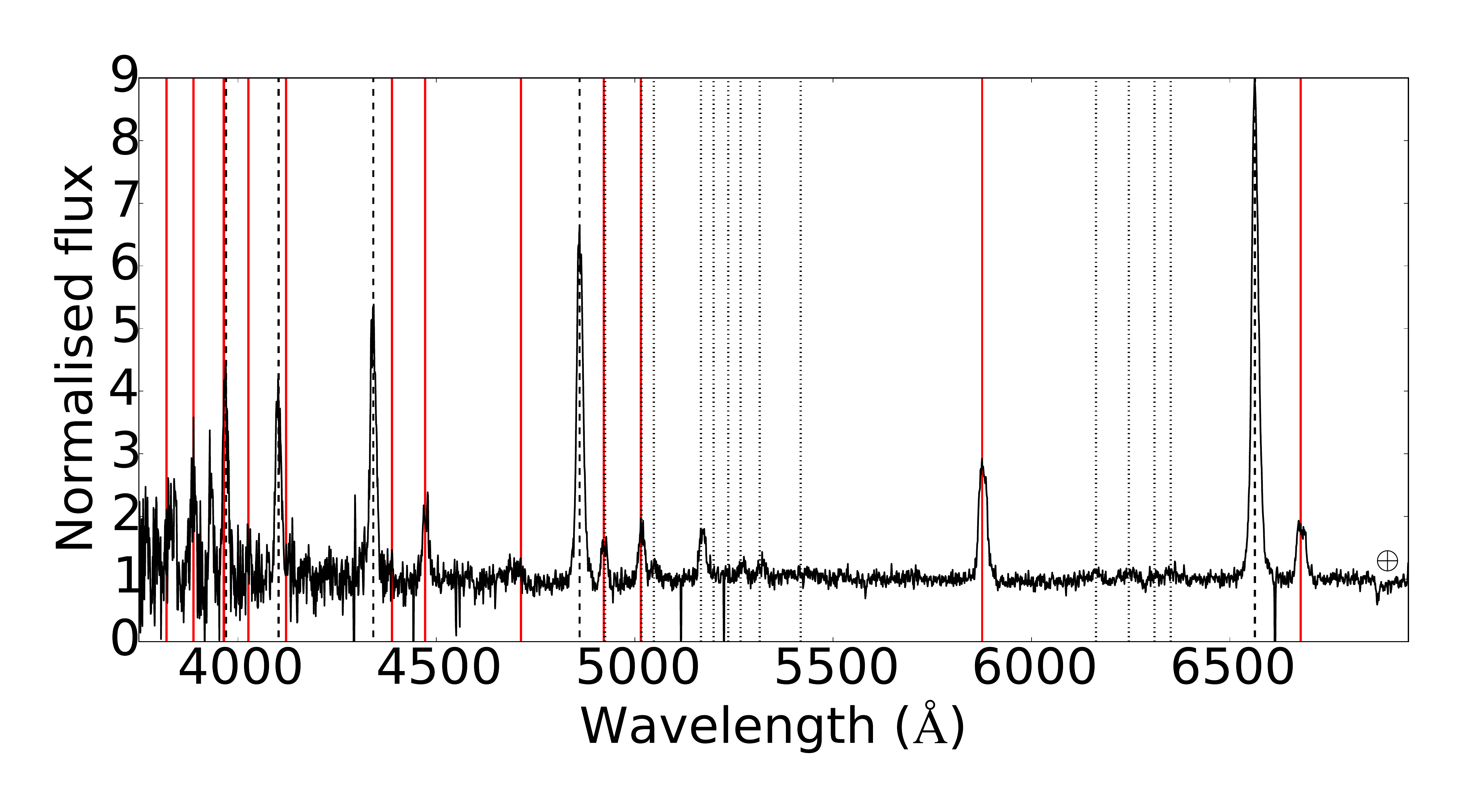}
  \caption{Continuum normalized SOAR spectrum of CX21. The H Balmer series are marked by dashed lines, He\,\textsc{i} lines are marked by red solid lines. We identify the emission lines near 5200\,\AA$\ $as Fe\,\textsc{ii} multiplets 42 and 49 (dotted lines).}
  \label{fig:cx21spec}
\end{figure}

We show the optical spectrum of CX21 in Figure \ref{fig:cx21spec}. 
The H Balmer series can be seen in emission up to H$\eta$  (dashed lines), together with He\,\textsc{i} $\lambda\lambda$3964,4471,4921,5015,5876,6678 (solid lines in Figure \ref{fig:cx21spec}). The morphology of these lines is diverse, including single and multi-peaked profiles. The Balmer lines decrease in EW with decreasing wavelength. For H\,$\alpha$, we measure an EW\,=\,136$\,\pm$\,5\,\AA$\ $and  an intrinsic FWHM of 893\,$\pm$\,5\,km s$^{-1}$. The strongest He\,\textsc{i} line ($\lambda5876$) has EW\,=\,45\,$\pm$\,2\,\AA$\ $and intrinsic FWHM\,=\,1170\,$\pm$\,20\,km s$^{-1}$. In addition, emission at $\lambda5169$ from the Fe\,\textsc{ii} multiplet 42 is clearly detected with EW\,=\,12.2\,$\pm$\,0.75\,\AA$\ $as well as weaker emission due to Fe\,\textsc{ii} multiplet 49 lines. Neither He\,\textsc{ii}\,$\lambda4686$ emission nor photospheric lines from the WD/donor star are obvious in the data. No DIBs are detected. The interstellar Na doublet falls on the He\,\textsc{i} $\lambda5876$ emission line. We establish EW\,$\leq$\,0.2\,\AA$\ $for the Na D2 $\lambda5896$ component, while we were unable to set a limit for the Na D1 line due to its location in the steep part of the He\,\textsc{i} line profile. The lack of interstellar features in the spectrum indicates that the source is nearby. We constrain $E(B-V)$\,$\leq$\,0.05 from the Na D2 EW using the calibration of \citet{Munari1997} and assuming a D2/D1 ratio of 2 (optically thin limit). Thereby we also constrain N$_{\rm H}$\,$\leq$\,3\,$\times$\,$10^{20}$ cm$^{-2}$. Using the 3D reddening map from \citet{Green2015}, we derive an upper limit on the distance of 0.5 kpc. Chandra detected 60 counts during the discovery observation, which yields an unabsorbed X-ray flux of 4.5\,$\times$\,10$^{-13}$ erg cm$^{-2}$ s$^{-1}$, and hence L$_X$\,$\leq$\,10$^{31}$ erg s$^{-1}$ for a distance of 500\,pc. The optical lightcurve of CX21 \citep{Britt2014} shows strong (1 mag amplitude) flickering on a time-scale of hours. We also find the IR counterpart, with J\,=17.28\,$\pm$\,0.05, H\,=\,16.85\,$\pm$\,0.1, K\,=\,16.44\,$\pm$\,0.15 mag to be variable with an amplitude of $\sim$\,0.5 mag in the UKIDDS Galactic Plane Survey \citep{Lucas2008}.

CX21 was classified as a CV with the optical light dominated by the accretion flow in \citet{Britt2014} on the basis of the strong optical flickering and the hardness ratio determined from ROSAT and Chandra observations. A disc dominated optical spectrum lacking He\,\textsc{ii} $\lambda4686$ can be explained if this GBS source is a (dwarf nova) CV below the period gap in quiescence. In this scenario, the contributions from the donor star and WD to the optical light are strongly veiled by the accretion disc and the spectrum shows strong H\,\textsc{i} and He\,\textsc{i} emission. In this regard, the emission line content in the spectrum of CX21 resembles that of the SU UMa-type CVs UV Per, VY Aql and V1504 Cyg \citep{Thorstensen1997}. The derived X-ray luminosity is also consistent with this interpretation \citep{Byckling2010}. Finally, the combination of a large FWHM and lack of orbital photometric modulation also suggests a short orbital period.

\subsubsection{CX118 = CXOGBS J173850.2--284808, a nova-like CV}
\label{subsec:cx118}
\begin{figure}
  \includegraphics[height=5cm, keepaspectratio]{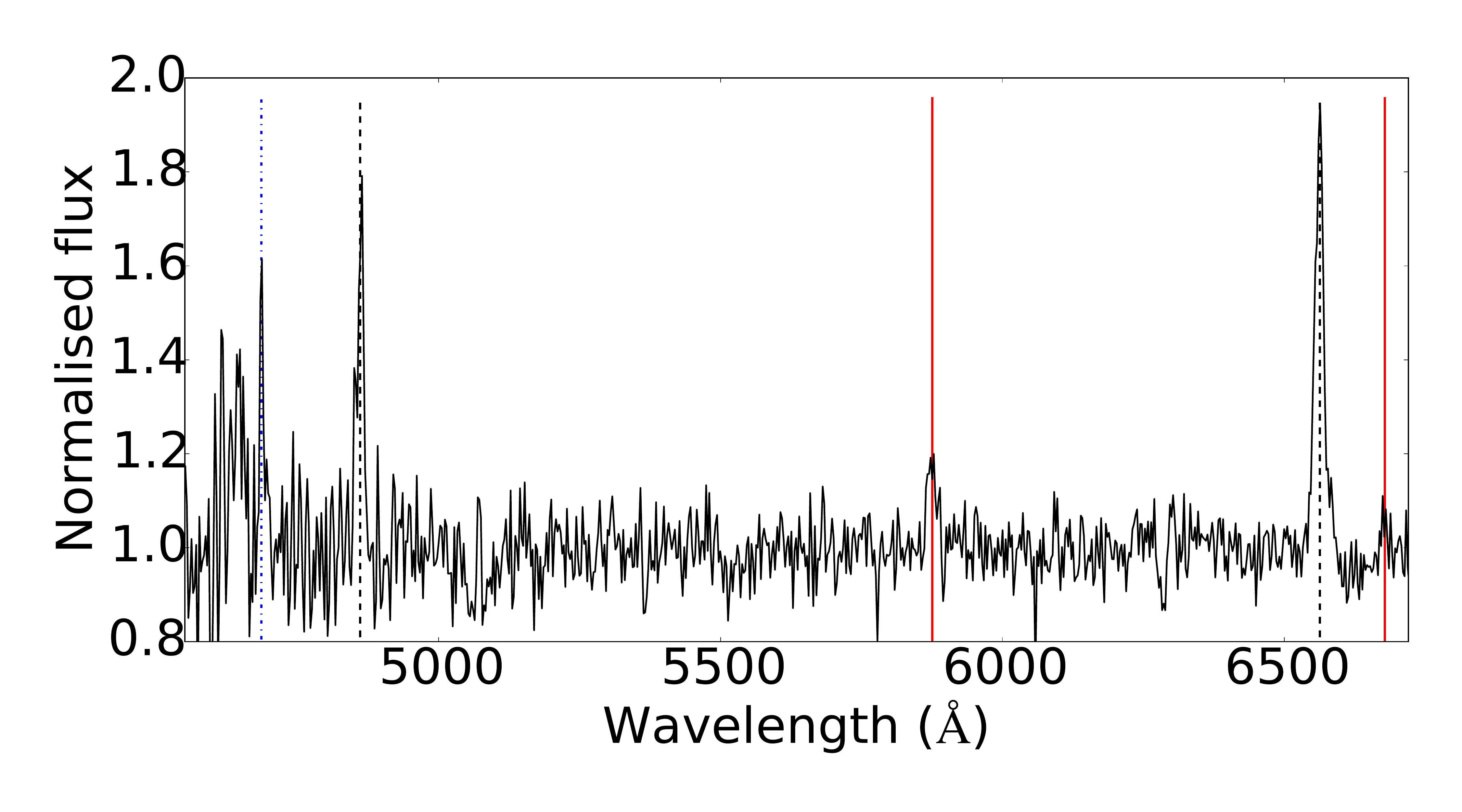}
 \label{fig:cx118spechydra}
  \includegraphics[height=5.1cm, keepaspectratio]{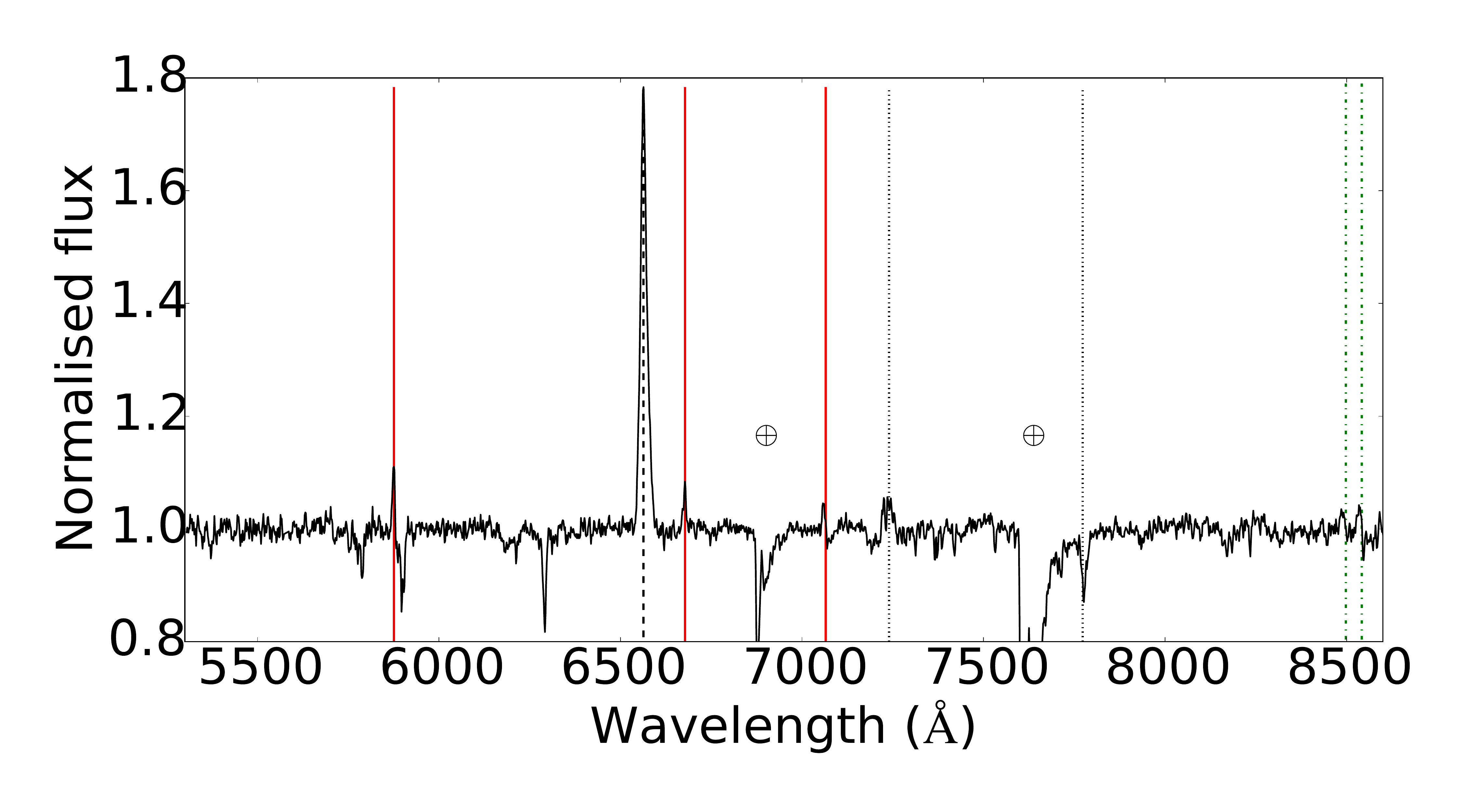}
    \caption{Continuum normalized Hydra (top) and FORS2 (bottom) spectra of CX118. Besides the prominent H Balmer emission lines (marked by dashed lines), there is evidence for He\,\textsc{i} emission at $\lambda\lambda$\,5575,6678 (solid lines) and He\,\textsc{ii} 4686\,\AA$\ $(blue, dash-dotted line). Two lines of the Ca\,\textsc{ii} triplet in emission are marked with green dash-dotted lines in the FORS2 spectrum. Dotted lines at $\lambda\lambda$7240,7773 originate in the accretion flow (see text).}
 \label{fig:cx118specfors}
\end{figure}

For CX118 we have two epochs of spectroscopic data, taken on 2008 July 1 using Hydra and on 2011 July 4 using FORS2 (Figure \ref{fig:cx118specfors}, top and bottom panels respectively). 

The Hydra observation covers the spectral range $\lambda\lambda$4100\,--\,6800. Both H\,$\alpha$ and H$\beta$ lines are found in emission and appear single-peaked. We measure EWs and FWHMs of 19.4\,$\pm$\,0.2\,\AA$\ $and 880\,$\pm$\,4\,km s$^{-1}$ for H\,$\alpha$ and 9.6\,$\pm$\,0.6\,\AA$\ $and 960\,$\pm$\,30 km s$^{-1}$ for H$\beta$. Applying a correction for the instrumental broadening yields an intrinsic FWHM\,=\,859\,$\pm$\,4\,km s$^{-1}$ for H\,$\alpha$ and 941\,$\pm$\,30\,km s$^{-1}$ for H$\beta$. Weaker He\,\textsc{i} $\lambda\lambda5876,6678$ emission lines are present while there is evidence for the He \textsc{ii} $\lambda4686$ emission line (EW\,$\leq$\,7\,\AA). The FORS2 observation covers the spectral range $\lambda\lambda$5300\,--\,8600 where we again detect single-peaked emission from H\,$\alpha$ and He\,\textsc{i} $\lambda\lambda5876,6678,7065$. The EW and intrinsic FWHM of H\,$\alpha$ are 16.98\,$\pm$\,0.15\,\AA$\ $and 940\,$\pm$\,3\,km s$^{-1}$. Redward of He\,\textsc{i} $\lambda6678$ we observe strong absorption at $\lambda7773$ due to O\,\textsc{i} which originates in the accretion flow \citep{Friend1988}. In addition, we detect broad emission at $\lambda7240$ that we tentatively identify as a blend of Fe\,\textsc{ii} lines from multiplet 73 and O\,\textsc{i}. Neither the Hydra nor the FORS2 spectra show signatures of photospheric lines from the donor star or a WD accretor. On the other hand, the data show the saturated interstellar Na D doublet as well as prominent DIBs at $\lambda5780$ and $\lambda6284$ with EWs of 0.43\,$\pm$\,0.02\,\AA$\ $and 1.51\,$\pm$\,0.02 \,\AA, respectively.

We obtain an $E(B-V)$\,=\,0.8 and 1.3 mag from the relationship between reddening and the EW for the $\lambda 5780$ and $\lambda6284$ DIBs \citep{Herbig1993, Cordiner2011}. This yields distance estimates of 1.2 and 1.8 kpc using the 3D reddening map by \citet{Green2015}. The distance obtained from the IR reddening map by \citet{Schultheis2014} is $\leq$\,2.5 kpc, therefore we estimate the uncertainty in the distance to be a factor of 2. Adopting $E(B-V)$\,=\,0.8 we obtain N${_{\rm H}}$\,$\sim$\,4.6\,$\times$\,$10^{21}$ cm$^{-2}$. From the 17 Chandra counts we calculate an unabsorbed X-ray flux of 1.9\,$\times$\,$10^{-13}$ erg cm$^{-2}$ s$^{-1}$ and thereby an X-ray luminosity of $L{_X}$\,=\,3\,$\times$\,$10^{31}$ erg s$^{-1}$ for a distance of 1.2 kpc or $L{_X}$\,=\,1.4\,$\times$\,$10^{32}$ erg s$^{-1}$ for a distance of 2.5 kpc.

The optical lightcurve of CX118 shows flickering with an amplitude of $\sim$\,0.6 mag \citep{Britt2014} that, together with our spectroscopy, shows that this GBS source has an optical spectrum dominated by emission from the accretion flow. Using the unabsorbed X-ray flux, $\frac{F_X}{F_{opt}}$ is 0.12. This is consistent with both a high accretion rate (nova-like) CV and a magnetic CV (e.g. \citeauthor{Britt2013} \citeyear{Britt2013}), although it is lower than for most magnetic CVs. This classification is also supported by the detection of He\,\textsc{ii} $\lambda4686$ emission. The broad single-peaked emission lines found in CX118 are typically observed in nova-like variables. In this class of high accretion rate CVs, even eclipsing systems do not show a double-peaked profile (see e.g. \citeauthor{Dhillon1992} \citeyear{Dhillon1992}, \citeauthor{Rodriguez2007a} \citeyear{Rodriguez2007a}, \citeyear{Rodriguez2007b}). A magnetic CV scenario seems unlikely as the spectrum of CX118 does not show line profiles composed of (at least) a narrow and a broad component, the former originating in the irradiated donor star (e.g. \citealt{Schwope1997}, \citealt{Schwarz2005}). These components can exhibit large radial velocities not observed in our two epochs of spectroscopy. We thus classify CX118 as a nova-like CV system.

\subsubsection{CX266 = CXOGBS J174620.7--314946, a chromospherically active star}
\label{sec:cx266}
The original X-ray error circle of CX266 (with radius 6\farcs49) contains two bright optical sources, one of which is saturated in the $i^{\prime}$-band photometry of \citet{Wevers2016}. We designate this source A, while the other object, which was identified as a photometric H\,$\alpha$ outlier in \citet{Wevers2016c}, is named source B. The re-analysis of deeper Chandra observations resulted in a significantly better X-ray source localisation at coordinates ($\alpha$, $\delta$)\,=\,(266.58649, --31.83014) with a 2\farcs49 error radius (see Appendix \ref{sec:appendix}). This reveals that source A is the true optical counterpart to CX266. Coincidentally, source B was identified by \citet{Wevers2016c} as an H$\alpha$ emission line candidate. 
\begin{figure} 
  \includegraphics[height=4.6cm, keepaspectratio]{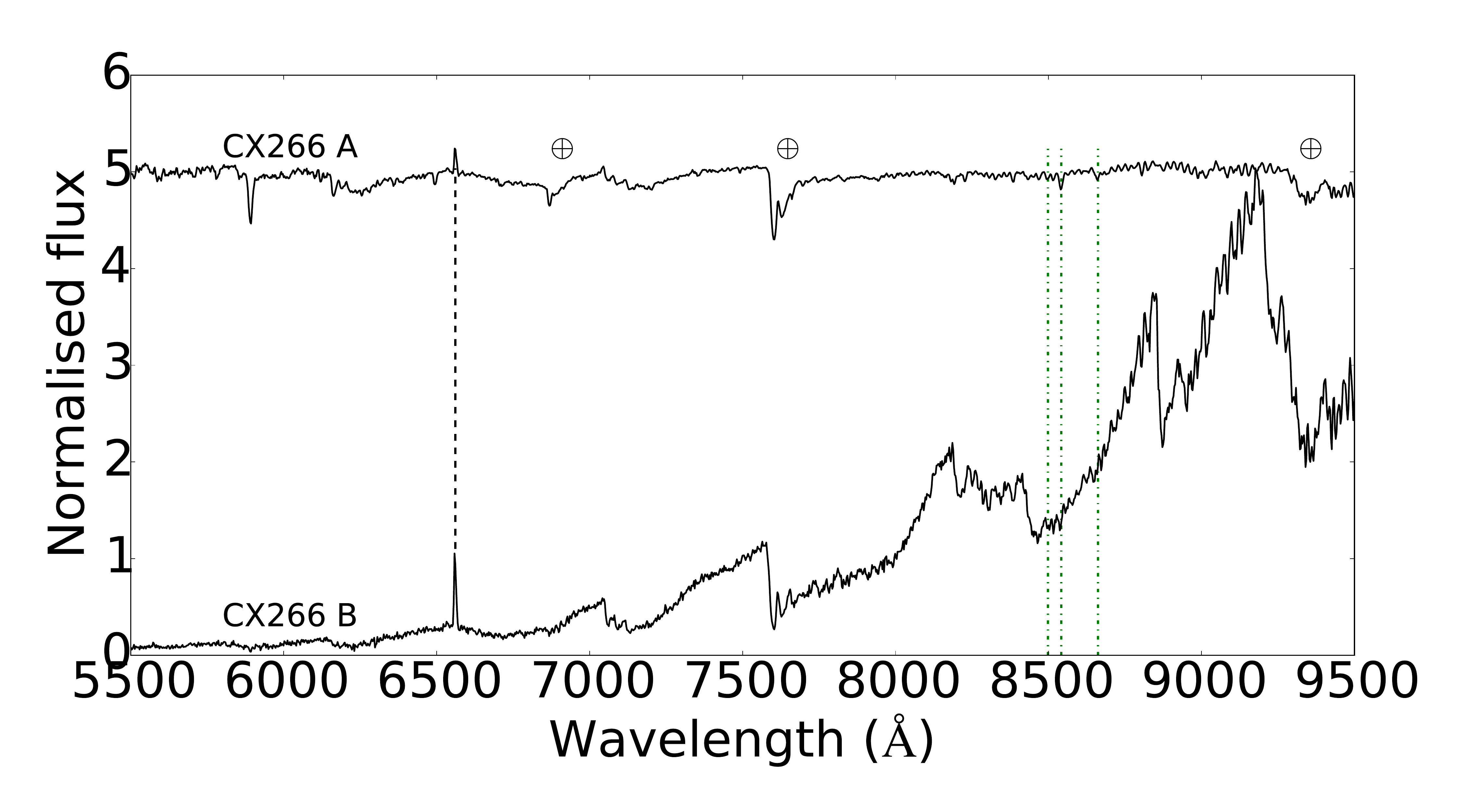}
  \caption{VIMOS spectra of CX266A and CX266B. Both sources show narrow H\,$\alpha$ emission, indicating chromospheric activity. The spectrum of source A is normalized to the continuum, while that of source B is normalized at 7500\,\AA. The latter displays molecular absorption bands redward of 7000\,\AA$\ $ indicative of a late M star. }
  \label{fig:cx266spec}
\end{figure}

The VIMOS spectra of both sources are shown in Figure \ref{fig:cx266spec}. 
Source A displays a weak H\,$\alpha$ emission feature, for which we measure EW\,=\,2.2\,$\pm$\,0.1\,\AA. Molecular absorption bands such as TiO\,$\lambda6200$ are visible in the spectrum, as well as the Ca\,\textsc{ii} IR triplet, indicating a late K spectral type. The presence or absence of DIBs at $\lambda\lambda5780,6284$ cannot be confirmed due to possible confusion with photospheric features at similar wavelengths. The X-ray to optical flux ratio of 0.005, together with the narrow emission lines, support a chromospherically active star interpretation. 

We also present the spectrum of source B because it was identified as a photometric outlier by \citet{Wevers2016c}. It shows strong H\,$\alpha$ emission together with TiO, ZrO and VO absorption bands in the red part of the spectrum. We measure an equivalent width of the H\,$\alpha$ emission line of EW\,=\,19.9\,$\pm$\,0.7\,\AA. Comparing the spectrum by eye to a series of templates, we infer a spectral type later than M7, mainly based on the depth of the TiO feature near 8400\,\AA. The lack of DIBs and the absence of the Na\ \textsc{i} D doublet in the optical spectrum indicate it is likely a dwarf star; if it were a giant, the distance would be large enough to expect detectable signatures from the DIBs. A M7V dwarf has an absolute $V$-band magnitude of 17.8 \citep{Pecaut2013}, therefore given that the source has $r^{\prime}$\,=\,19.2 we infer a distance of $\sim$\,20 pc to source B (using the colour transformation of \citeauthor{Jester2005} \citeyear{Jester2005}). The non-detection in the Chandra observations implies an upper limit to the X-ray flux of $\sim$\,1\,$\times$\,10$^{-14}$ \flx \citep{Jonker2011}, which provides an upper limit to the X-ray to optical flux ratio of 0.005. We thus spectroscopically confirm the emission line nature of this object as a nearby active star. 

\subsubsection{CXB12 = CXOGBS J175233.2-293944, a candidate symbiotic binary/qLMXB}
\label{sec:cxb12}
\begin{figure} 
  \includegraphics[height=5.5cm, keepaspectratio]{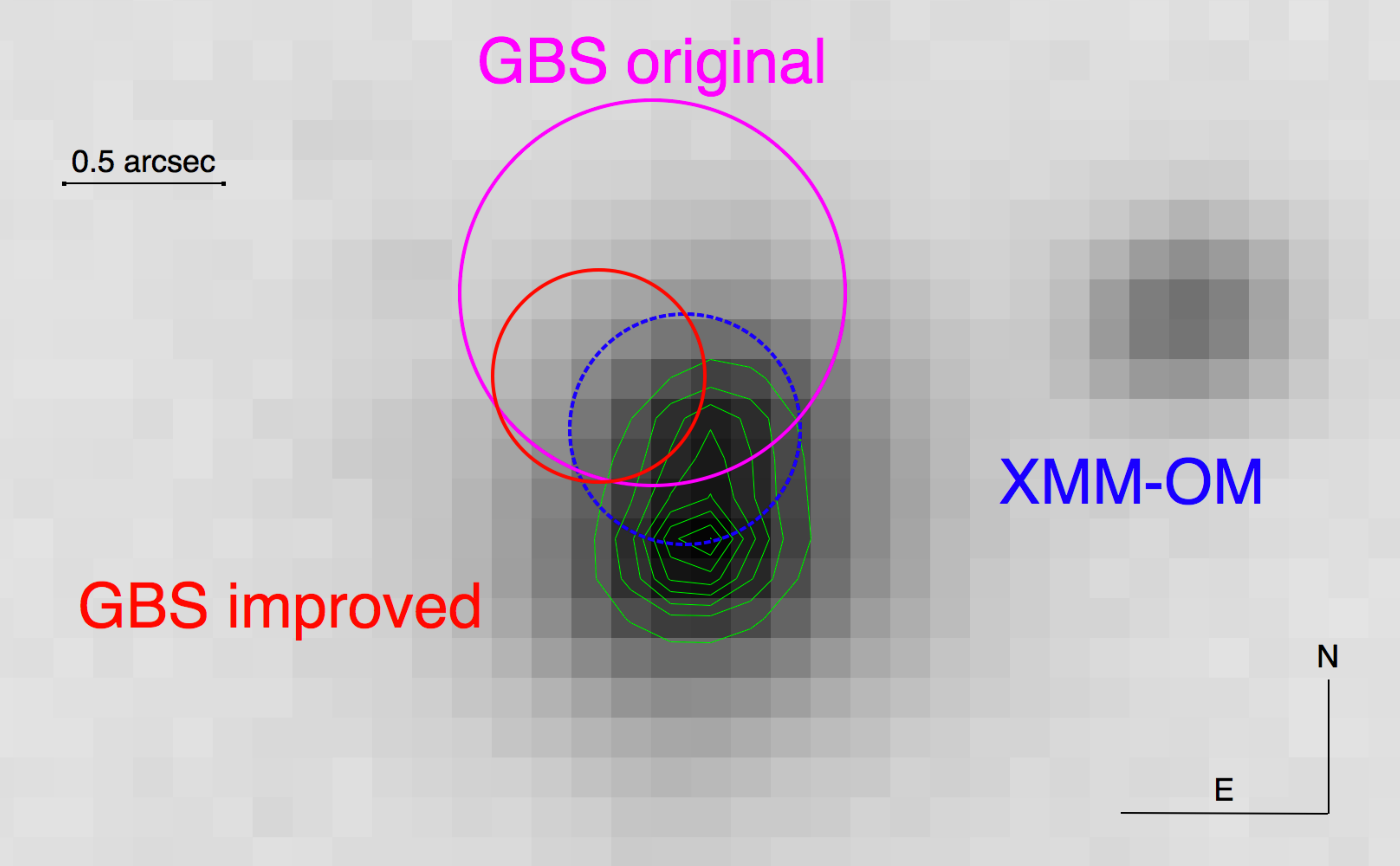}
  \caption{Finding chart of CXB12 in the $R$-band, and overlaid in solid circles the original (magenta) and improved (red) X-ray source positions. The dashed blue circle represents the XMM-OM source localisation. }
  \label{fig:cxb12finder}
\end{figure}

\begin{figure} 
  \includegraphics[width=0.5\textwidth, keepaspectratio]{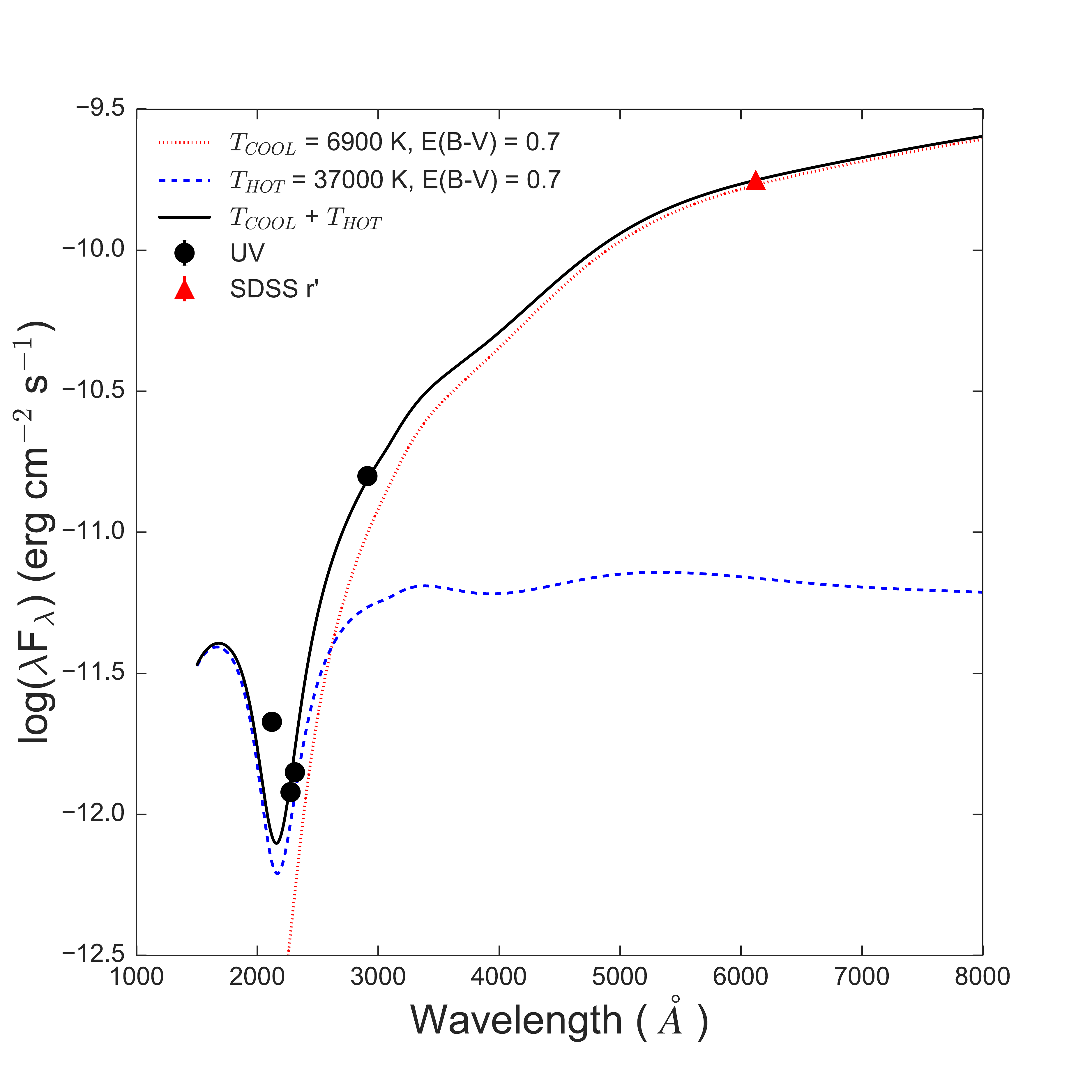}
  \caption{SED of CXB12. The error bars are smaller than the data points. The two blackbody components are shown as dashed and dotted lines, while the sum is plotted as a solid line.}
  \label{fig:cxb12sed}
\end{figure}

The optical ($R$-band) acquisition image of CXB12, taken with FORS2 (Figure \ref{fig:cxb12finder}), shows an elongated intensity profile, suggesting that the source is a blend of 2 objects. An interesting feature is the detection of an UV counterpart in the XMM serendipitous UV source catalogue \citep{Page2012}, with Vega magnitudes of UVW2 ($\lambda2120$)\,=\,16.74\,$\pm$\,0.06, UVM2($\lambda2310$)\,=\,17.17\,$\pm$\,0.05 and UVW1($\lambda2910$)\,=\,14.57\,$\pm$\,0.01 mag. The UV\,--\,optical SED of CXB12 is shown in Figure \ref{fig:cxb12sed}. The UVM2 filter is located on the 2200\,\AA\ extinction bump, which means that this is a lower limit for the intrinsic source brightness. This explains the lower flux in that filter compared to the bluer UVW2 filter. 

The original X-ray error circle of CXB12 has a radius of 1.16 arcsec \citep{Jonker2014}. The analysis of a deeper Chandra observation yields a better X-ray localisation ($\alpha$, $\delta$)\,=\,(268.13885, --29.66250) with a 0\farcs67 error circle. The new source position is consistent with the UV position, and pinpoints the X-ray source to the fainter component of the optical blend. We perform PSF fitting on the FORS2 $R$-band image to determine the flux ratio of the two optical components. We estimate that the brightness ratio is about 2:1, with the fainter component of the blend being consistent with the X-ray and UV positions. We use this brightness ratio to calculate the $r^{\prime}$-band magnitude of the two sources from the apparent brightness of the (unresolved) blend from the 1\,s DECam images (Table \ref{tab:outlierctparts}). We estimate that $r^{\prime}_{\text{faint}}$\,=\,12.35\,$\pm$\,0.1 mag and $r^{\prime}_{\text{bright}}$\,=\,11.5\,$\pm$\,0.1 mag. 

Because we were unaware of the blended nature of the optical counterpart, the FORS2 spectrum of CXB12 was taken with the slit tilted to the parallactic angle. Unfortunately, the dispersion direction is parallel to the elongation of the blend, and the optical spectrum is dominated by the bright source. From a visual comparison to template stars, we infer a spectral type of early to mid-K. 

Although the UV position is formally consistent with both components of the blend, the observed SED cannot be fit with a single temperature blackbody model consistent with a K-type star (with a typical temperature of T\,$\sim$\,4500K). We conclude that the bright optical source is most likely an interloper and the UV and X-ray source are associated to the faint component in the blend. We use the value of $r^{\prime}_{\text{ctpart}}$\,=\,12.35\,$\pm$\,0.1 mag to model the UV\,--\,optical part of the SED. 

We only use UV and $r^{\prime}$-band optical data in the SED, as all other archival data are of insufficient image quality to resolve the two components. We fit single and two-component blackbody models reddened by the interstellar medium to the data using a Monte Carlo approach, varying the temperature and reddening to find the best fit. {We do not perform synthetic photometry but rather compare the model fluxes directly to the observations. This is justified because the aim is not to derive accurate stellar parameters but to illustrate that a single blackbody model does not reproduce the observed SED.} The best-fitting model is a two-blackbody model, with a T\,$\sim$\,6900\,$\pm$\,700 K component and a hotter component of T\,$\sim$\,37000\,$\pm$\,15000 K. We infer a reddening in the range $E(B-V)$\,$\sim$\,0.4\,--\,0.9 from the Monte Carlo trials, translating to N$_{\rm H}$\,=\,2.3\,--\,5.2\,$\times$\,10$^{21}$\,cm$^{-2}$. 

\begin{figure} 
  \includegraphics[width=0.5\textwidth, keepaspectratio]{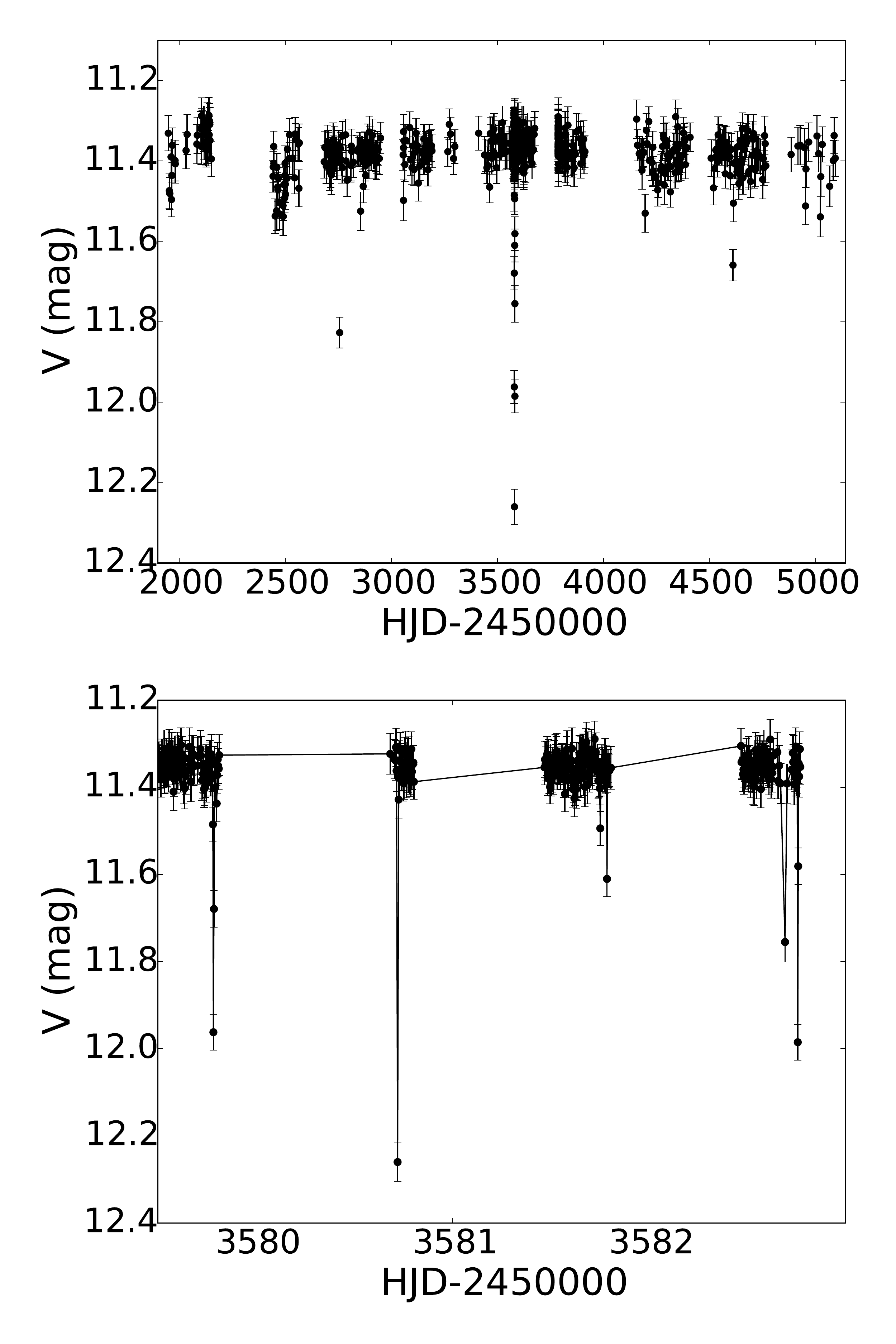}
  \includegraphics[width=0.5\textwidth, keepaspectratio]{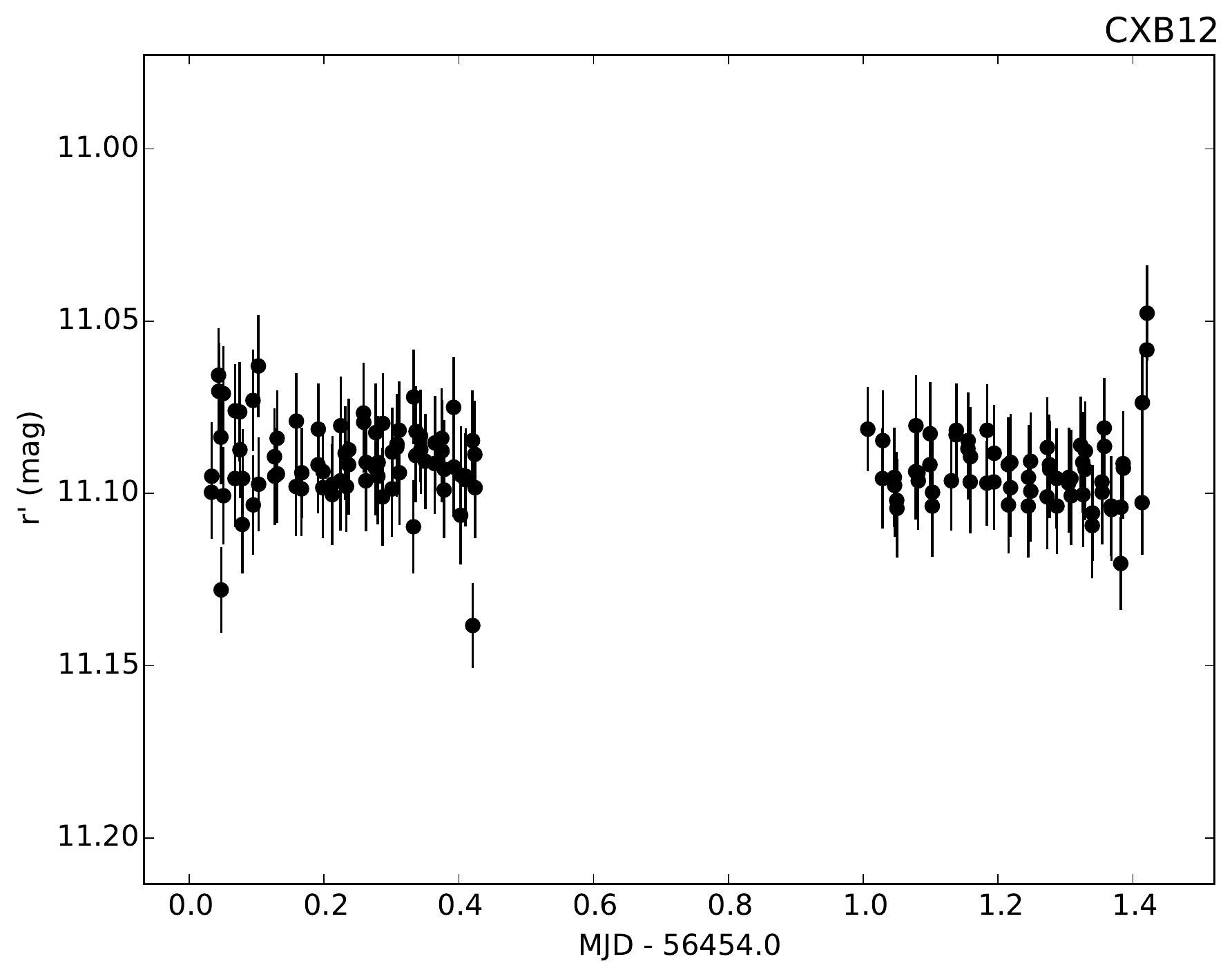}
  \caption{Top: V-band lightcurve of CXB12 from ASAS. The middle panel zooms in on the potential eclipses around HJD\,=\,2453580. Most potential eclipses consist of single datapoint outliers. Bottom: DECam $r^{\prime}$-band lightcurve. No outliers are observed, suggesting they are either spurious datapoints or were missed due to the short time coverage.}
  \label{fig:CXB12asas}
\end{figure}

The publicly available V-band lightcurve from the All Sky Automated Survey (ASAS, \citeauthor{Pojmanski1997} \citeyear{Pojmanski1997}) is shown in Figure \ref{fig:CXB12asas}. We note that the blend is unresolved in these observations. The typical cadence is 1 to a few days, and on these timescales there is evidence of small scale variability with an amplitude of 0.2 mag. The error budget for each measurement is $\sim$\,0.05\,--\,0.1 mag. There are significant drops in brightness (up to 1 mag) that could be interpreted as evidence for eclipses, but this cannot be firmly established because of the low cadence. In our high cadence ($\sim$\,20 min) DECam $r^{\prime}$-band lightcurve (with an error budget of $\sim$\,0.01 mag), spanning two nights, we do not see eclipses nor low amplitude variability. 

We fitted the extracted XMM--{\it Newton} spectrum of CXB12 using {\sc xspec} version 12.9.0n \citep{ar1996}. Due to the low number of source counts (659), we performed fits with both $\chi^2$ and Cash statistics \citep{Cash1979}. We focus on chi-squared results below, but the C-statistic results were consistent. We binned the PN spectrum with 25 counts per bin, and the merged MOS spectrum with 20 counts per bin. We ignored data above 10 keV, and allowed a (small) constant offset between the PN and MOS spectra. 

An absorbed power law gives an adequate description (reduced $\chi^2$\,=\,1.3 for 26 degrees of freedom (d.o.f.), null hypothesis probability of 13 per cent) of the spectrum, with a plausible N$_{\rm H}$\,=\,4.9\,$\times$\,10$^{21}$ cm$^{-2}$. We note that some structure remains in the residuals of this fit, and the slope of the power law is high (3.0\,$\pm$\,0.5). More extensive modelling of the X-ray spectrum is presented in Appendix \ref{sec:apxrayfitting}, but no firm conclusions can be drawn from these fits. We do note that the inferred flux does not change by more than $\sim$\,25 per cent between different models.

The unabsorbed 0.5\,--\,10 keV flux from the XMM\,--\,{\it Newton} observation, assuming the power law fit, is F$_{\rm X}$\,=\,2\,$\times$\,10$^{-13}$ \flx, while the unabsorbed X-ray flux inferred from the Chandra observation in the same energy range is 4.3\,$\times$\,10$^{-13}$ \flx. The X-ray detections by XMM-{\it Newton} and Chandra were dated on 2005 September 18 and 2011 November 2, respectively, suggesting that the source is a persistent X-ray emitter.
We convert the interstellar reddening estimates obtained from the SED into a distance range of 1.5\,--\,4 kpc to CXB12 using the 3D reddening map by \citet{Green2015}. Alternatively, using the IR reddening map by \citet{Schultheis2014} results in a distance of 2\,--\,4 kpc. This implies that the absolute magnitude of the optical source, with $r^{\prime}$\,=\,12.35 mag, is in the range --0.5 to --3.5 (corrected for reddening), indicative of a giant star. If we convert the extinction obtained from the IR reddening map in the K-band from \citet{Schultheis2014} to 2175\,\AA$\ $assuming the extinction law of \citet{Cardelli1989}, we find the extinction at 2175\,\AA$\ $to range between 3.5 mag for 2 kpc and 7.8 mag for 4 kpc.

Assuming a distance to CXB12 of 2 kpc with the power law fit yields an unabsorbed X-ray luminosity of L$_X$\,=\,9\,$\times$\,10$^{31}$ erg s$^{-1}$. Alternatively, using the X-ray flux obtained from the Chandra observation yields an unabsorbed X-ray luminosity of L$_X$\,=\,2\,$\times$\,10$^{32}$ erg s$^{-1}$. From the UVW2 measurement we infer an absorbed UV luminosity of L$_{UV}$\,$\sim$\,10$^{33}$ erg s$^{-1}$. An extinction of 3.5 mag implies that the intrinsic flux is 25 times higher than the observed one, which yields an UV luminosity of L$_{UV}$\,=\,2.5\,$\times$\,10$^{34}$ erg s$^{-1}$ at 2 kpc.

CXB12 is an X-ray source located in a relatively unobscured line of sight, at a distance of 2\,--\,4 kpc. A bright interloper is located less than 1 arcsec from the multiwavelength counterpart. The observed SED shows excess emission in the UV, suggesting the presence of a hot, compact object in addition to a cool component. We derive an absolute magnitude for the optical counterpart of the order M$_{r^{\prime}}$\,=\,--1 (assuming the distance inferred from the SED spectral fitting), indicative of a giant companion. 

The presence of an UV excess combined with a giant companion leaves open several possibilities regarding the nature of CXB12, including an RS CVn system, a symbiotic binary or a (q)LMXB. Given the temperature of the hot component we obtain from the SED fitting (T\,$\sim$\,37000\,$\pm$\,15000 K), we can rule out an RS CVn scenario as the class definition requires the presence of an F--G type star. The minimum inferred temperature for the hot component, taking into account the error budget, is $\sim$\,22000 K, which would correspond to a B spectral type. In addition, the luminosity inferred from the Chandra observation is a factor 2 higher than typically found in RS CVn (e.g. \citeauthor{Dempsey1993b} \citeyear{Dempsey1993b} find X-ray luminosities up to 7\,$\times$\,10$^{31}$ erg s$^{-1}$). The high UV luminosity is also hard to explain in this scenario.

On the other hand, a compact remnant such as a WD or NS are compatible with the SED fitting results. The X-ray and UV luminosity are consistent with a long period CV in orbit with a giant companion (symbiotic binary) in quiescence. In this respect, the absence of photometric optical variability is unusual but not unheard of if the donor dominates the optical light (e.g. \citeauthor{Mukai2016} \citeyear{Mukai2016}, \citeauthor{Sokoloski2017} \citeyear{Sokoloski2017}). Finally, the system properties could also be explained in a quiescent LMXB scenario given the X-ray and UV luminosities. In this case, the absence of photometric variability could be explained by the companion dominating the optical light. 

We conclude that based on the available observations, we cannot establish a firm classification for CXB12. Follow-up observations of this intriguing object are required to confirm the nature of the system. 

\subsubsection{CXB34 = CXOGBS J174728.9--321441, a state-changing polar}
\begin{figure} 
  \includegraphics[height=5cm, keepaspectratio]{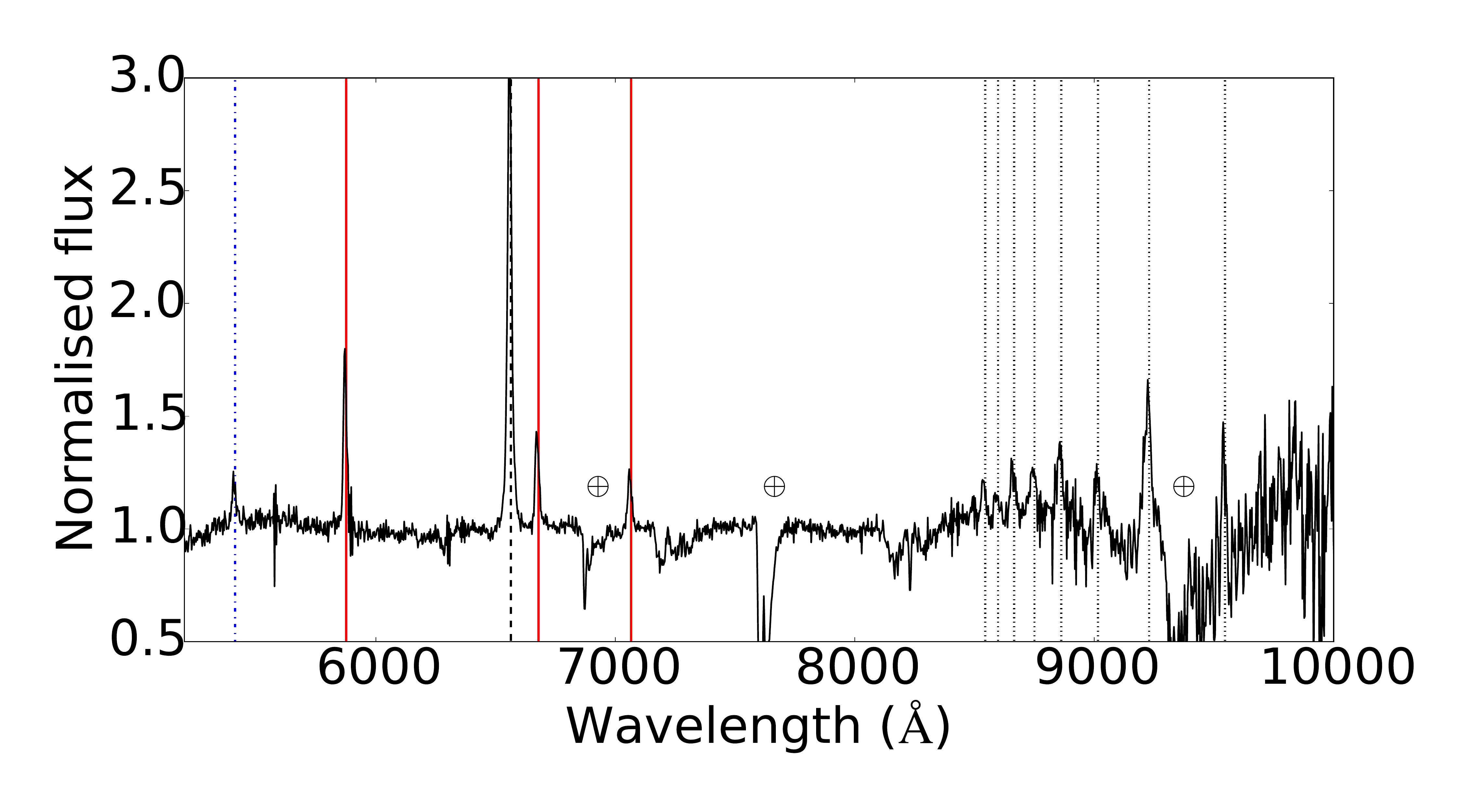}
  \caption{Continuum normalized OSIRIS spectrum of CXB34. H\,$\alpha$ is marked by a dashed line, while He\,\textsc{i} lines are marked by solid lines. H Paschen transitions are identified by dotted lines, and the dash-dotted line is He\,\textsc{ii} $\lambda5412$.}
  \label{fig:cxb34}
\end{figure}

\begin{figure} 
\centering
  \includegraphics[height=5cm, keepaspectratio]{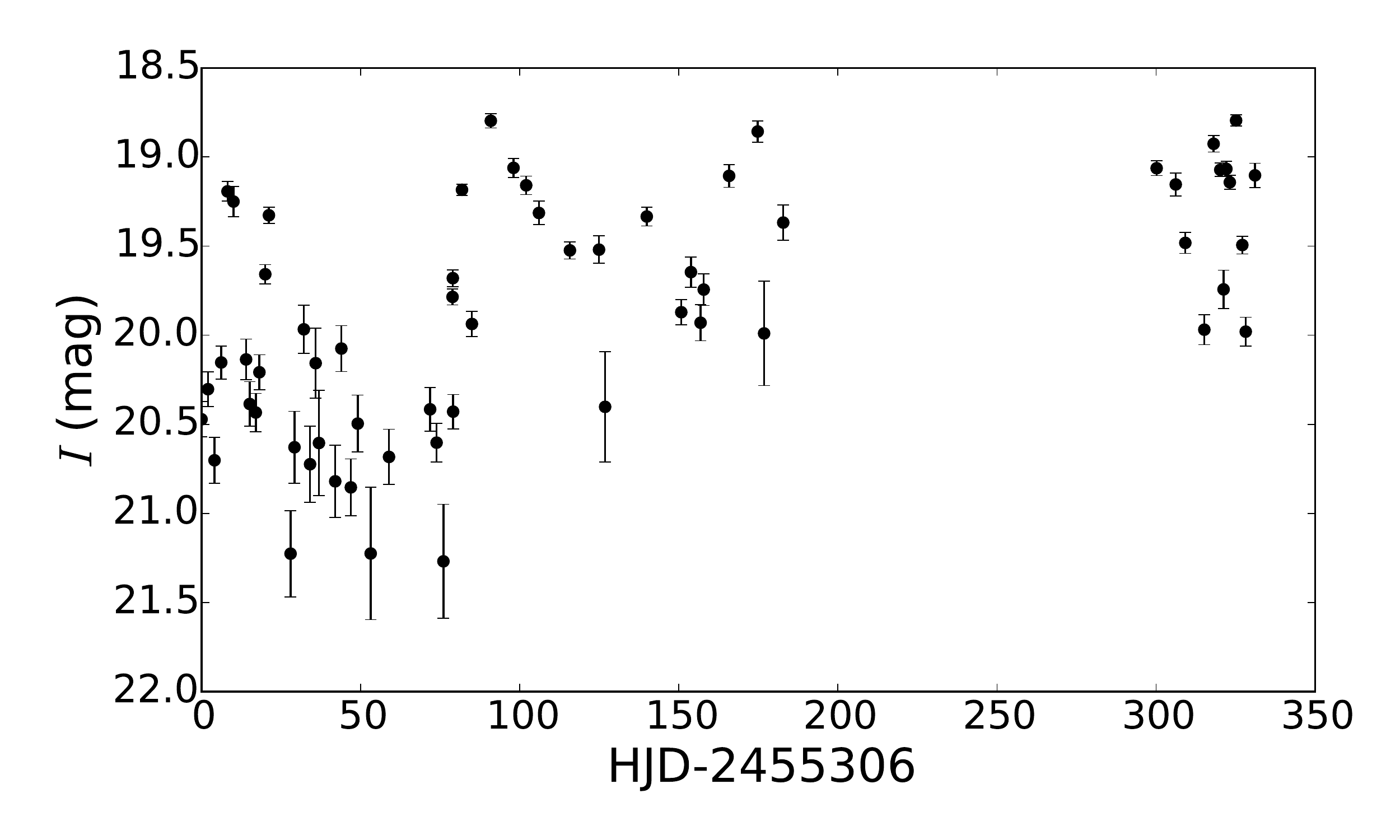}
  \caption{OGLE lightcurve of CXB34. Strong flickering $\sim$\,1 mag as well as a change from an optically faint to bright state are observed.}
  \label{fig:cxbogle}
\end{figure}

CXB34 stands out as a blue outlier to field stars in an optical colour-colour diagram \citep{Wevers2016c}.  We show the averaged GTC spectrum in Figure \ref{fig:cxb34}. 
Single-peaked emission lines are visible, in particular H\,$\alpha$ and He\,\textsc{i} $\lambda\lambda$\,5876,6678,7065. We measure an H\,$\alpha$ EW\,=\,48.36\,$\pm$\,0.42\,\AA$\ $and intrinsic FWHM\,=\,873\,$\pm$\,11 km s$^{-1}$. H\,$\alpha$ is followed in strength by He\,\textsc{i} $\lambda$\,5876 with EW\,=\,13.5\,$\pm$\,0.8\,\AA$\ $and (intrinsic) FWHM\,=\,786\,$\pm$\,17 km s$^{-1}$. H Paschen lines P15 $\lambda8545$ up to P$\epsilon$ $\lambda9546$ are also observed in emission, with P$\zeta$ $\lambda9229$ being the strongest (EW\,=\,25\,$\pm$\,2\,\AA). We also detect He\,\textsc{ii} emission at $\lambda5412$ for which we measure an EW\,=\,3.2\,$\pm$\,0.3\,\AA$\ $and an intrinsic FWHM\,=\,937\,$\pm$\,8 km s$^{-1}$. We do not detect absorption features from a WD or late type donor star. No DIBs are visible in the spectrum. The Na D interstellar doublet is partially blended with He\,\textsc{i} $\lambda5876$. We constrain the EW of the Na D2 line to be $\leq$\,0.3\,\AA. \\The OGLE survey lightcurve \citep{Udalski2015} of CXB34, shown in Figure \ref{fig:cxbogle}, reveals that the system changes between a low state (around $I$\,=\,21 mag) and a high state around $I$\,=\,19 mag. Strong optical flickering with an amplitude $\sim$\,1 mag, as well as several short brightening episodes are also observed. The spectra were taken in the high state (as measured from the acquisition image). 

As for CX21 (Section \ref{subsec:cx21}), we use the Na D2 EW to estimate a reddening $E(B-V)$\,$\leq$\,0.05 and thereby N$_{\rm H}$ to be less than 3\,$\times$\,10$^{20}$ cm$^{-2}$. In addition, we constrain the distance to CXB34 using the limit on $E(B-V)$, the 3D reddening map by \citet{Schultheis2014} and the transformations in \citet{Schlegel1998}. We find that the source should be less than 0.5 kpc from Earth. From the 16 counts detected with Chandra we derive an unabsorbed X-ray flux of F$_X$\,=\,1.3\,$\times$\,10$^{-13}$ erg cm$^{-2}$ s$^{-1}$. The X-ray luminosity is therefore L$_X$\,$\leq$\,4\,$\times$\,10$^{30}$ erg s$^{-1}$ for a distance of $\leq$\,500 pc.

The photometric and spectroscopic characteristics of CXB34 imply that the optical light is most likely fully dominated by the accretion flow. Transitions between a faint and bright optical state ($\sim$\,2 mag amplitude or larger) are observed in certain types of CVs, specifically some non-magnetic nova-likes (so-called VY Scl stars) and the strongly magnetic polars \citep{Honeycutt2004, Kafka2005}. $\frac{F_X}{F_{opt}}$ in CXB34 ranges between 3.3 at minimum light to 0.5 at maximum brightness, consistent with both interpretations \citep{Britt2013}. However, the presence of He\,\textsc{ii} $\lambda$5412 suggest it is most likely a magnetic (polar) CV \citep{Warner1995}. We also note that the absolute magnitude (M$_{\rm I}$\,=\,10.2 at 500 pc) expected for CXB34 at its observed maximum brightness is low for a nova-like (e.g. fig. 4.16 in \citeauthor{Warner1995} \citeyear{Warner1995}), while the low X-ray luminosity is consistent with a faint magnetic CV (e.g. \citeauthor{Cool1993} \citeyear{Cool1993}; \citeauthor{Grindlay1995} \citeyear{Grindlay1995}; \citeauthor{Watson2016} \citeyear{Watson2016}). We also remark that an amplitude difference of 2 mag, although small, could be consistent with a DN outburst in a long period CV or an IP.
\begin{figure} 
\centering
  \includegraphics[height=6cm, keepaspectratio]{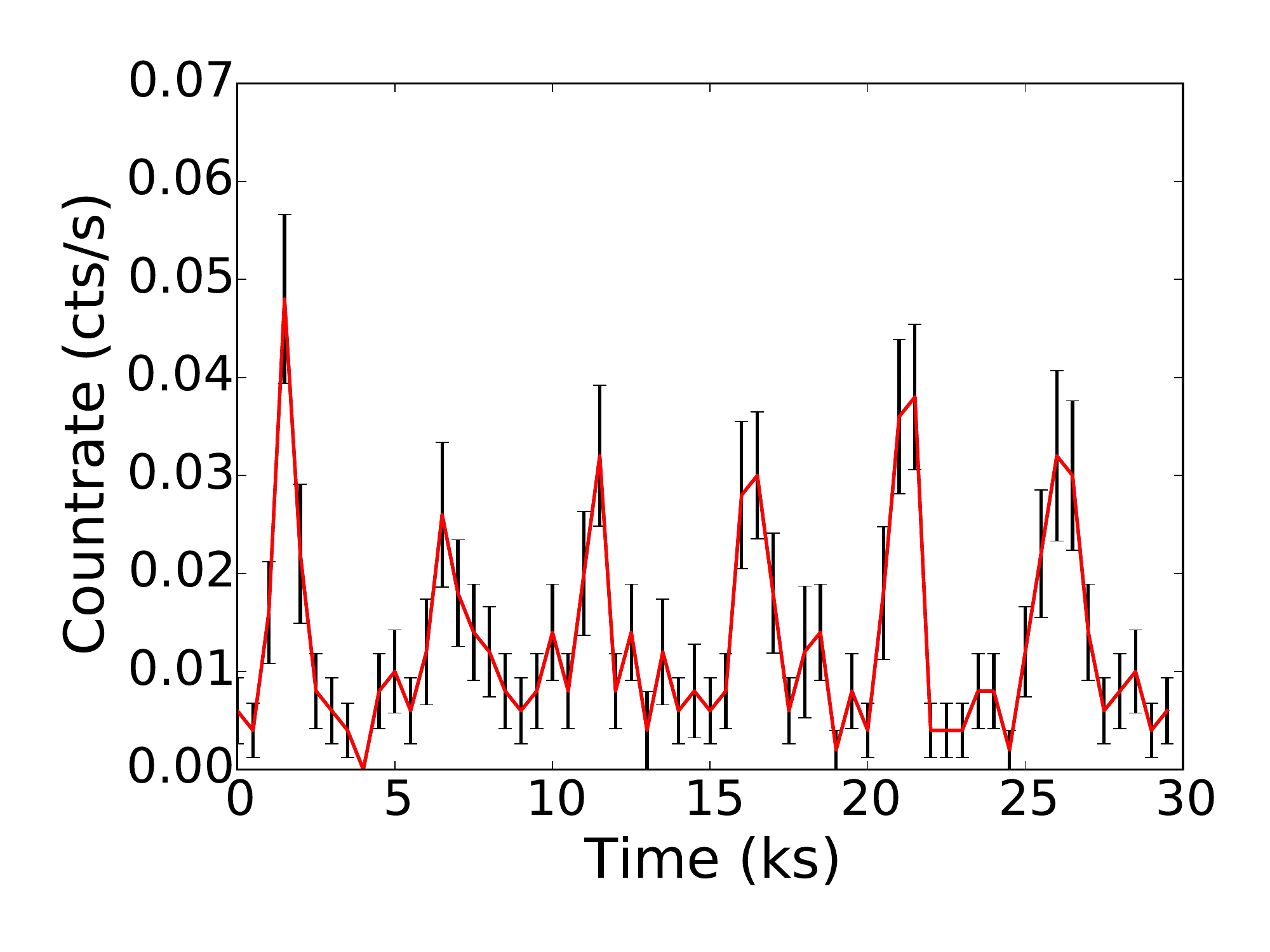}
    \includegraphics[height=6cm, keepaspectratio]{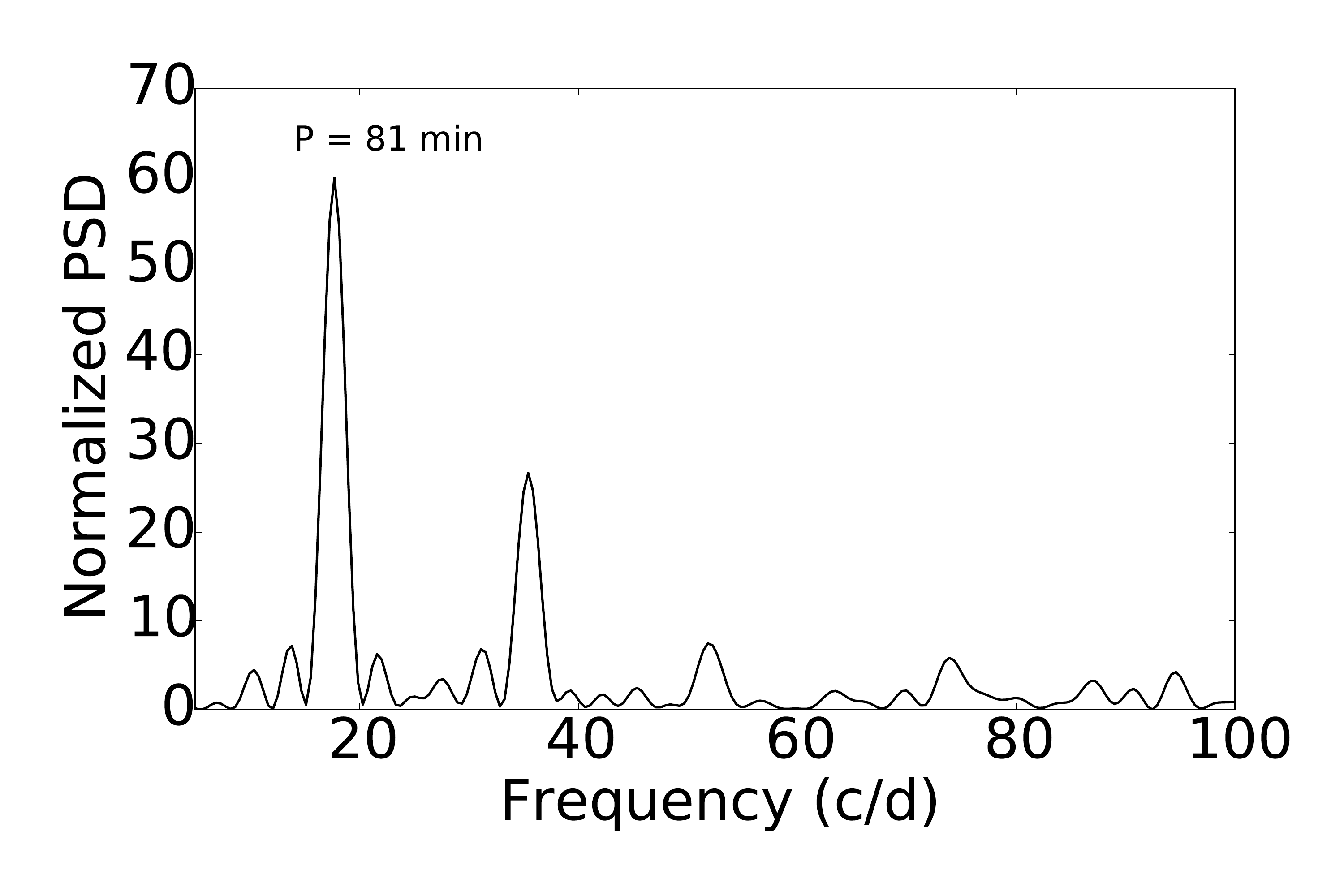}
  \caption{Top: XMM EPIC-PN lightcurve of CXB34, binned into 500s segments. The signal is shown in red, the errors in black. A periodically modulated signal is observed with a period of $\sim$\,5 ks. Bottom: Lomb-Scargle periodogram of the lightcurve shown in the top panel. The peak with the most power is located at P\,=\,81\,$\pm$\,2 minutes (approximately 4900 s).}
  \label{fig:cxb34xmmlc}
\end{figure}

To unambiguously determine the nature of this system, we turn to the archival XMM--{\it Newton} observation. We show the EPIC-PN lightcurve, binned in 500\,s segments, in the top panel of Figure \ref{fig:cxb34xmmlc}. It shows periodically modulated brightness variations which we interpret as X-ray pulsations. A Lomb-Scargle periodogram is shown in Figure \ref{fig:cxb34xmmlc} (bottom). We find the strongest peak at a period of 81\,$\pm$\,2 min, although we caution that the data only covers 6 cycles. Longer X-ray observations should be obtained to confirm this result. The peak around 40.5 min is a harmonic of the aforementioned periodicity. An 81 minute orbital/spin period is in line with that observed in synchronous (polar) CVs \citep{Norton2004}. While a few transitional objects (asynchronous polars or synchronous IPs) have P$_{\text{spin}}$\,$\sim$\,4900 s, these systems are rare (e.g. fig. 1 in \citeauthor{Norton2004} \citeyear{Norton2004}). Regarding common IPs, none are observed to have such a long spin period. If the periodicity is interpreted as the orbital period of an IP, we would expect to see power in the periodogram revealing the spin period at $\sim$\,8 minutes. We also note that the tentative orbital period coincides with the period minimum for CVs \citep{Gaensicke2009}, indicating that the system is close to or beyond period bounce.

We classify CXB34 as a synchronous magnetic CV belonging to the AM Her subclass (polar) based on the X-ray periodicity. The detection of a change between a low and high accretion states has been reported in other AM Her binaries, e.g. in MR Ser (P$_{\rm orb}\sim$113 min, \citeauthor{Schwope1993} \citeyear{Schwope1993}), Bl Hydri (P$_{\rm orb}\sim$113 min, \citeauthor{Schwope1995} \citeyear{Schwope1995}) and PT Per (P$_{\rm orb}\sim$82 min, \citeauthor{Watson2016} \citeyear{Watson2016}).

\subsubsection{CXB42 = CXOGBS J175332.6--294648: a weak-lined T Tauri star with UX Ori-like variability}
\begin{figure} 
\centering
  \includegraphics[height=4.6cm, keepaspectratio]{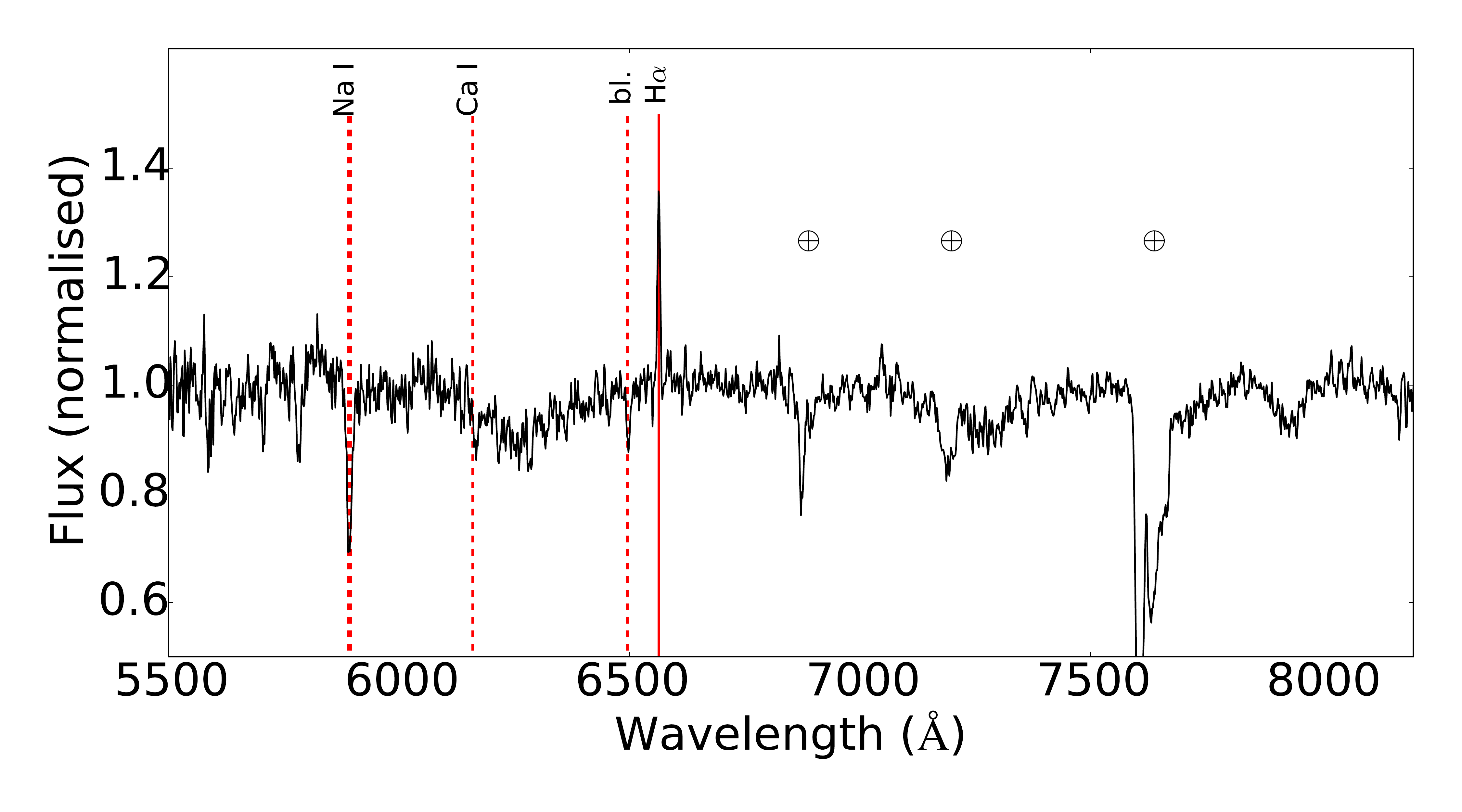}
  \caption{FORS2 spectrum of the optical counterpart to CXB42. We have marked typical stellar features as well as the position of H\,$\alpha$ with dashed and dotted lines, respectively. }
  \label{fig:cxb42spec}
\end{figure}

\begin{figure} 
\centering
  \includegraphics[height=6cm, width=0.43\textwidth]{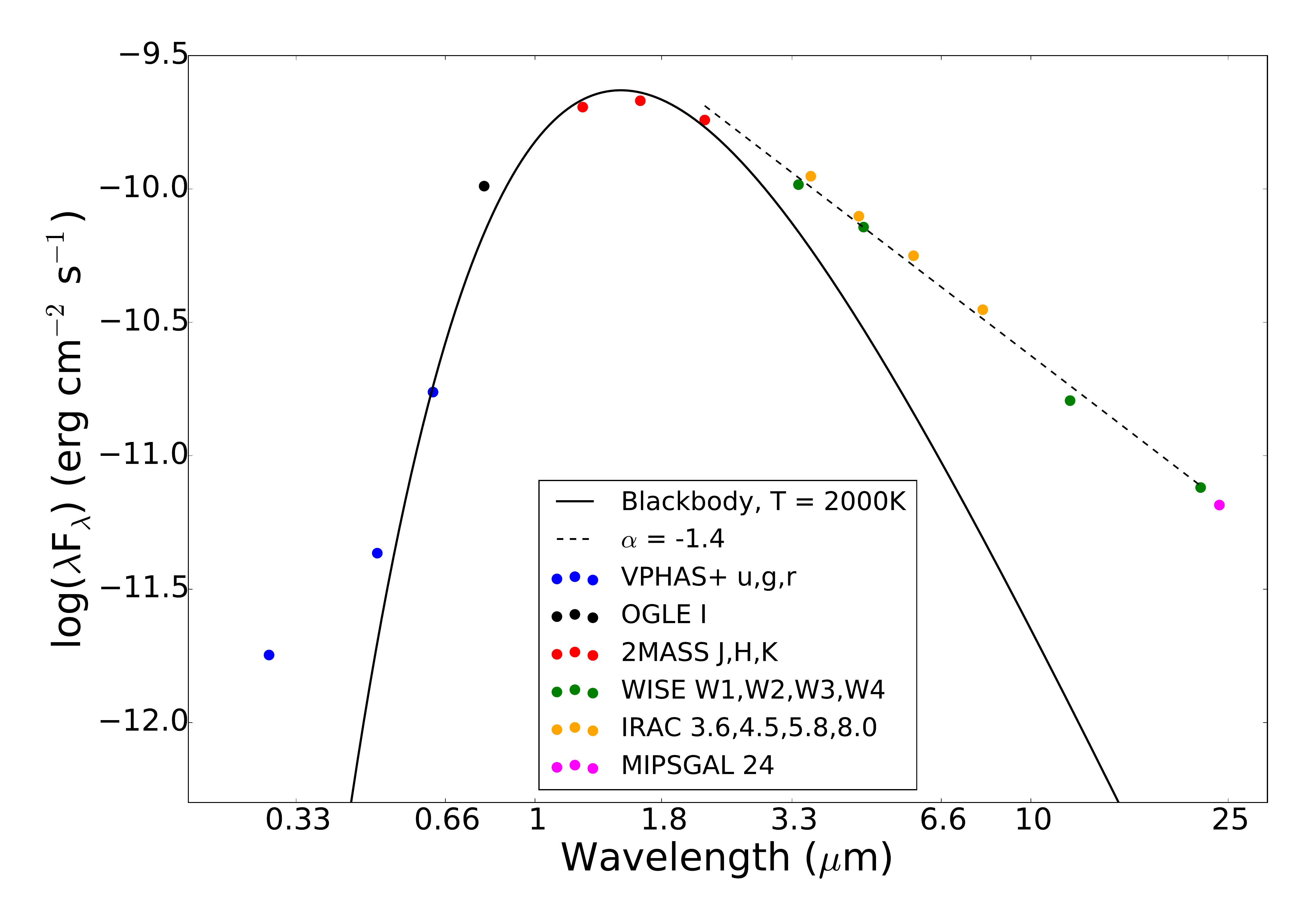}
  \caption{Optical and IR (reddened) SED of CXB42, compiled from the literature. The error bars are smaller than the data points. The solid line indicates a 2000 K blackbody. We fit the slope of the 2\,--\,22 $\mu$m part of the SED and find $\alpha$\,=\,--1.4, implying a class II YSO. }
  \label{fig:cxb42sed}
\end{figure}

CXB42 is associated with a bright optical (I\,=\,12.4) and IR (J\,=\,10.69\,$\pm$\,0.02, H\,=\,9.86\,$\pm$\,0.03, K\,=\,9.27\,$\pm$\,0.02) counterpart. We have obtained an optical spectrum with FORS2, which we show in Figure \ref{fig:cxb42spec}. 
The spectrum shows unresolved H\,$\alpha$ emission with EW\,=\,2.4\,$\pm$\,0.1\AA. The Ca\,+\,Fe blend at $\lambda6495$ and the TiO band at $\lambda6200$ are also visible, pointing towards a mid K spectral type (adopted from a visual comparison to template stars). We compile all the photometric measurements available in the literature and create the SED, shown in Figure \ref{fig:cxb42sed}. We note that the photometry was not taken simultaneously and variability may influence the measurements. We fit a blackbody to the observed SED without applying a reddening correction, and find that a temperature of T\,=\,2000 K best fits the data. Assuming this blackbody represents the reddened stellar contribution (which can explain why the temperature does not seem to agree with the spectral type inferred from the spectrum), we also deduce the presence of an IR excess component starting in the K$_S$-band and a potential blue excess. The u$^{\prime}$-band detection implies that the source is relatively close to Earth. We fit the slope of the 2\,--\,22 $\mu$m IR excess (shown as the dashed line in Figure \ref{fig:cxb42sed}) and find $\alpha$\,=\,--1.4. Chandra detected 15 counts in the GBS observations, yielding an absorbed X-ray flux of F$_X$\,=\,1.2\,$\times$\,10$^{-13}$\,erg cm$^{-2}$ s$^{-1}$.
\begin{figure} 
\centering
  \includegraphics[height=5cm, width=0.4\textwidth]{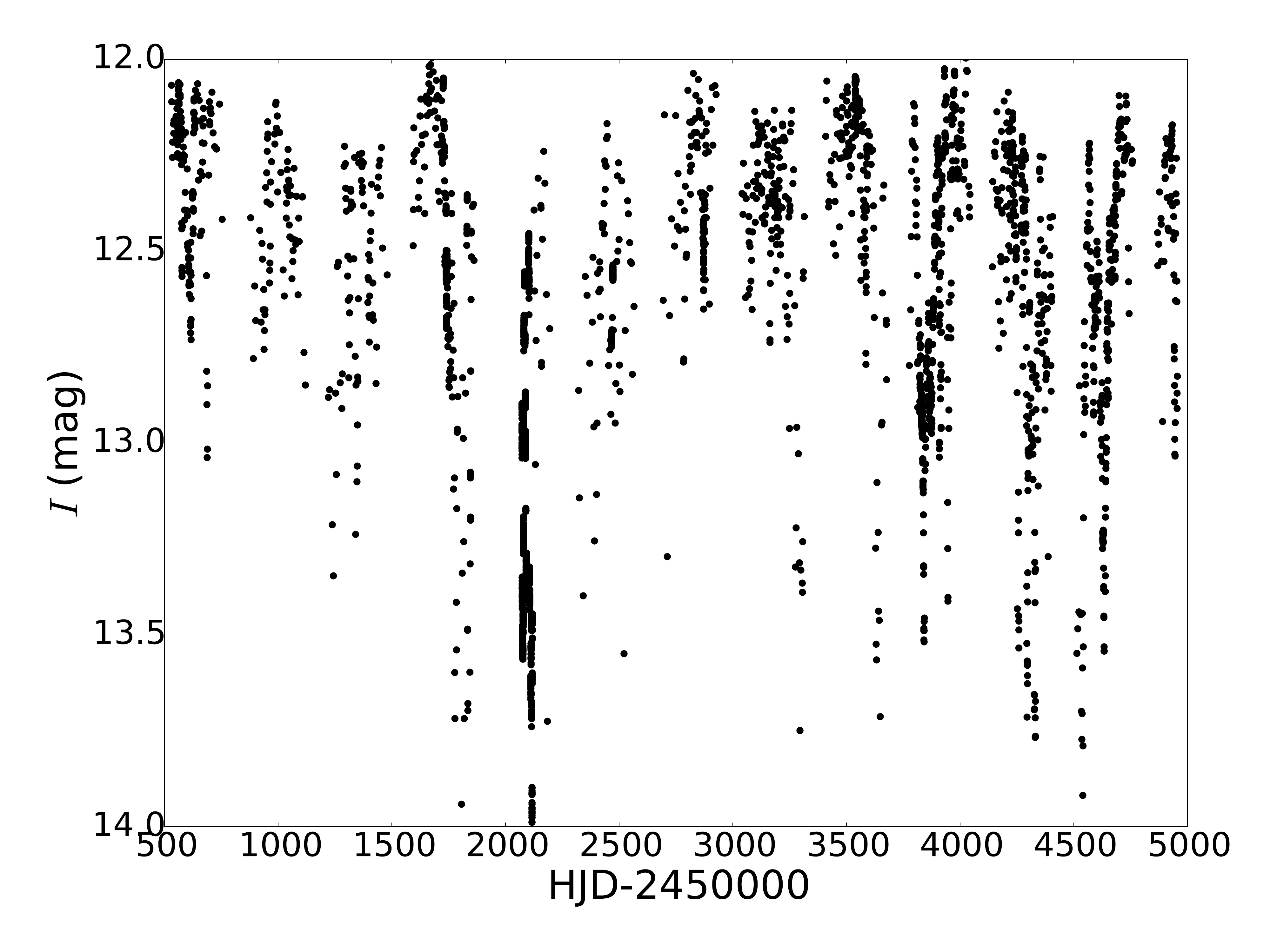}
  \includegraphics[height=5cm, width=0.4\textwidth]{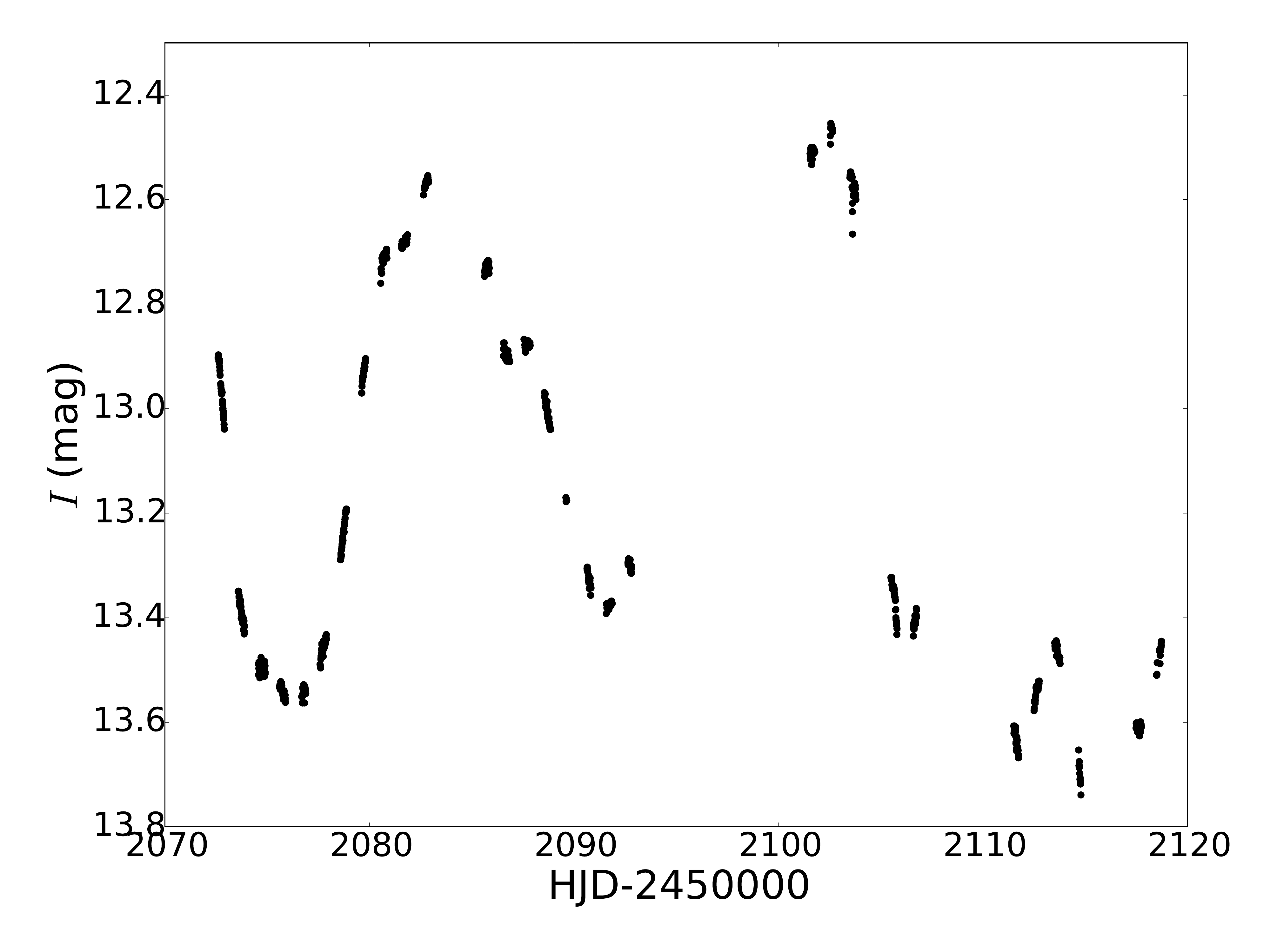}
  \caption{Top: OGLE lightcurve of CXB42. The error bars are smaller than the data points. Deep, irregular dips are observed, as well as a seasonal variation of the mean magnitude of the system. The latter feature is typical for lightcurves in UX Ori like systems. Bottom: zoom of a densely sampled region of the lightcurve.}
  \label{fig:cxb42lc}
\end{figure}
\begin{figure} 
\centering
  \includegraphics[height=6cm, width=0.43\textwidth]{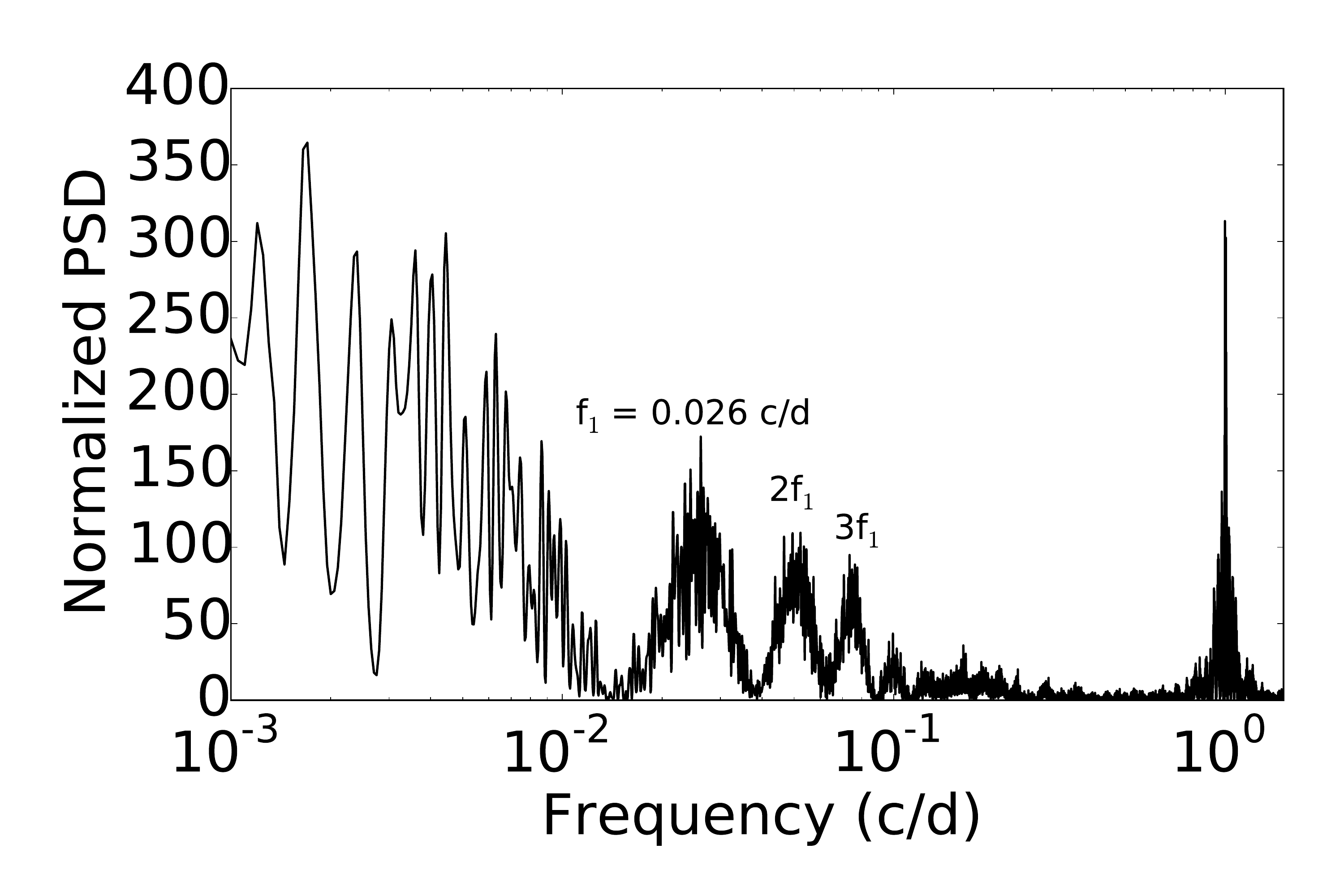}
  \caption{Lomb-Scargle periodogram of the lightcurve of CXB42. There is a quasi-periodic signal at $f_1$\,=\,0.026 c/d; harmonics of this periodicity are also visible and marked. The power at lower frequency is likely associated to the long term variations in the lightcurve. The narrow peak at 1 c/d is introduced by the sampling.}
  \label{fig:cxb42ls}
\end{figure}

We present the OGLE $I$-band lightcurve in Figure \ref{fig:cxb42lc}. The data span 12 bulge observing seasons, and the lightcurve shows deep, irregularly spaced dips of up to 1.6 mag in amplitude (75 per cent of the stellar light). Lower amplitude variations of the mean magnitude from season to season are also obvious. In the bottom panel of Figure \ref{fig:cxb42lc} we zoom into a high cadence part of the lightcurve for this object. The observed dimming event is smooth and symmetric, lasting for about 9 days; more erratic variability is also visible. A LS periodogram of the lightcurve is presented in Figure \ref{fig:cxb42ls}. Besides the narrow peak at 1 c/d, which is introduced by the sampling of the lightcurve, there are several peaks at frequencies that are integer multiples of $f_1$\,$\sim$\,0.026 c/d or 38 days. The two peaks at 2\,$f_1$ and 3\,$f_1$ are most likely harmonics of the dominant periodicity, which is a sign of non sinusoidal periodic variations. The broadness of the peaks in the periodogram indicates that the variations are quasi-periodic. The power at lower frequencies is likely related to the long term variations in the lightcurve.

The IR counterpart to CXB42 was identified as a YSO candidate by \citet{Marton2016} on the basis of the IR colours. Here we confirm the YSO nature of this object based on spectroscopic and photometric observations. First, \citet{Marton2016} showed that the IRAC IR colours are consistent with those of class II YSOs \citep{Herbig1989} according to \citet{Robitaille2008}. Second, the IR part of the SED follows a power-law with a slope between 2\,--\,22\,$\mu$m of $\alpha$\,=\,--1.4, roughly consistent with the class II definition of \citet{Greene1994}. Finally, the estimated spectral type, combined with the low measured EW of the H\,$\alpha$ emission line (EW\,=\,--2.4\,\AA) are consistent with a weak-lined T Tauri star (WTTS, usually defined as having K1 or later spectral type and EW(H\,${\alpha}$)\,$\leq$\,10\AA, \citeauthor{Herbst1999} \citeyear{Herbst1999}). The X-ray detection of the system is also consistent with a WTTS interpretation \citep{Telleschi2007}. The X-ray emission is thought to be produced in the chromosphere of the pre-main sequence star \citep{Kastner2004}.

The occurrence of frequent deep, irregular dips in the lightcurve suggest that the extinction occurs in the circumstellar rather than the interstellar medium. This kind of variability is typical for UX Ori-like variable stars (see \citeauthor{Grinin1998} \citeyear{Grinin1998} for example lightcurves). However, the nature of the irregular dips is not completely understood. One possible explanation is that they are due to clumpy circumstellar material in the proto-planetary disc passing through our line of sight \citep{Herbst2012}. Other models infer variable accretion, in combination with warped or misaligned discs to explain the variability \citep{Kesseli2016}. In this scenario, the quasi-periodicities found in the periodogram could be related to the orbital or precession period of the material in the disk, although the exact nature remains unclear. UX Ori-like variability has so far been observed in early and classical T Tauri stars (see \citeauthor{Herbst1994} \citeyear{Herbst1994} for a definition), which are still in their main accretion phase. In contrast, WTTS are generally thought to have very low residual accretion rates (as indicated by the weak H\,$\alpha$ emission), i.e. they are in a more evolved stage of their pre-main sequence evolution. \citet{Herbst1999} have claimed that UX Ori-like variability only occurs in early type YSOs; however, \citet{Lada2006} find that $\sim$\,10 per cent of WTTS are surrounded by an optically thick disc. \citet{Ansdell2016} found a dozen UX Ori-like systems using Kepler data, some of which are spectroscopically consistent with WTTSs. 

We remark that CXB42 bears striking resemblances to the T Tauri UX Ori-like system RZ Psc \citep{Hoffmeister1931}. The presence of an IR excess, the photometric variability and the X-ray emission are all similar between the two objects. For RZ Psc, which is well characterized but not well understood (\citeauthor{Grinin2010} \citeyear{Grinin2010}), several scenarios have been put forward to explain the photometric variability. These range from the presence of an asteroid belt \citep{dewit2013} to magnetic propeller effects \citep{Grinin2015} or disc-magnetosphere interactions \citep{Shulman2015}. Despite the large amount of available data, no theoretical framework is at present able to unambiguously identify the geometry of the system, nor the origin of the photometric variability. 

In conclusion, we classify CXB42 as an analog to the post UX Ori system RZ Psc. An in-depth observational study of CXB42, including high-resolution spectroscopy, spectropolarimetry and multi band lightcurves may allow for progress in understanding the UX Ori phenomenon.

\subsubsection{CXB43 = CXOGBS J175039.6--302056, unknown nature}
\begin{figure}
\centering
  \includegraphics[height=5cm, keepaspectratio]{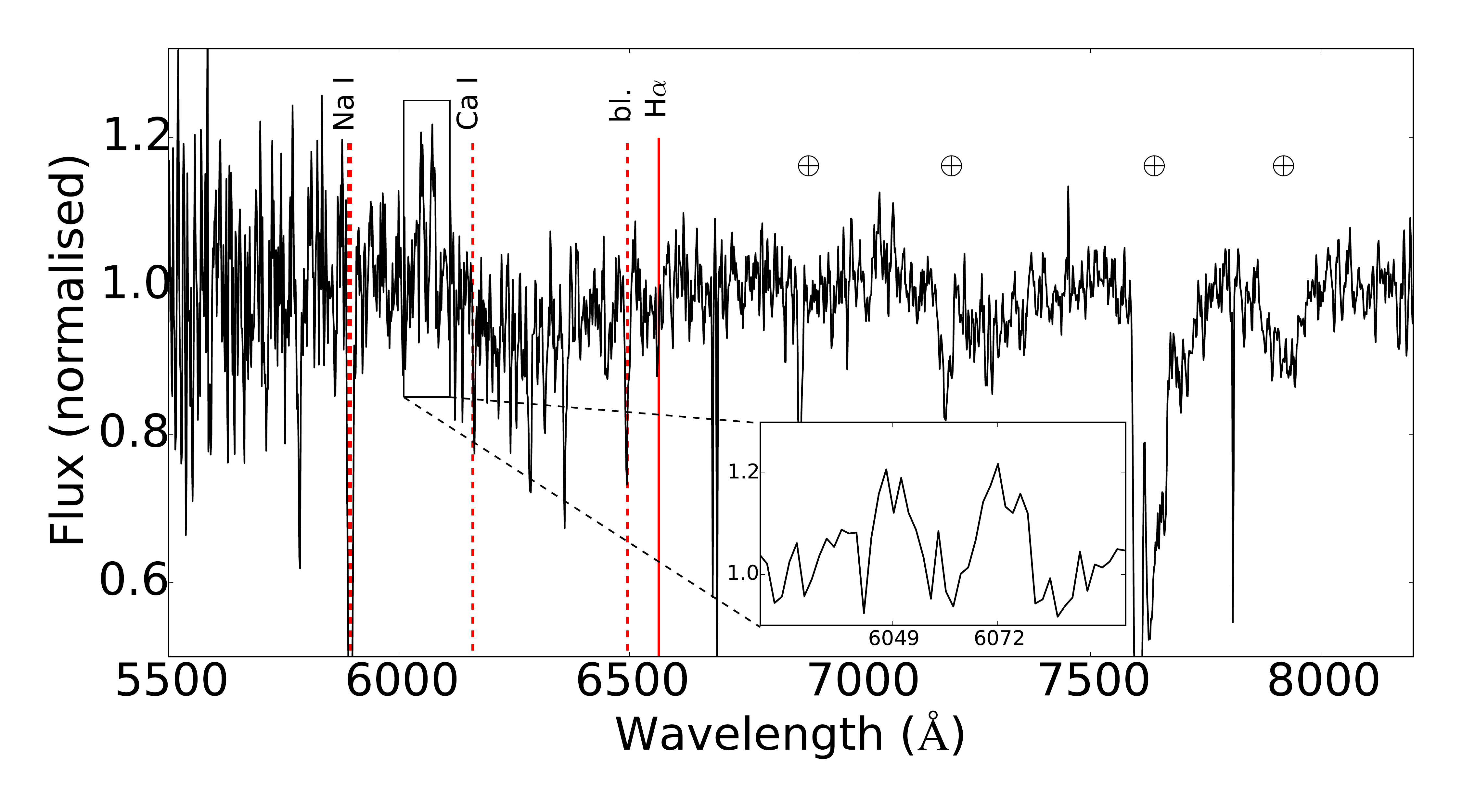}
  \caption{FORS2 spectrum of the optical counterpart to CXB43. We have marked typical stellar features as well as the position of H\,$\alpha$ with dashed and solid lines, respectively. The inset shows a zoom of the region containing the two double-peaked emission lines. }
  \label{fig:cxb43spec}
\end{figure}

\begin{figure} 
\centering
  \includegraphics[height=5cm, width=0.4\textwidth]{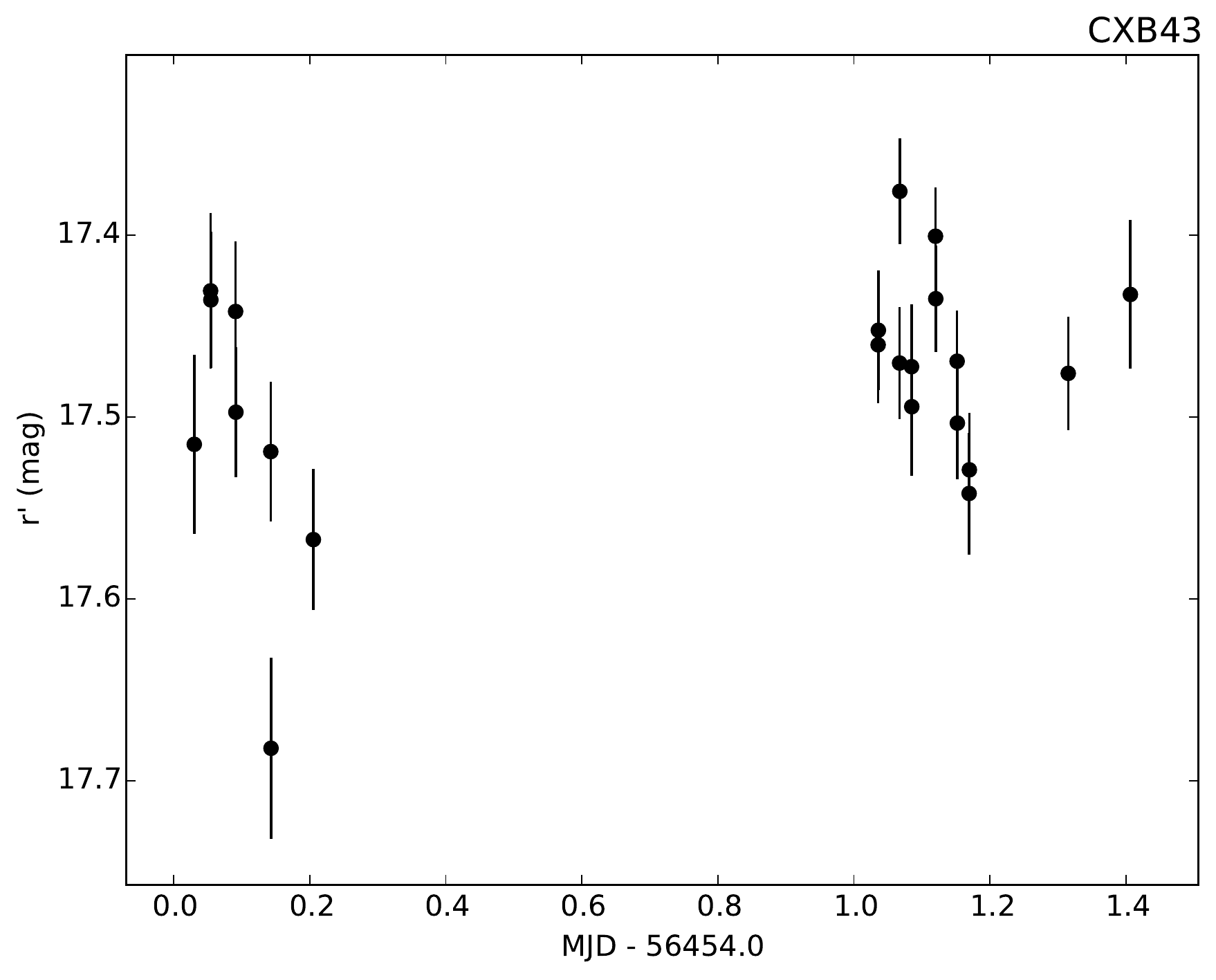}
  \caption{DECam lightcurve of CXB43, spanning 2 nights. Flickering with amplitude $\sim$\,0.2 mag is visible.}
  \label{fig:cxb43lc}
\end{figure}

The optical counterpart to CXB43 was detected by \citet{Wevers2016} at $r^{\prime}$\,=\,18.50\,$\pm$0.05, $i^{\prime}$\,=\,16.59\,$\pm$\,0.05 and H\,$\alpha$\,=\,17.62\,$\pm$\,0.05. The acquisition image of the spectroscopic observation, taken in 0.6 arcsec conditions, reveals that the source could be a blend of two objects. We show the spectrum, obtained with FORS2 on 2016 April 4, in Figure \ref{fig:cxb43spec} and the DECam lightcurve, obtained on 2013 June 11 and 12, in Figure \ref{fig:cxb43lc}. The source was detected at $r^{\prime}$\,=\,17.45 mag in the DECam images, 0.85 mag brighter than the earlier detection (in 2006) presented by \citet{Wevers2016}. This indicates that the source is optically variable, although we caution that it is unresolved in the DECam data and we cannot completely rule out that another source is causing the apparent variability if it is blended. Assuming the variability is related to the X-ray counterpart, the presence of flickering with an amplitude of $\sim$\,0.2 mag in the DECam data suggests a significant contribution from the accretion flow to the optical light. However, the FORS2 spectrum lacks H or He emission lines (Figure \ref{fig:cxb43spec}). We identify Ca\,\textsc{i} and blended Fe\,\textsc{i} absorption lines in the data. The strong Na\,\textsc{i}\,D (EW\,=\,8.5\,$\pm$\,0.2 \AA) absorption feature likely has two contributions, from the companion star and the interstellar medium. We also see two (double-peaked) emission lines at $\lambda\lambda$6049,6072. We measure EWs of 1.5\,$\pm$\,0.3 \AA\ for the $\lambda$6049 feature and 2.4\,$\pm$\,0.2 \AA\ for the $\lambda$6072 feature. We also measure a FWHM\,=\,448\,$\pm$\,35 km s$^{-1}$ and 566\,$\pm$\,63 km s$^{-1}$ respectively. We determine a FWHM spectral resolution of $\sim$150 km s$^{-1}$ at H\,$\alpha$ using skylines, indicating that these lines are resolved. We carefully checked that no artefacts or cosmic rays are present in the raw 2D spectra. We caution that we used a wide slit (1.5 arcsec) in 0.6 arcsec seeing conditions, so the line positions may be affected by centroiding issues if the source was not well centred. However, we inspected the through-slit images and do not find evidence for this.

To constrain the distance to the system, we measure the EW of the DIB at $\lambda6284$ as 1.6\,$\pm$\,0.2 \AA. The DIB at $\lambda5780$ is blended with a stellar absorption line and hence not suitable for a distance determination. Based on the measured EW of the DIB at $\lambda6284$ we derive an $E(B-V)$\,=\,1.4 using \citet{Cordiner2011}, which converts to a distance of $\sim$\,5.5 kpc using the reddening map of \citet{Schultheis2014}. This distance estimate is consistent with the absolute magnitude of a late K giant companion at that distance, considering a reddening of A$_{r^{\prime}}$\,=\,4 mag. 
An IR counterpart is indeed detected in the UKIDDS Galactic Plane Survey at J\,=\,13.53, H\,=\,12.42, K$_S$\,=\,12.06 mag. At a distance of 5.5 kpc, the observed K$_S$-band magnitude implies an absolute magnitude of M$_K$\,=\,--0.3, again consistent with a late K giant. The source is variable in the K$_S$-band with an amplitude of $\sim$\,0.2 mag \citep{Lucas2008}.

The 15 counts detected by Chandra imply an unabsorbed X-ray flux of F$_X$\,=\,2\,$\times$\,10$^{-13}$ erg cm$^{-2}$ s$^{-1}$ (assuming N$_{\rm H}$\,=\,8\,$\times$\,10$^{21}$ cm$^{-2}$). At 5.5 kpc this gives an intrinsic X-ray luminosity of L$_X$\,=\,7\,$\times$\,10$^{32}$ erg s$^{-1}$. From the $i^{\prime}$-band magnitude in \citet{Wevers2016} we derive $\frac{F_X}{F_{\text{opt}}}$\,=\,0.05 at minimum light.  

Regarding the interpretation of this source, we advance two possibilities. Firstly, the system properties are consistent with a symbiotic binary scenario in quiescence. The absence of nebular emission lines in the optical spectrum is not uncommon among quiescent symbiotic systems \citep{Hynes2014, Mukai2016}. The two emission lines cannot be easily explained in this scenario. 

Alternatively, we have tried to identify the two emission lines in an AGN scenario. There are pairs of emission lines with a similar ($\sim$\,23\,\AA) separation, e.g. O\,\textsc{iii} $\lambda\lambda4932,4960$ or H$\gamma$ and O\,\textsc{iii} $\lambda4364$, that could be redshifted and thus explain their presence in the optical spectrum originating in a background AGN. However, we are not able to identify more emission and/or absorption lines assuming these identifications and the implied redshifts. Moreover the lines mentioned above, when originating from an AGN, are not double-peaked, making this scenario unlikely.

We conclude that an AGN scenario can be ruled out, but follow-up observations are needed to verify the nature of the unidentified emission lines and establish a firm identification of CXB43. 

\subsection{Active single and binary stars}
\label{sec:cxb49}
\begin{figure*} 
  \includegraphics[height=9cm,keepaspectratio]{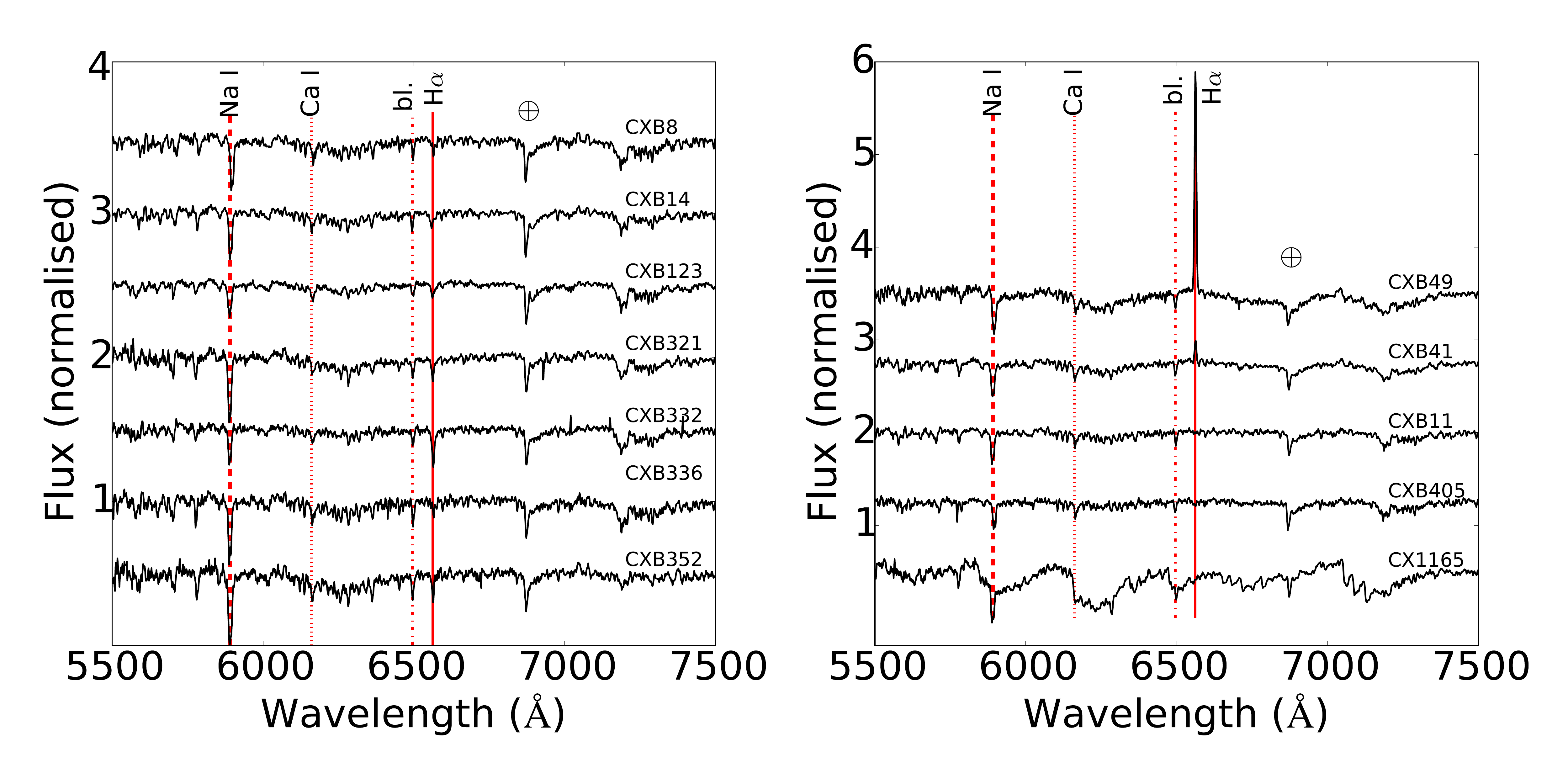}
  \caption{FORS2 spectra of bright optical counterparts to X-ray sources. The left panel shows sources that display stellar H\,$\alpha$ absorption. The right panel shows active systems with H\,$\alpha$ visible in emission (CXB41 and CXB49). The stellar H\,$\alpha$ absorption feature of CXB11 and CXB405 is likely filled in by an emission line.}
  \label{fig:activestarsfors}
\end{figure*}

\begin{figure} 
  \includegraphics[width=0.5\textwidth,keepaspectratio]{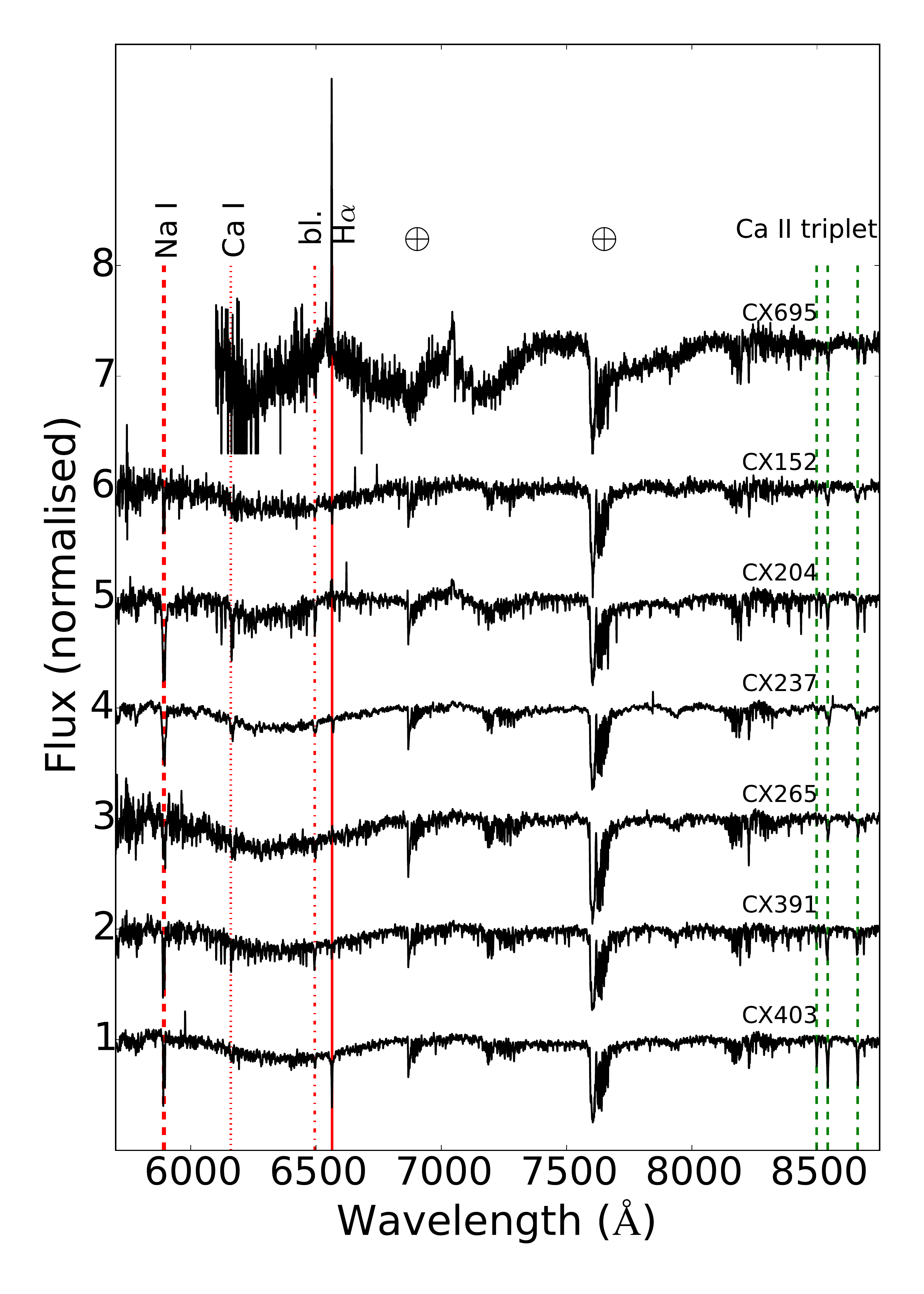}
  \caption{X-shooter spectra of bright optical counterparts to X-ray sources. We omitted the bluest part of the CX695 spectrum because of the low SNR. }
  \label{fig:activestarsxshooter}
\end{figure}

In addition to the systems classified above, we have also identified 18 active single/binary stars. We present the FORS2 spectra in Figure \ref{fig:activestarsfors} and the visible arm of the X-shooter spectra in Figure \ref{fig:activestarsxshooter}.
Some spectra show signs of H\,$\alpha$ in emission (CX204, CX265, CX695, CXB41, CXB49), while other sources show narrow (photospheric) H\,$\alpha$ absorption. The X-ray to optical flux ratio of these sources ranges between 0.001\,--\,0.01 (Table \ref{tab:outlierctparts}), consistent with the proposed active star interpretation. 

We inspect the region around the Calcium triplet at $\lambda\lambda8498,8542,8662$ in the X-shooter spectra, and find that 4 sources (CX204, CX237, CX391 and CX695) display double-peaked and/or asymmetric absorption line profiles. Periodicity information for these 4 sources is available from the literature (Table \ref{tab:outlierctparts}). We thus classify these systems as active binary stars.

Regarding CXB49, we resolve two optical sources in the acquisition image. Using the improved X-ray position of CXB49 at ($\alpha$, $\delta$)\,=\,(267.37037, --31.30674), with an uncertainty of 0\farcs73. CXB49 is thus classified as an active star. We refer to Appendix \ref{sec:appendix} for a detailed finding chart of this system.

Finally, the spectra of CXB11 and CXB405 do not display H\,$\alpha$ emission nor absorption as is generally expected for G/K type stars (as their spectra suggest). The lack of a stellar H\,$\alpha$ absorption feature can be explained if it is filled in by H\,$\alpha$ emission due to chromospheric activity. The X-ray to optical flux ratios of 0.002 for both CXB11 and CXB405 are consistent with this interpretation. An alternative, but less likely explanation for the absence of the H\,$\alpha$ spectral feature is that these systems contain a compact object but the optical light is dominated by the companion star. In this scenario, H\,$\alpha$ emission originating in the accretion disc fills in the absorption line from the donor star (see e.g. \citeauthor{Wu2015} \citeyear{Wu2015}).

\section{Summary}
\label{sec:summary}
We combine optical time-resolved photometric and spectroscopic observations with archival multiwavelength data to classify 26 GBS X-ray sources. Our sample contains 4 photometrically selected H\,$\alpha$ emitters and 1 blue outlier compared to field stars \citep{Wevers2016c}. Regarding these outliers, we classify CX21 as a quiescent short orbital period CV and CX118 as a nova-like CV. CXB34 is a magnetic CV (polar) with verified X-ray pulsations showing a change in accretion state, while CX266 and CX695 are chromospherically active stars. We confirm the presence of H\,$\alpha$ emission in the spectra of all the photometrically selected H\,$\alpha$ emission line sources from \citet{Wevers2016c}. 

In addition, we present spectra of 21 optically or NIR bright ($r^{\prime}$\,$\leq$\,18, H\,$\leq$\,14) or variable counterparts obtained as a queue filler programme at the VLT. CXB12 is a bright UV and optical source, but at present no solid classification can be given. We identify CXB42 as a weak-lined T Tauri star that exhibits UX Ori-like photometric variability. The presence of two double-peaked emission lines at $\lambda\lambda6049,6072$ in the spectrum of CXB43 can not be readily explained in any scenario. We note that the system properties could be consistent with a symbiotic binary star in quiescence. Follow-up observations are required to firmly establish the nature of this object. CXB11 and CXB405 do not display any spectral feature near H\,$\alpha$, suggesting the stellar H\,$\alpha$ absorption is filled in. This could be due to chromospheric activity, or alternatively (but less likely) by emission originating in an accretion disc. Among the remaining bright optical counterparts, 4 sources (CX204, CX695, CXB41 and CXB49) show evidence for the presence of narrow H\,$\alpha$ emission due to chromospheric activity, while 4 are likely binaries (CX204, CX237, CX391 and CX695) based on the Ca\,\textsc{ii} triplet morphology in the spectrum and periodicities reported in the literature. The X-ray to optical flux ratios of all these bright counterparts are consistent with an active star interpretation. 

%%%%%%%%%%%%%%%%%%%%%%%%%%%%%%%%%%%%%%%%%%%%%%%%%%%%%%%%%%%%%%%%%%%%%%%%%%%%%%%%%%%%%%%%%%%%%%%%%%%%%%%%%%%%%%%%%%%%%%%%%%%%%%%%%%%%%%%%%%%%%%%%%%%%%%%%%%%%%%%%%%%%%%%%%%%%%%%%%%%%%%%%%%%%%%%%%%%%%%%%%%%%%%%%%%%%%%%%%%
%%%%%%%%%%%%%%%%%%%%%%%%%%%%%%%%%%%%%%%%%%%%%%%%%%%%%%%%%%%%%%%%%%%%%%%%%%%%%%%%%%%%%%%%%%%%%%%%%%%%%%%%%%%%%%%%%%%%%%%%%%%%%%%%%%%%%%%%%%%%%%%%%%%%%%%%%%%%%%%%%%%%%%%%%%%%%%%%%%%%%%%%%%%%%%%%%%%%%%%%%%%%%%%%%%%%%%%%%%
\section*{Acknowledgements}
The authors would like to thank the referee, Christian Knigge, for his useful remarks and for pointing out some mistakes. His report improved the manuscript considerably.

PGJ and ZKR acknowledge support from European Research Council Consolidator Grant 647208. COH acknowledges support from an NSERC Discovery Grant, and Discovery Accelerator Supplement. DMS acknowledges Fundaci\'on La Caixa for the financial support in the form of a PhD contract. JC and DMS acknowledge support by the Spanish Ministerio de Economia y Competitividad under grant AYA2013-42627. JC also acknowledges support by the Leverhulme Trust through the Visiting Professorship Grant VP2-2015-046. RIH, CBJ, and JW acknowledge support from the National Aeronautics and Space Administration through Chandra Award Numbers AR3-14002X and AR5-16004X issued by the Chandra X-ray Observatory Center, which is operated by the Smithsonian Astrophysical Observatory for and on behalf of the National Aeronautics Space Administration under contract NAS8-03060. Based on observations collected at the European Organisation for Astronomical Research in the Southern Hemisphere under ESO programmes 085.D-0441(A), 085.D-0441(C), 087.D-0596(D), 091.D-0062(A), 097.D-0845(A) and 097.D-0845(C). 
The OGLE project has received funding from the National Science Centre, Poland, grant MAESTRO 2014/14/A/ST9/00121 to AU. Based in part on observations at Cerro Tololo Inter-American Observatory, National Optical Astronomy Observatory (2008B-0954, PI: Jonker), which is operated by the Association of Universities for Research in Astronomy (AURA) under a cooperative agreement with the National Science Foundation. Based on observations obtained at the Southern Astrophysical Research (SOAR) telescope, which is a joint project of the Minist\'{e}rio da Ci\^{e}ncia, Tecnologia, e Inova\c{c}\~{a}o (MCTI) da Rep\'{u}blica Federativa do Brasil, the U.S. National Optical Astronomy Observatory (NOAO), the University of North Carolina at Chapel Hill (UNC), and Michigan State University (MSU). Based on observations made with the Gran Telescopio Canarias (GTC), installed in the Spanish Observatorio del Roque de los Muchachos of the Instituto de Astrof\'isica de Canarias, in the island of La Palma. This work is based on observations obtained with XMM-–{\it Newton}, an ESA science mission with instruments and contributions directly funded by ESA and NASA. This research has made use of the NASA/ IPAC Infrared Science Archive, which is operated by the Jet Propulsion Laboratory, California Institute of Technology, under contract with the National Aeronautics and Space Administration. We thank Tom Marsh for developing the software package \textsc{molly}.

%%%%%%%%%%%%%%%%%%%%%%%%%%%%%%%%%%%%%%%%%%%%%%%%%%%%%%%%%%%%%%%%%%%%%%%%%%%%%%%%%%%%%%%%%%%%%%%%%%%%%%%%%%%%%%%%%%%%%%%%%%%%%%%%%%%%%%%%%%%%%%%%%%%%%%%%%%%%%%%%%%%%%%%%%%%%%%%%%%%%%%%%%%%%%%%%%%%%%%%%%%%%%%%%%%%%%%%%%%
%%%%%%%%%%%%%%%%%%%%%%%%%%%%%%%%%%%%%%%%%%%%%%%%%%%%%%%%%%%%%%%%%%%%%%%%%%%%%%%%%%%%%%%%%%%%%%%%%%%%%%%%%%%%%%%%%%%%%%%%%%%%%%%%%%%%%%%%%%%%%%%%%%%%%%%%%%%%%%%%%%%%%%%%%%%%%%%%%%%%%%%%%%%%%%%%%%%%%%%%%%%%%%%%%%%%%%%%%%

\bibliographystyle{mnras.bst}
\bibliography{bibliography_halpha.bib}
\appendix
\section{Additional X-ray spectral fitting of CXB12}
\label{sec:apxrayfitting}
\begin{figure} 
  \includegraphics[height=7cm, keepaspectratio]{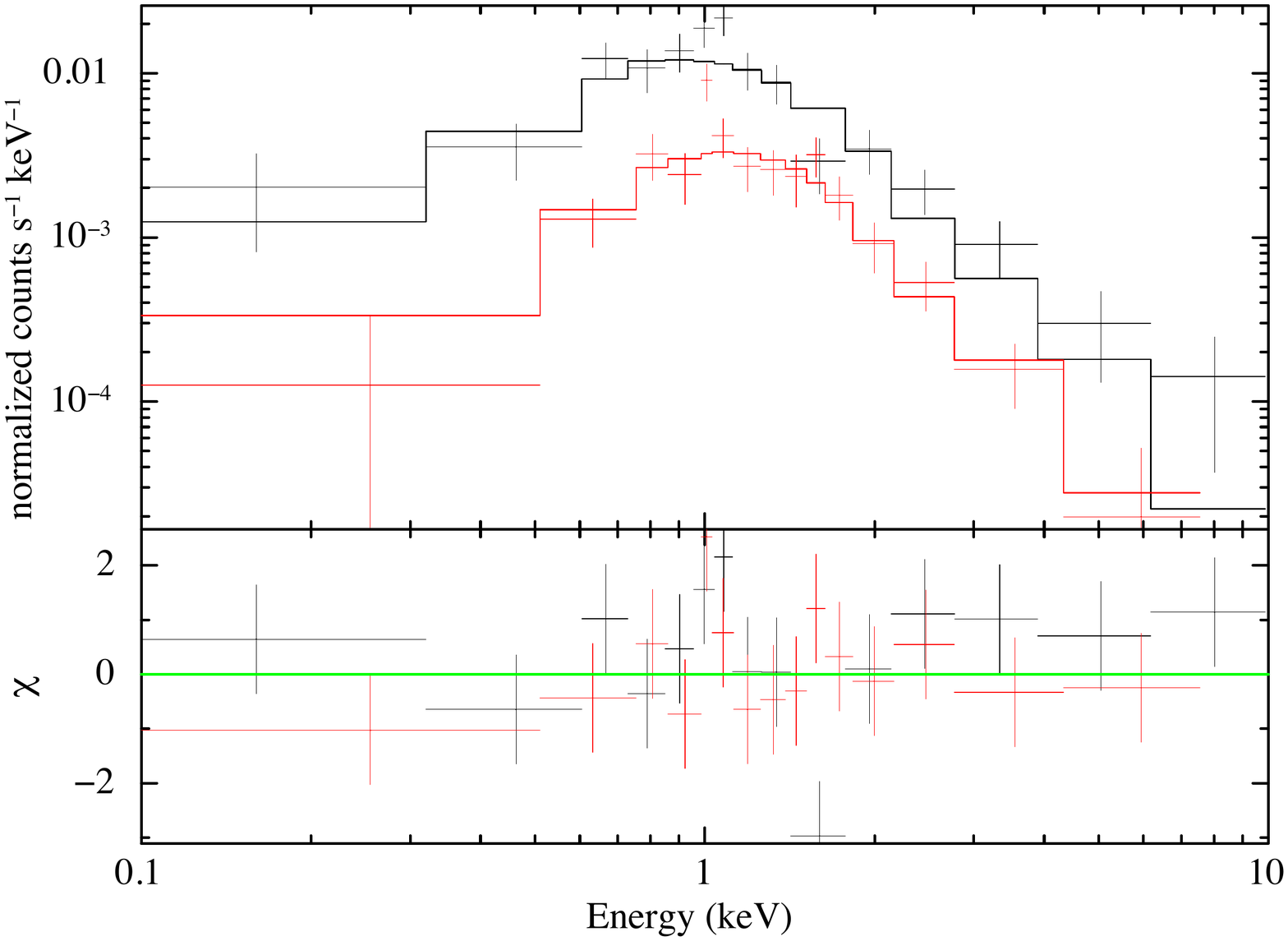}
  \caption{XMM PN and MOS spectra of CXB12. The fit shown is a two-APEC model, with temperatures of 0.3 and 1.2 keV respectively. The positive residuals around 1 keV could be Fe L-shell emission from a hot plasma, indicative of an RS CVn scenario. Their significance is such that we treat this as an indication only, and not as real evidence.}
  \label{fig:cxb12xrayspec}
\end{figure}

We note that the power law fit to the XMM--{\it Newton} spectrum has two peculiar features which motivated us to test different spectral models. Firstly, the photon index is unusually large (3.0\,$\pm$\,0.5); secondly, there are suggestive positive residuals in the spectrum at 1 keV (Figure \ref{fig:cxb12xrayspec}), where Fe L-shell lines are common in the emission from thermal plasma with temperatures of a few keV. 

This motivated us to try APEC thermal plasma models\footnote{http://atomdb.org}. Single-temperature solar abundance APEC models give unacceptable fits (reduced $\chi^2$\,=\,2.26, for 26 d.o.f.), as the 1 keV line complex is too strong. This line complex could be diluted either by reducing the abundances or by having the APEC model only provide part of the flux (the rest could be provided by a second, higher temperature APEC model, or by e.g. a power-law). In the following two paragraphs we explore these ideas further.

A two-APEC spectral model finds temperatures of $0.3^{+1.2}_{-0.1}$ keV and $1.9^{+0.6}_{-0.5}$ keV, with N$_{\rm H}$\,=\,4$^{+5}_{-3}$\,$\times$\,10$^{21}$\,cm$^{-2}$. The emission measure for the lower temperature component is roughly half of the contribution of the higher temperature component. These properties are consistent with the X-ray emission from RS CVn stars (e.g. \citeauthor{Dempsey1993a} \citeyear{Dempsey1993a}, who find typical temperatures of 0.2 and 1.4 keV for RS CVn stars), and gives a reduced $\chi^2$\,=\,1.01, for 24 d.o.f. The observed and unabsorbed 0.5\,--\,10 keV fluxes for this model are 9\,$\times$\,10$^{-14}$ and 1.7\,$\times$\,10$^{-13}$ \flx, respectively. 

A model including an APEC\,+\,power-law is marginally inconsistent with the column density constraint, as it finds N$_{\rm H}$\,=\,2.2$^{+1.4}_{-1.0}$\,$\times$\,10$^{21}$\,cm$^{-2}$, with photon index of 2.1$^{+0.8}_{-0.7}$ and APEC temperature of 1.3\,$\pm$\,0.3 keV. A variable-abundance single APEC model finds a metal abundance of 0.21$^{+0.27}_{-0.14}$, temperature of 1.5$^{+0.4}_{-0.3}$ keV, and reduced $\chi^2$\,=\,1.18, but the N$_{\rm H}$ appears rather low, at 2.0$^{+0.9}_{-0.6}$\,$\times$\,10$^{21}$\,cm$^{-2}$. We also test a model of a neutron star atmosphere (NSATMOS, \citeauthor{Heinke2006} \citeyear{Heinke2006}) plus power-law, as seen in quiescent low-mass X-ray binaries (e.g. \citeauthor{Campana1998} \citeyear{Campana1998}); we find an acceptable fit (reduced $\chi^2$\,=\,1.21, for 25 d.o.f.), with N$_{\rm H}$\,=\,5.4\,$\times$\,10$^{21}$\,cm$^{-2}$, NS temperature 10$^{5.9}$ K, and photon index 2.2. We note that this model does not readily explain the residuals around 1 keV. 

\section{Finding charts}
\label{sec:appendix}
\begin{figure*}
\centering
\begin{minipage}{0.3\textwidth}
  \centering
  \includegraphics[width=\linewidth]{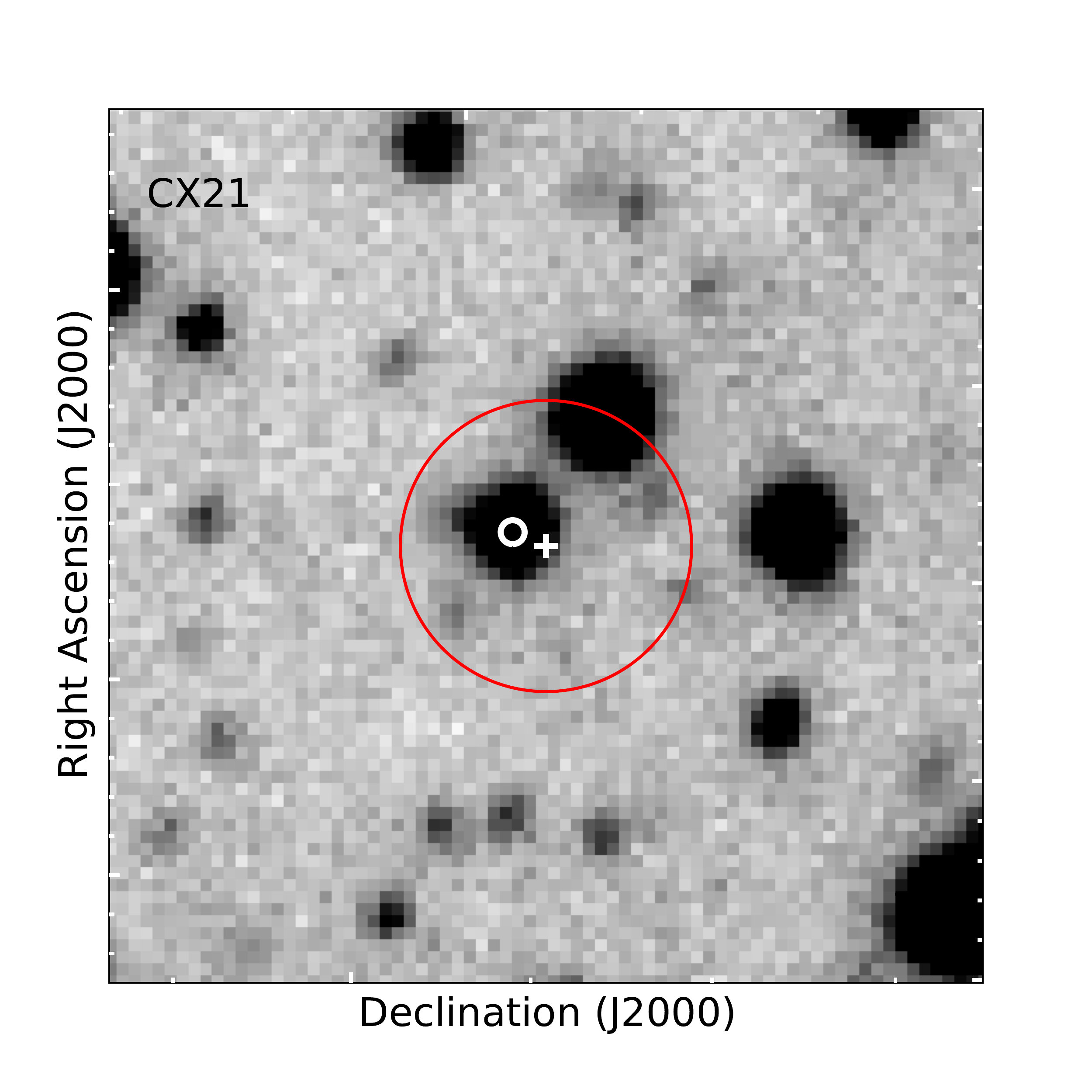}
  %\captionof{figure}{A figure}
  %\label{fig:test1}
\end{minipage}%
\begin{minipage}{.3\textwidth}
  \centering
  \includegraphics[width=\linewidth]{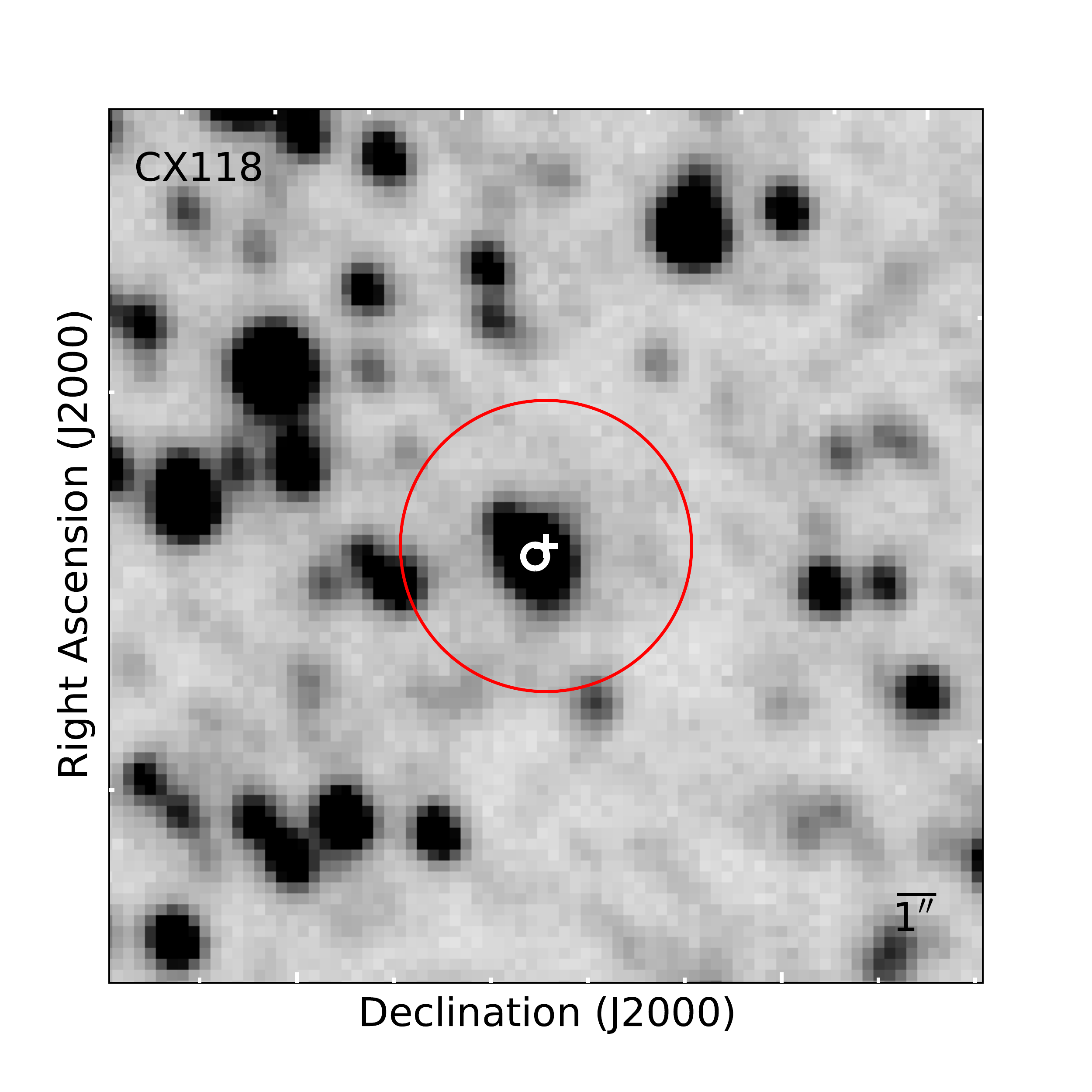}
\end{minipage}
\begin{minipage}{0.3\textwidth}
  \centering
  \includegraphics[width=\linewidth]{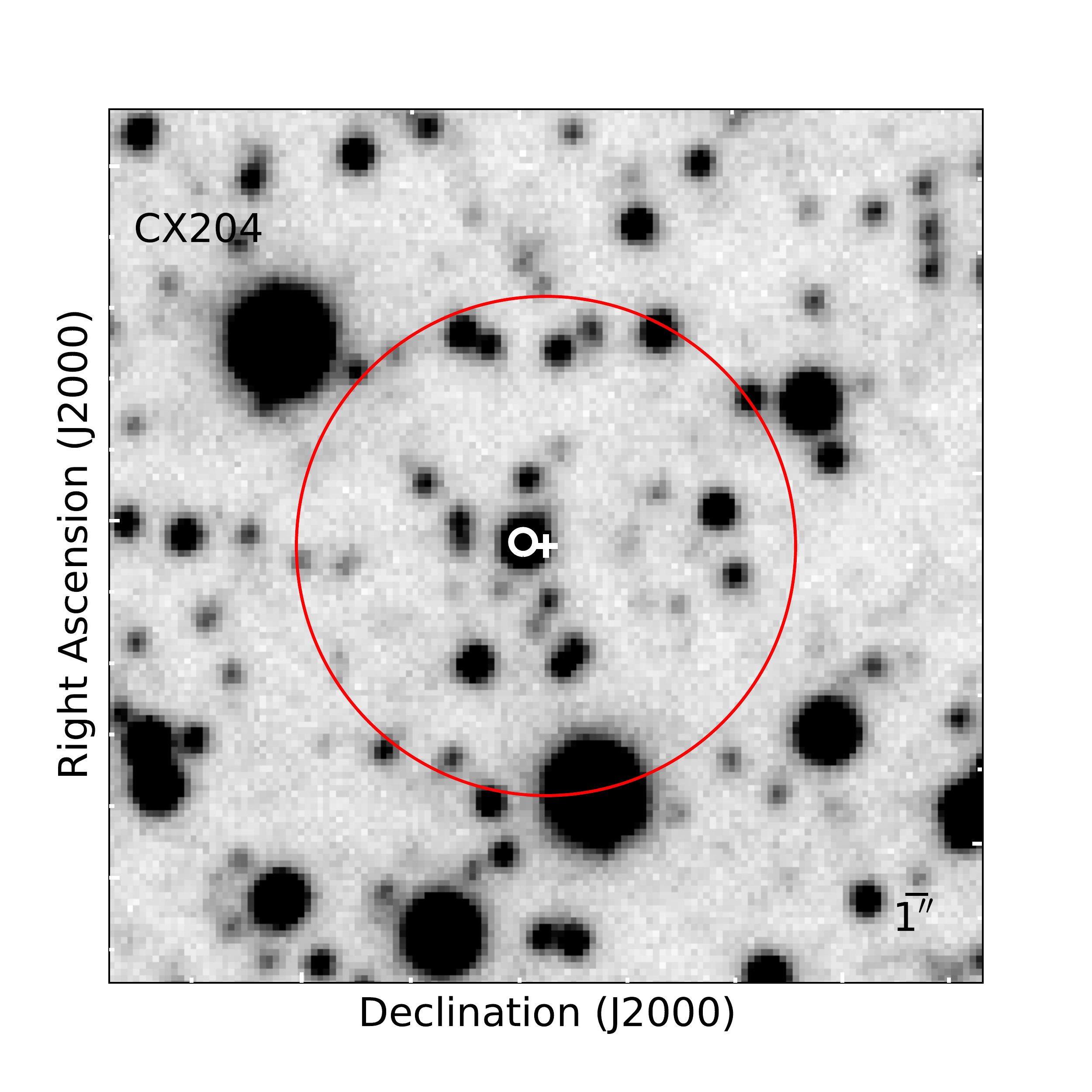}
\end{minipage}

\begin{minipage}{.3\textwidth}
  \centering
  \includegraphics[width=\linewidth]{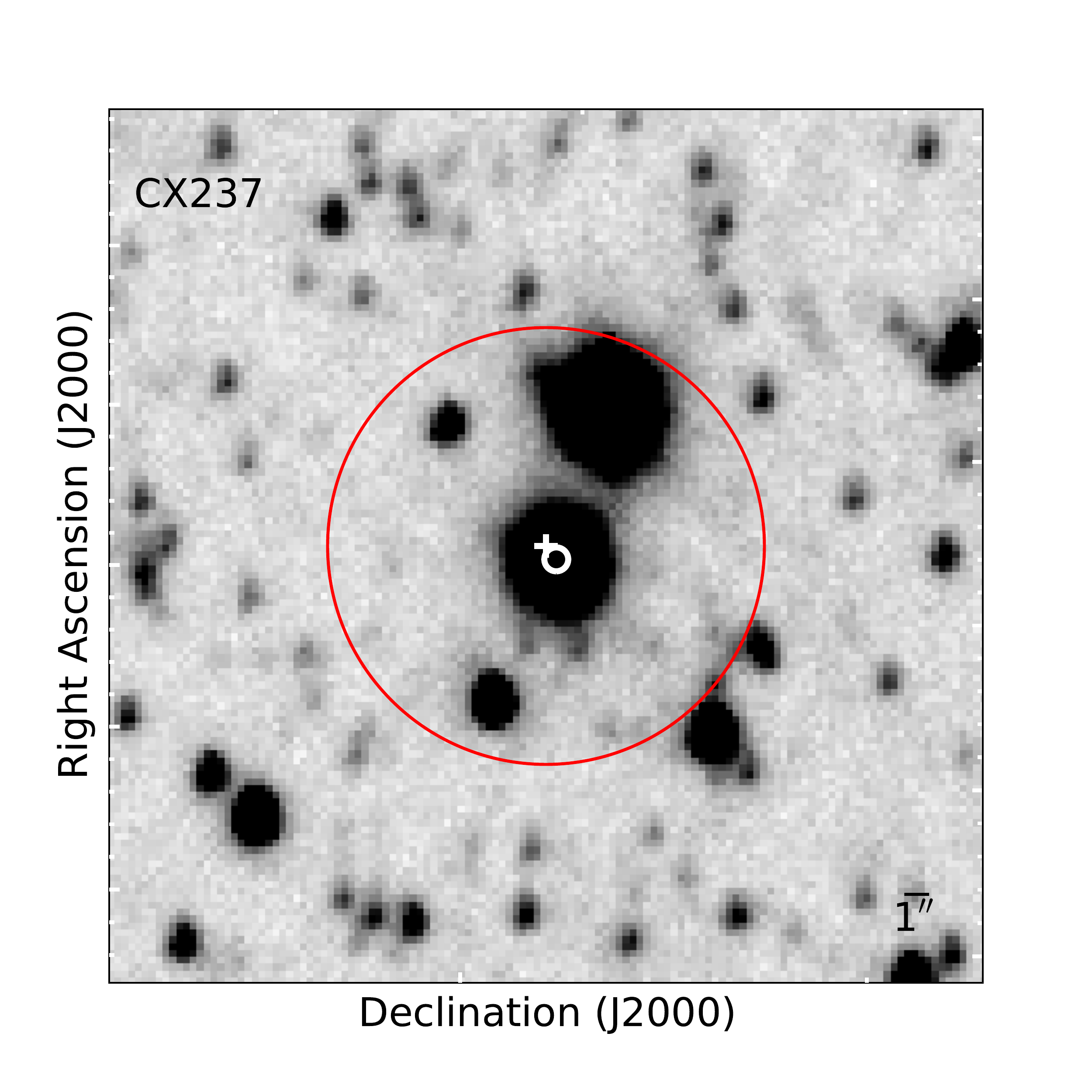}
\end{minipage}
\begin{minipage}{0.3\textwidth}
  \centering
  \includegraphics[width=\linewidth]{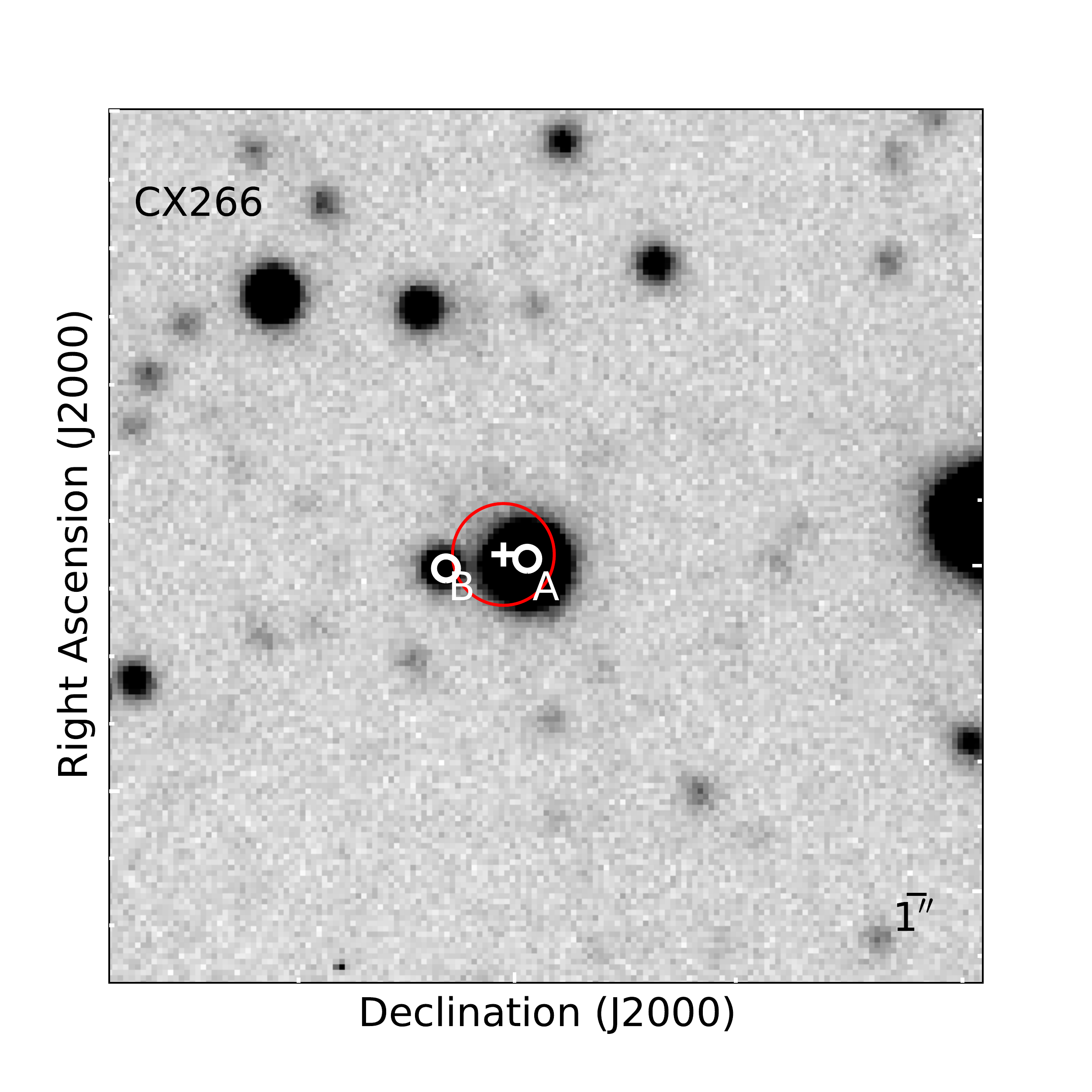}
\end{minipage}%
\begin{minipage}{.3\textwidth}
  \centering
  \includegraphics[width=\linewidth]{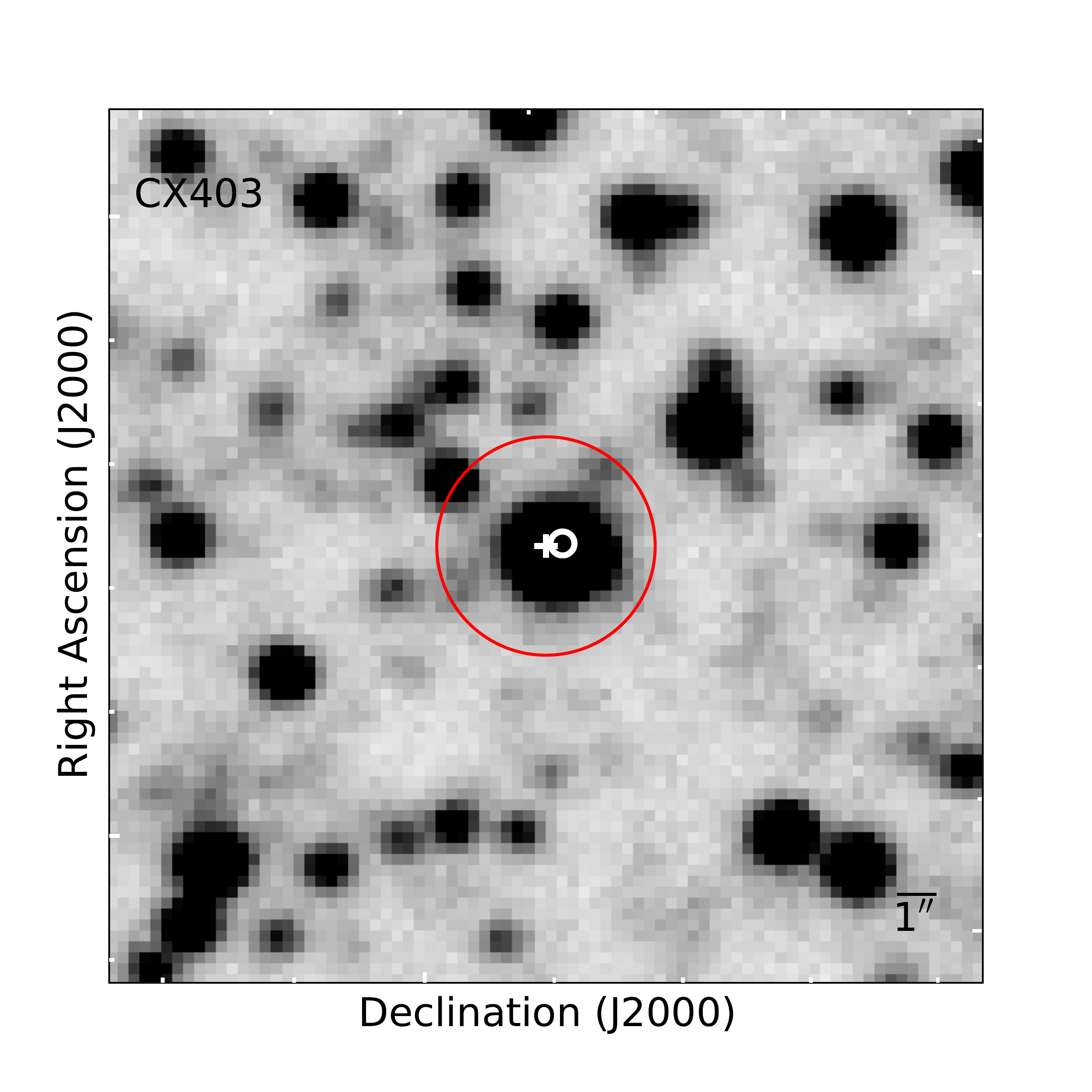}
\end{minipage}

\begin{minipage}{0.3\textwidth}
  \centering
  \includegraphics[width=\linewidth]{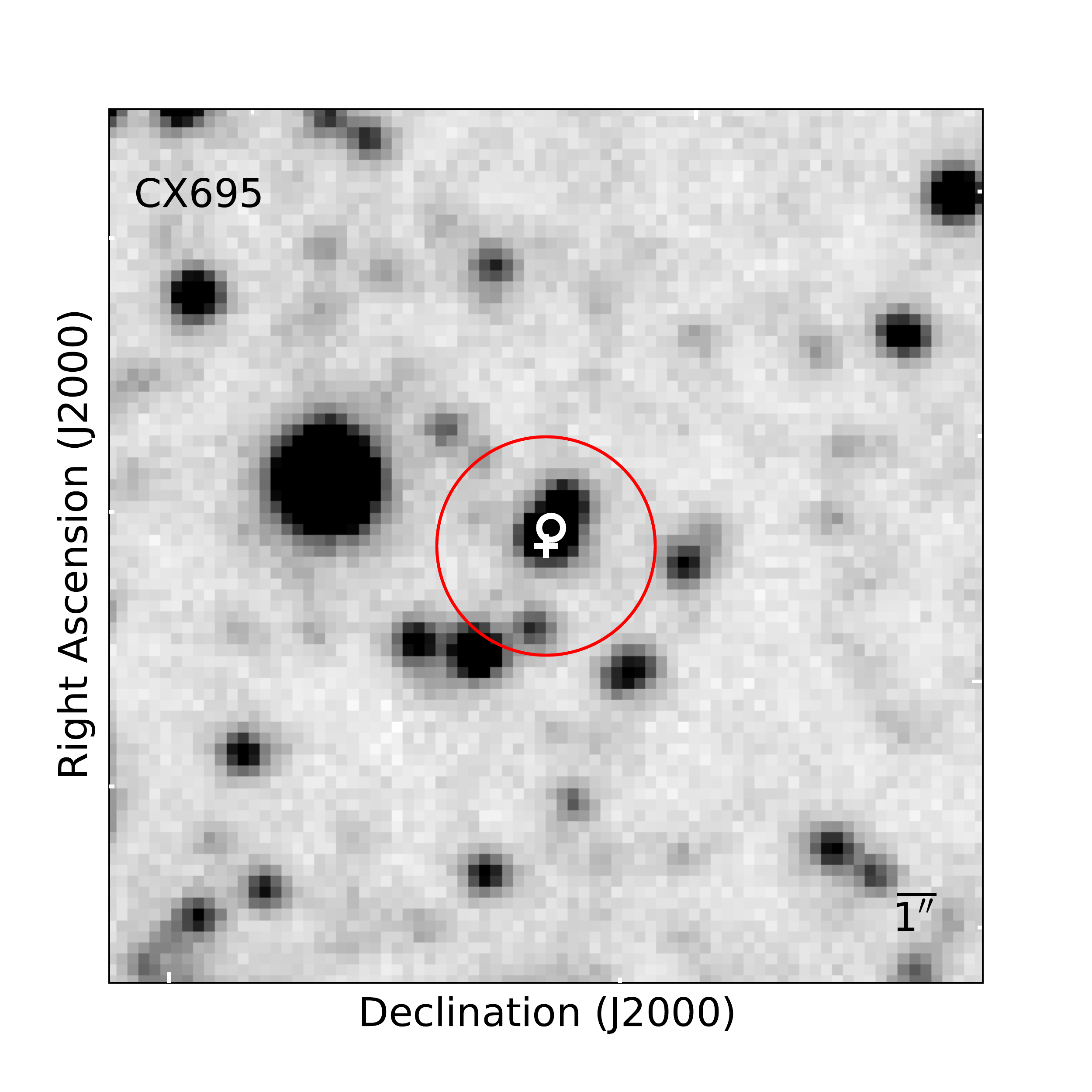}
\end{minipage}%
\begin{minipage}{.3\textwidth}
  \centering
  \includegraphics[width=\linewidth]{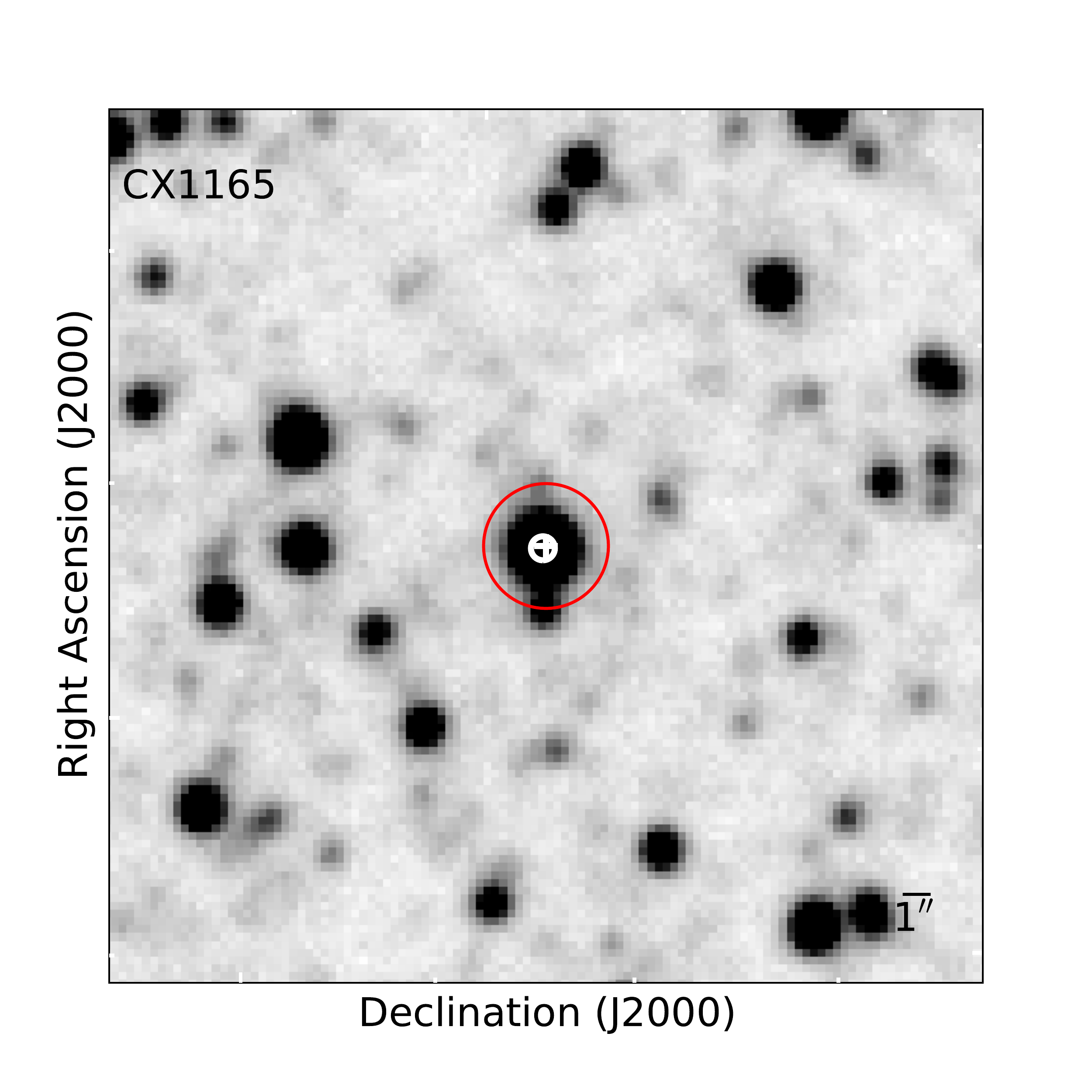}
\end{minipage}
\begin{minipage}{.3\textwidth}
  \centering
  \includegraphics[width=\linewidth]{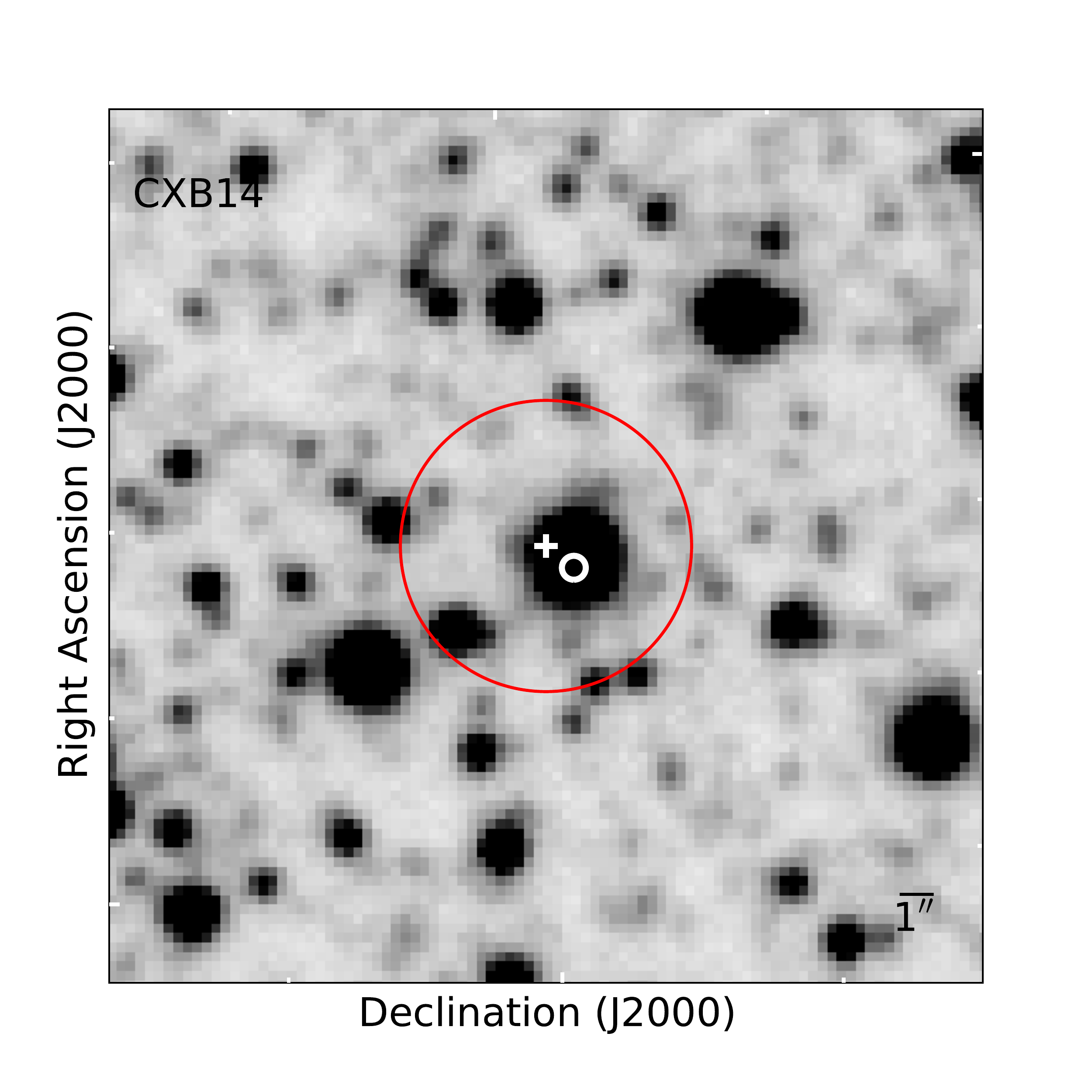}
\end{minipage}

\begin{minipage}{.3\textwidth}
  \centering
  \includegraphics[width=\linewidth]{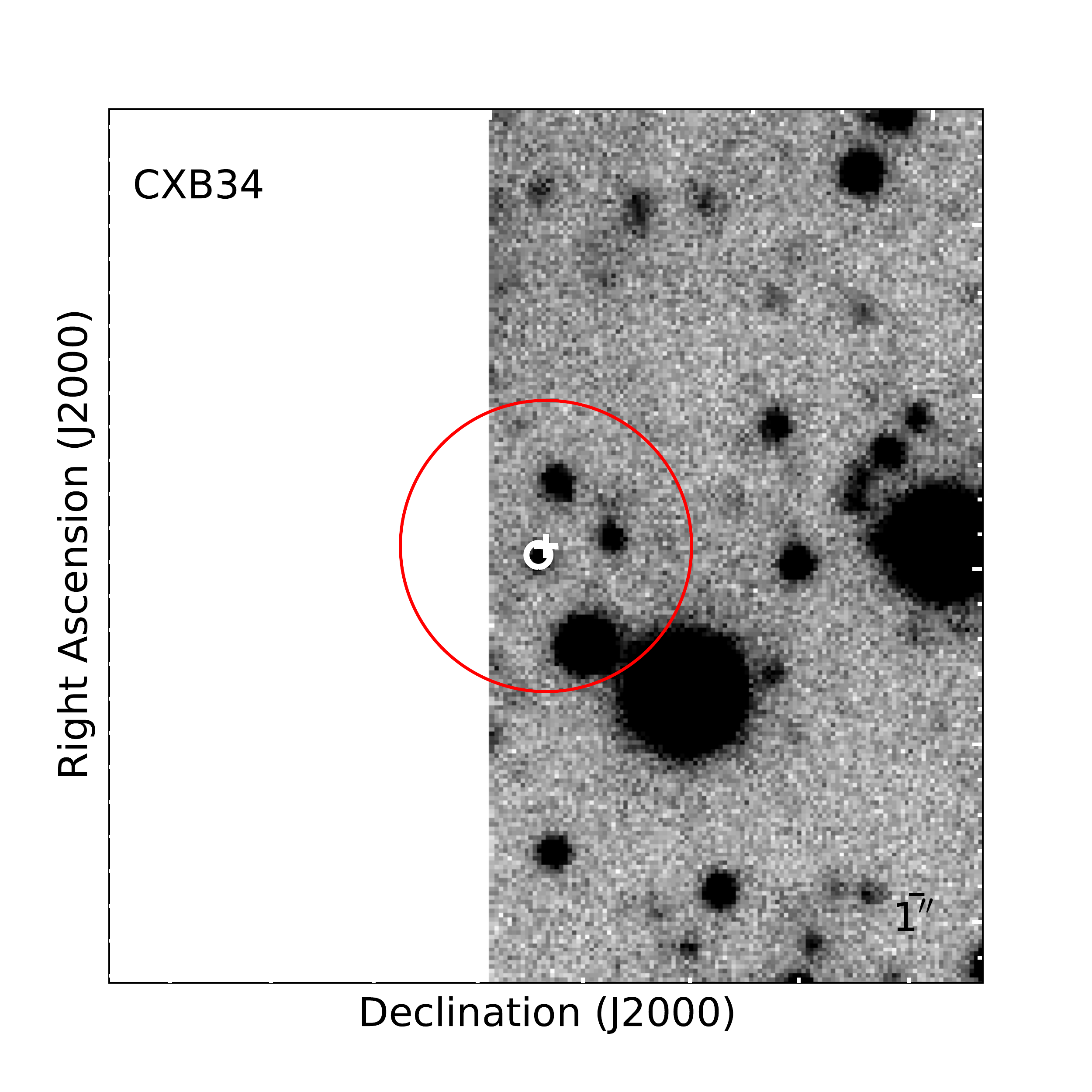}
\end{minipage}
\begin{minipage}{0.3\textwidth}
  \centering
  \includegraphics[width=\linewidth]{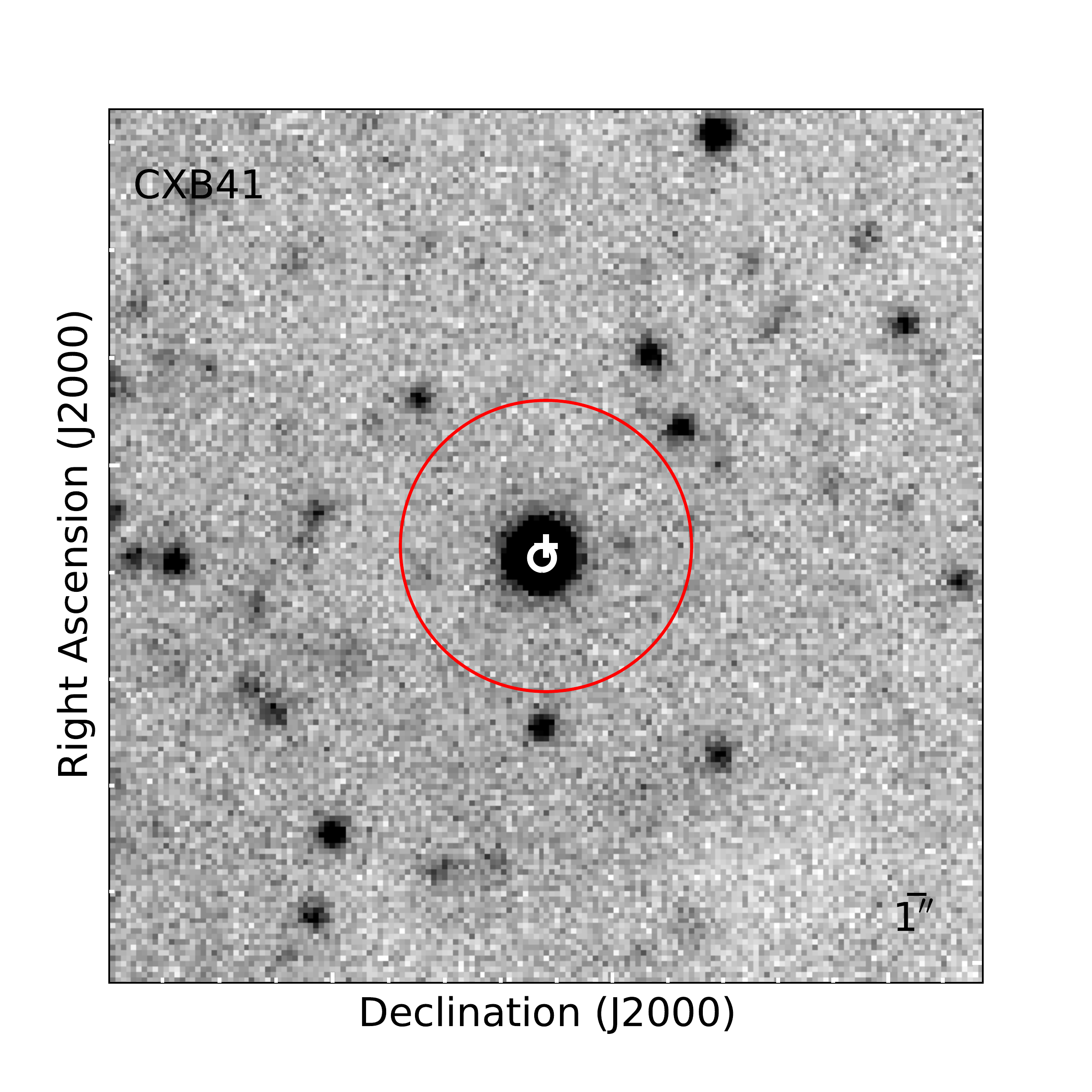}
\end{minipage}%
\begin{minipage}{.3\textwidth}
  \centering
  \includegraphics[width=\linewidth]{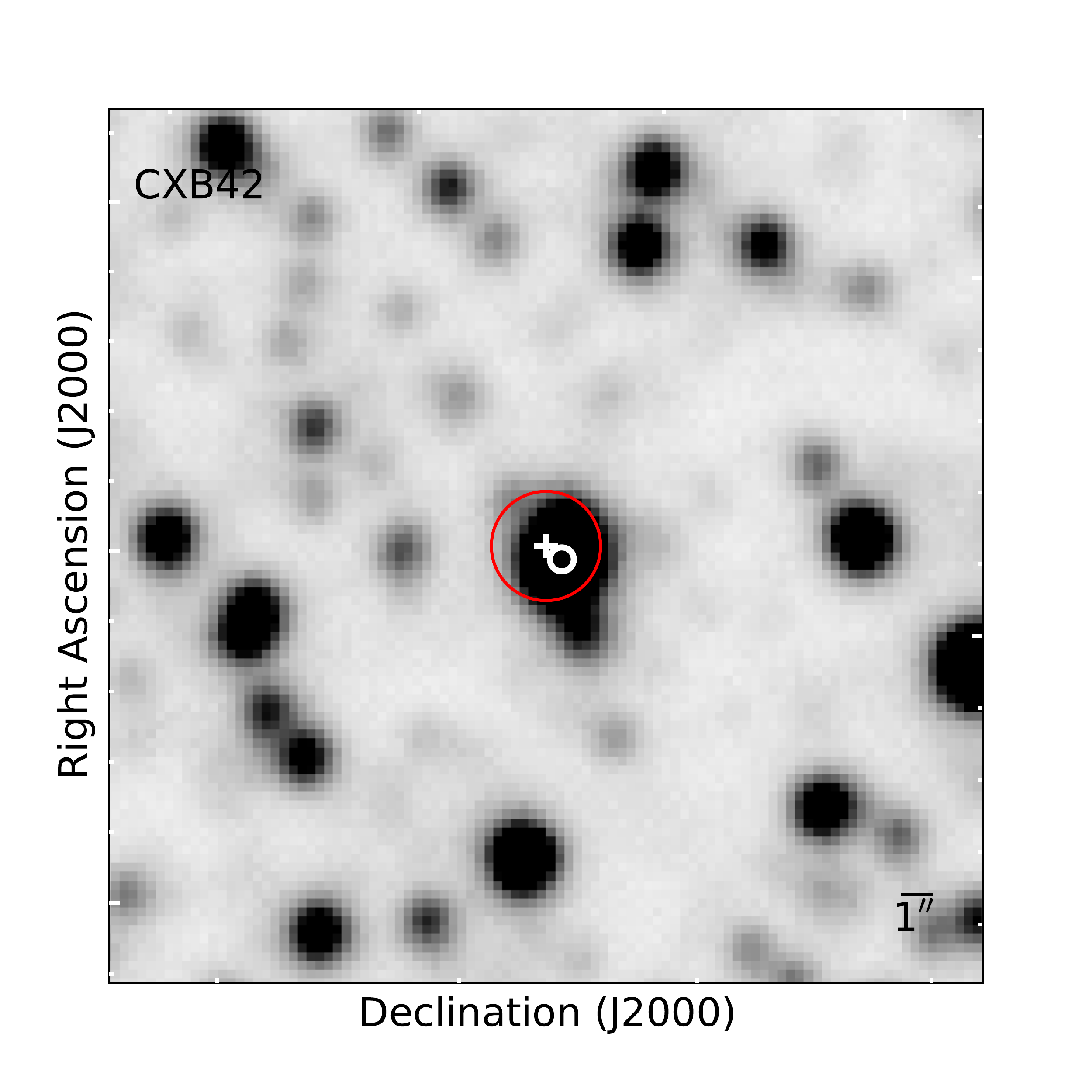}
\end{minipage}
\caption{Finding charts of the optical counterparts in the $r^{\prime}$-band. East is up, and North is to the right. The red circle indicates the X-ray error circle, while the white cross marks the best-fit X-ray position and the white circle the optical counterpart of which we obtained the spectrum.}
\end{figure*}

\begin{figure*}
\centering
\begin{minipage}{0.3\textwidth}
  \centering
  \includegraphics[width=\linewidth]{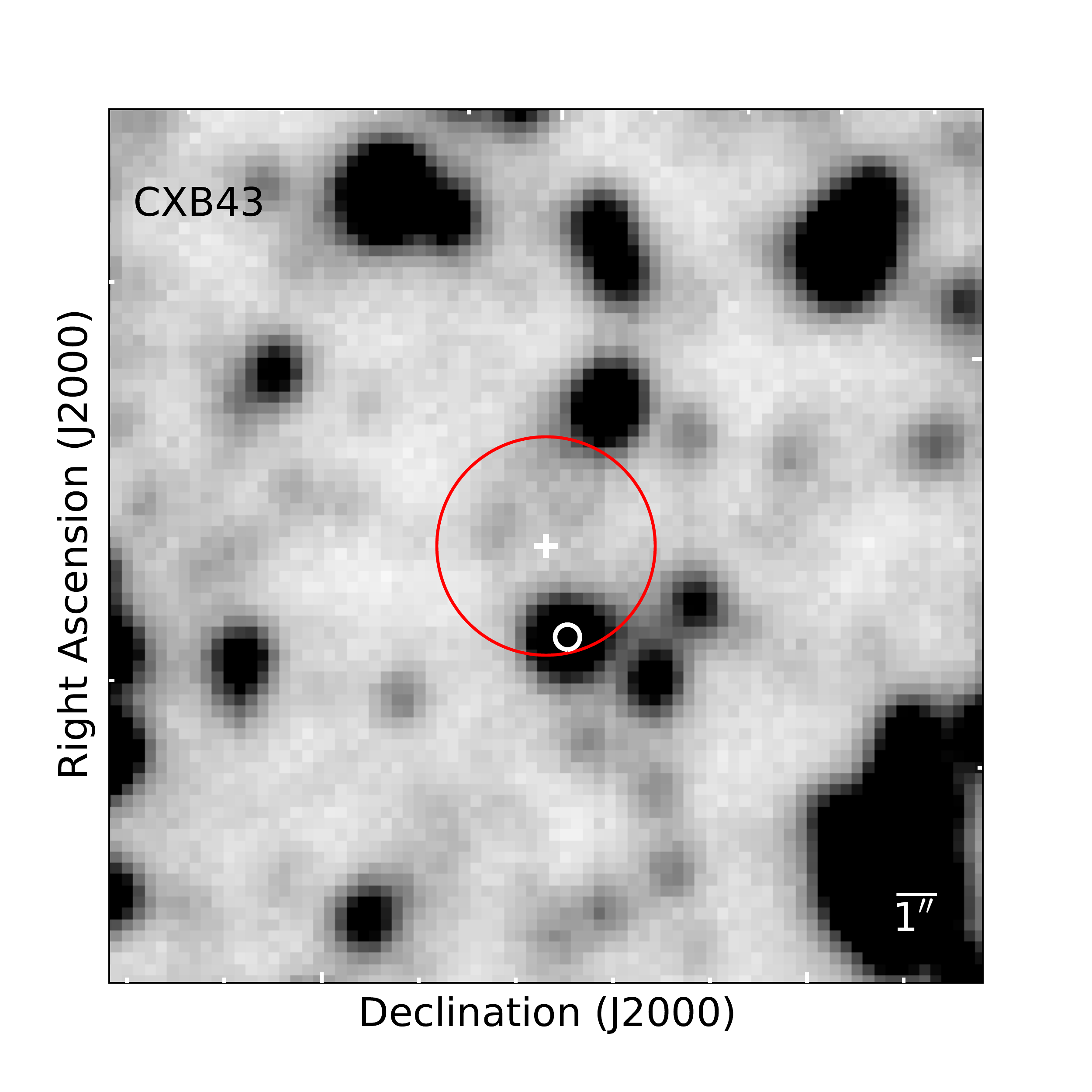}
\end{minipage}
\begin{minipage}{.3\textwidth}
  \centering
  \includegraphics[width=\linewidth]{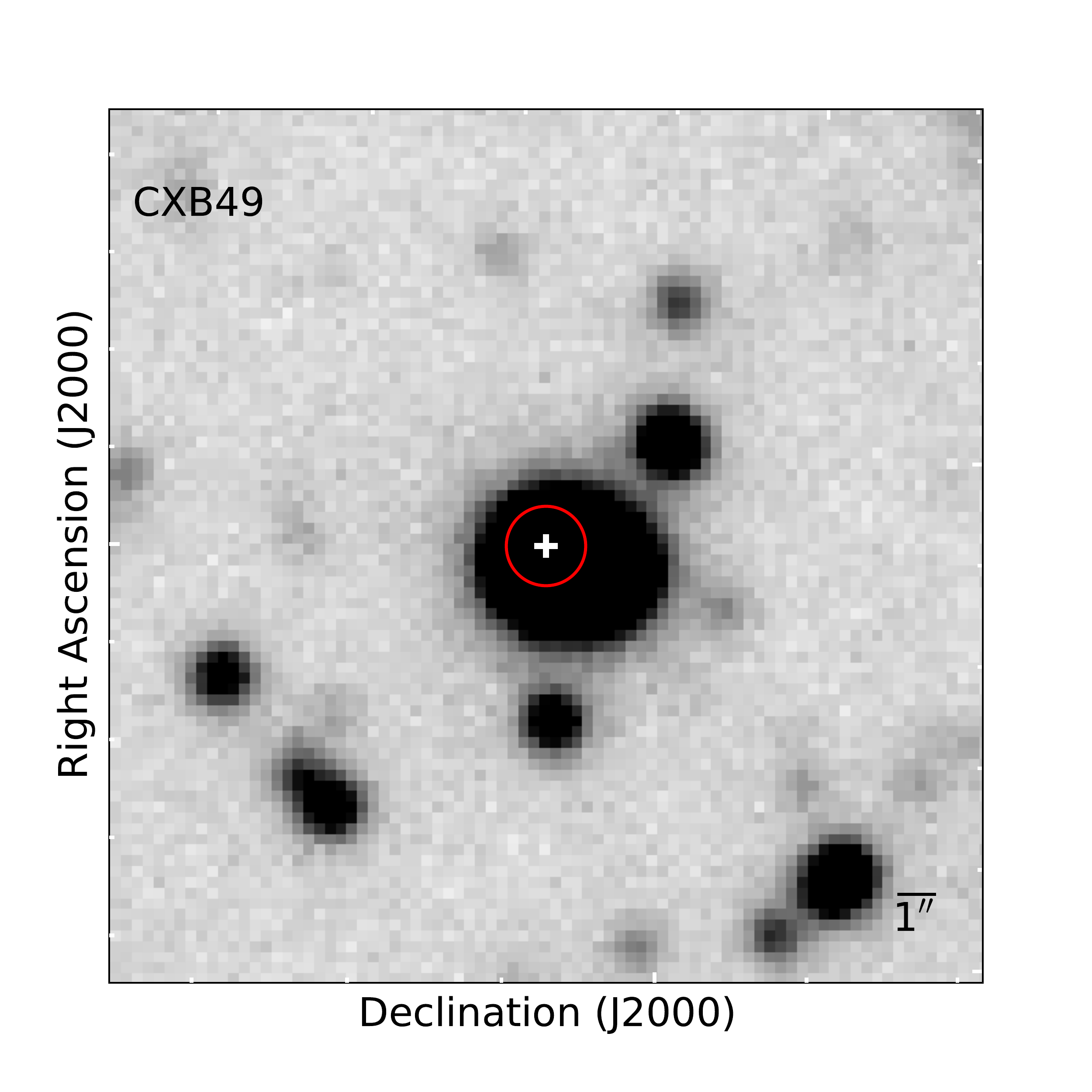}
\end{minipage}
\begin{minipage}{.3\textwidth}
  \centering
  \includegraphics[width=\linewidth, height=4.3cm]{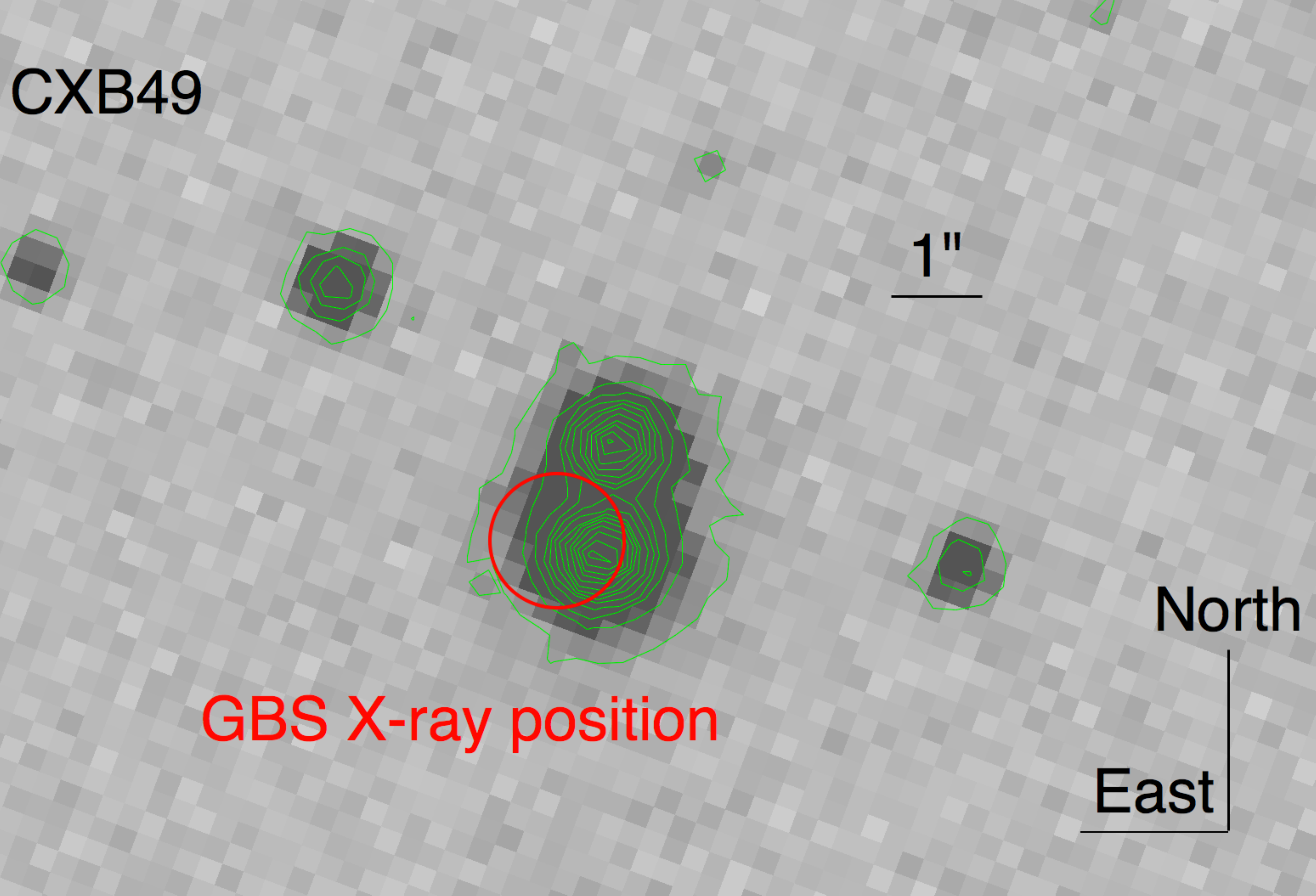}
\end{minipage}

\begin{minipage}{0.3\textwidth}
  \centering
  \includegraphics[width=\linewidth]{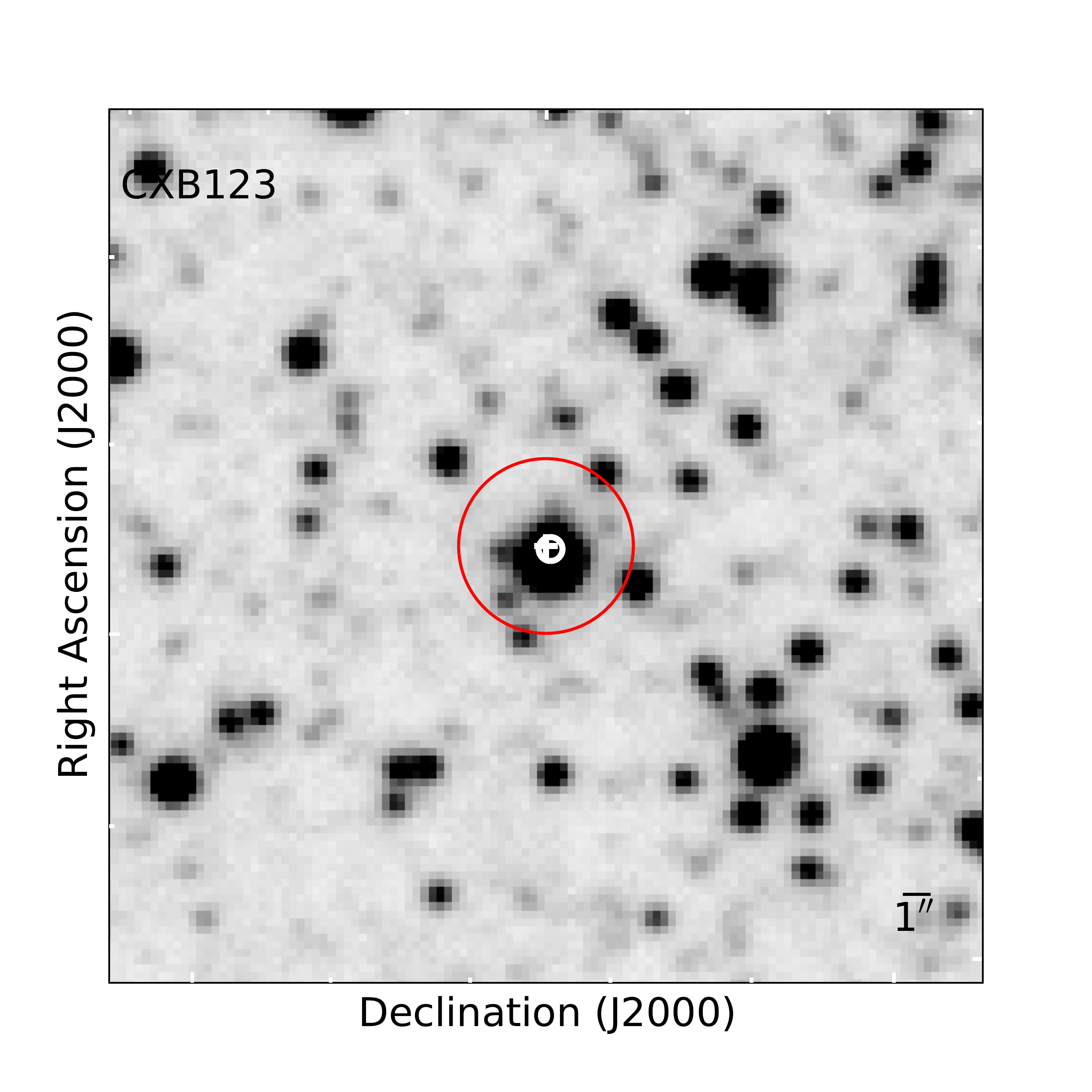}
\end{minipage}%
\begin{minipage}{.3\textwidth}
  \centering
  \includegraphics[width=\linewidth]{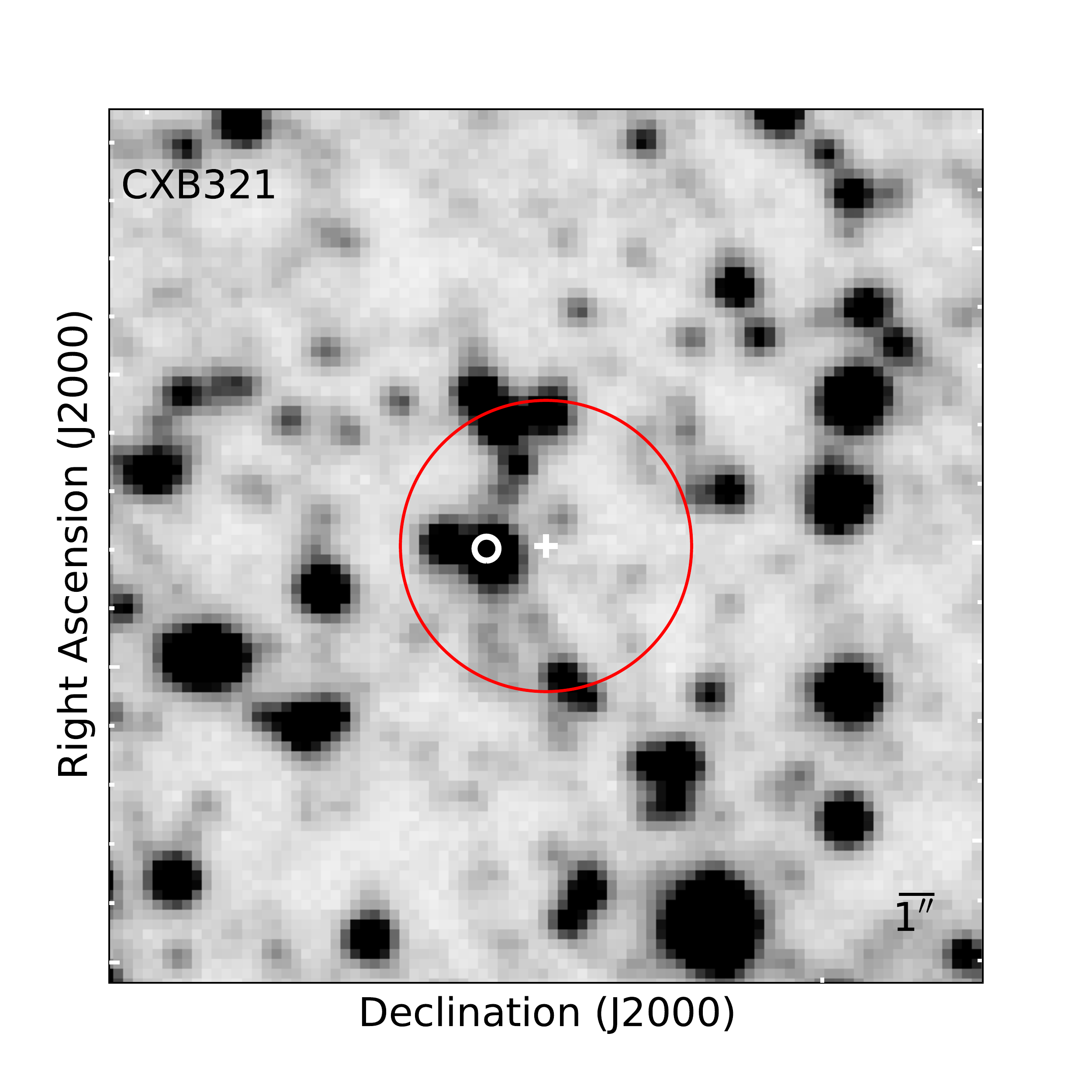}
\end{minipage}
\begin{minipage}{.3\textwidth}
  \centering
  \includegraphics[width=\linewidth]{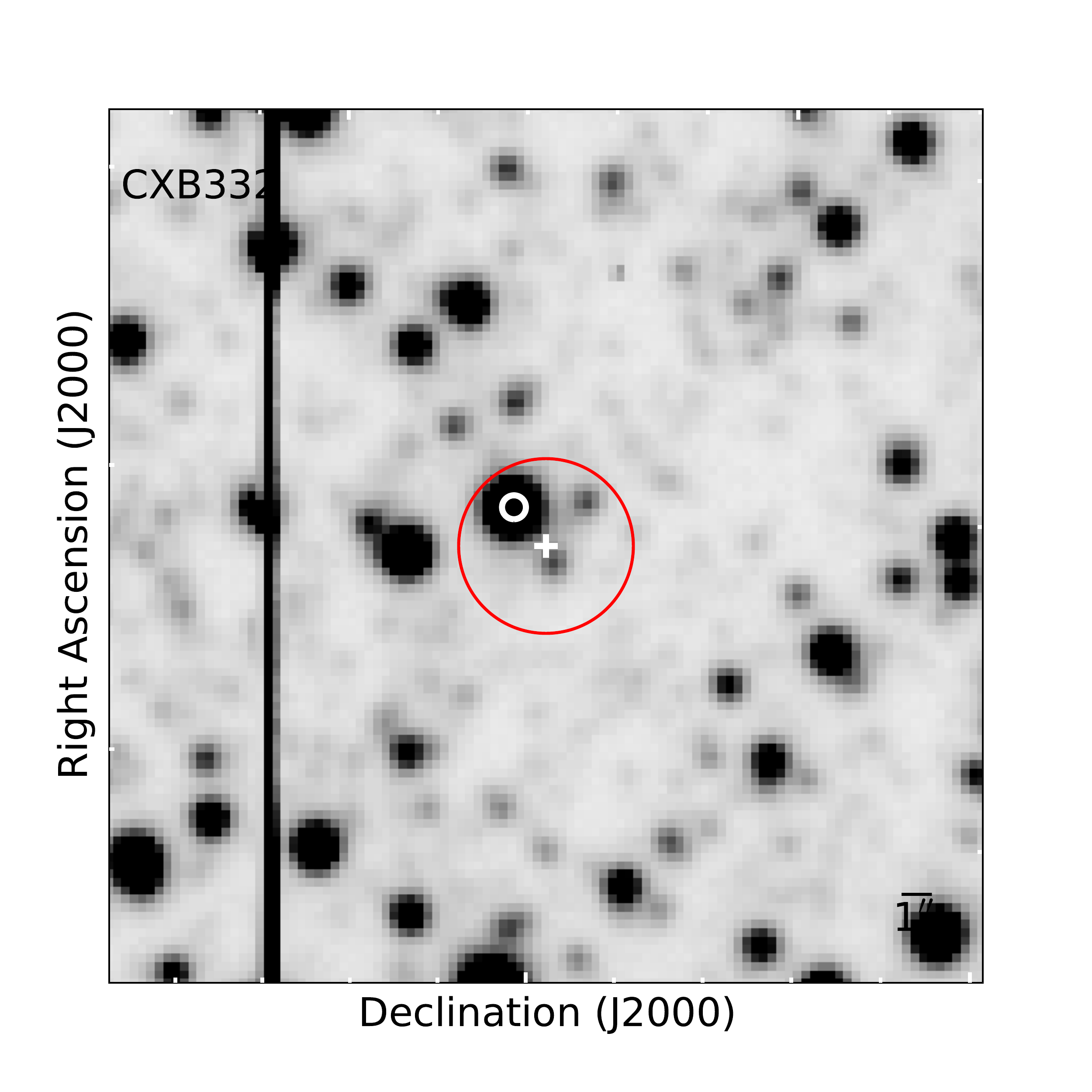}
\end{minipage}

\begin{minipage}{0.3\textwidth}
  \centering
  \includegraphics[width=\linewidth]{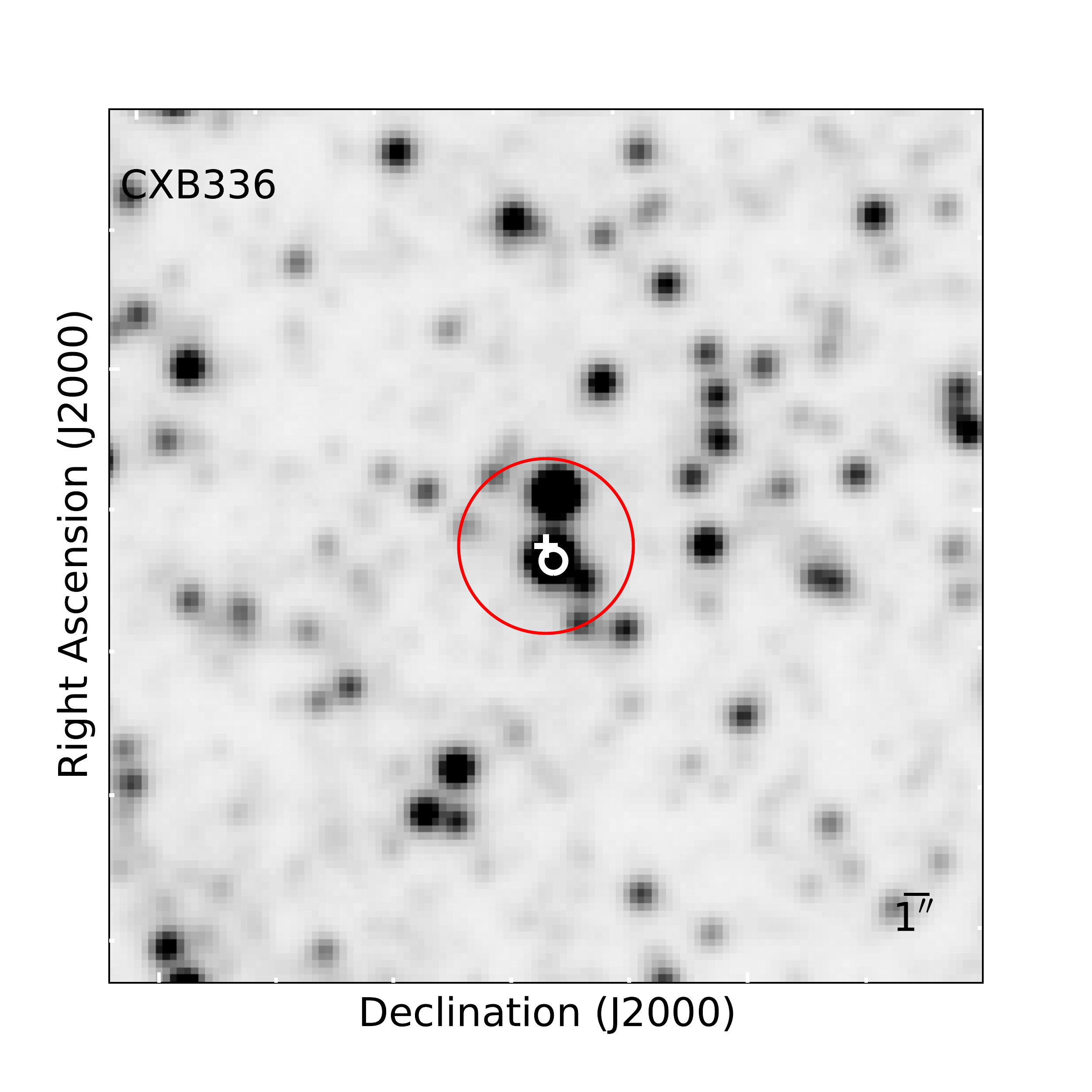}
\end{minipage}%
\begin{minipage}{.3\textwidth}
  \centering
  \includegraphics[width=\linewidth]{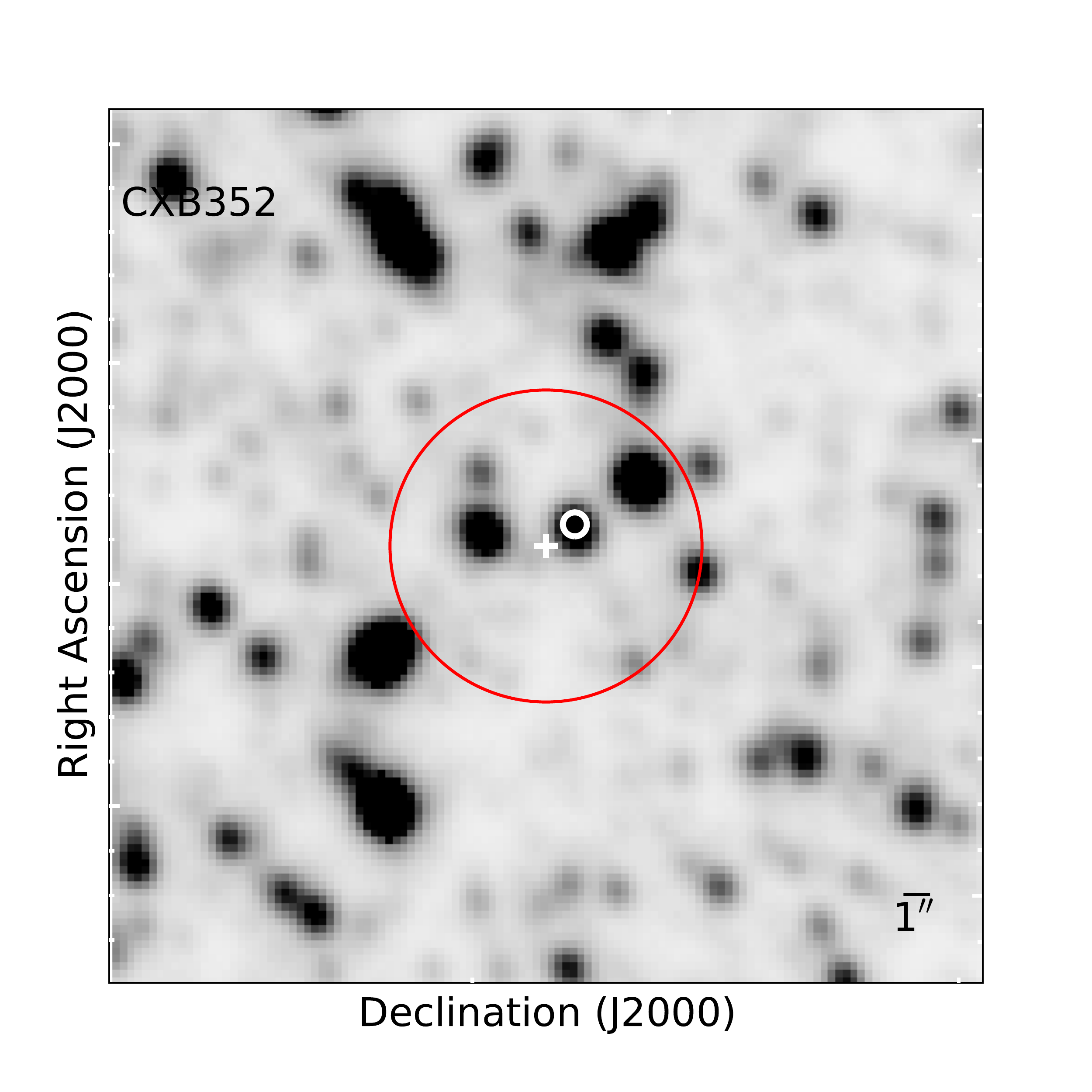}
\end{minipage}
\caption{Finding charts of the counterparts in the $r^{\prime}$-band. East is up, and North is to the right unless it is indicated on the figure. A red circle indicates the X-ray error circle, while the white cross marks the best-fit X-ray position and the white circle the optical counterpart of which we obtained the spectrum.}
\end{figure*}
\label{lastpage}

\end{document}